\documentclass[12pt,a4paper]{article}

\usepackage[T1]{fontenc}
\usepackage[latin1]{inputenc}
\usepackage[nosort]{cite}
\usepackage{color}
\usepackage[bulletsep]{collref}
\usepackage{graphicx}
\usepackage{bbm}
\usepackage{amsmath}
\usepackage{amssymb}
\usepackage{physics}
\usepackage{mathtools}
\usepackage{subfig}
\usepackage{verbatim}
\usepackage{multirow}
\usepackage{tikz}
\usepackage{circuitikz}
\usepackage{amsthm}
\usepackage{fancyhdr}
\usepackage{wrapfig}
\usepackage{hyperref}
\usepackage{dsfont}
\usepackage{delimset} 
\usepackage{shuffle} 
\usepackage{mathtools}
\usepackage{nicefrac}
\usepackage{enumitem}

\makeatletter \@addtoreset{equation}{section} \makeatother

\makeatletter
\let\old@startsection=\@startsection
\let\oldl@section=\l@section
\renewcommand{\@startsection}[6]{\old@startsection{#1}{#2}{#3}{#4}{#5}{#6\mathversion{bold}}}
\renewcommand{\l@section}[2]{\oldl@section{\mathversion{bold}#1}{#2}}
\makeatother

\makeatletter
\let\old@makecaption=\@makecaption
\def\@makecaption{\small\old@makecaption}
\makeatother

\let\oldPhi=\Phi
\let\oldPsi=\Psi
\let\oldGamma=\Gamma
\let\oldDelta=\Delta
\let\oldSigma=\Sigma
\let\oldTheta=\Theta
\let\oldPi=\Pi
\let\oldUpsilon=\Upsilon
\renewcommand{\Phi}{\mathnormal{\oldPhi}}
\renewcommand{\Psi}{\mathnormal{\oldPsi}}
\renewcommand{\Gamma}{\mathnormal{\oldGamma}}
\renewcommand{\Sigma}{\mathnormal{\oldSigma}}
\renewcommand{\Delta}{\mathnormal{\oldDelta}}
\renewcommand{\Theta}{\mathnormal{\oldTheta}}
\renewcommand{\Pi}{\mathnormal{\oldPi}}
\renewcommand{\Upsilon}{\mathnormal{\oldUpsilon}}

\newcommand{\superN}{\mathcal{N}}

\newcommand{\sign}{\mathop{\mathrm{sign}}}
\renewcommand{\Re}{\mathop{\mathrm{Re}}}

\newcommand{\gen}[1]{\mathrm{#1}}
\newcommand{\levo}[1]{ \gen{\widehat #1}}
\newcommand{\levz}[1]{ \gen{ #1}}

\newcommand{\Eval}{s} 
\makeatletter
\newlength{\apb@width}
\newcommand{\autoparbox}[2][c]{\settowidth{\apb@width}{#2}\parbox[#1]{\apb@width}{#2}}
\newcommand{\includegraphicsbox}[2][]{\autoparbox{\includegraphics[#1]{#2}}}
\makeatother

\ifx\genfrac\sdflkaj

\else

\fi
\newcommand{\sfrac}[2]{{\textstyle\frac{#1}{#2}}}

\makeatletter
\def\mr@ignsp#1 {\ifx\:#1\@empty\else #1\expandafter\mr@ignsp\fi}%
\newcommand{\multiref}[1]{\begingroup
\xdef\mr@no@sparg{\expandafter\mr@ignsp#1 \: }%
\def\mr@comma{}%
\@for\mr@refs:=\mr@no@sparg\do{\mr@comma\def\mr@comma{,}\ref{\mr@refs}}%
\endgroup}
\makeatother

\newcommand{\hypref}[2]{\ifx\href\asklfhas #2\else\href{#1}{#2}\fi}
\newcommand{\Secref}[1]{Section~\multiref{#1}}
\newcommand{\secref}[1]{Section~\multiref{#1}}
\newcommand{\Appref}[1]{Appendix~\multiref{#1}}
\newcommand{\appref}[1]{Appendix~\multiref{#1}}
\newcommand{\Tabref}[1]{Table~\multiref{#1}}

\newcommand{\Figref}[1]{Figure~\multiref{#1}}

\ifx\href\asklfhas\newcommand{\href}[2]{#2}\fi

\newcommand{\be}{\begin{eqnarray}}
\newcommand{\ee}{\end{eqnarray}}

\newcommand*\FF[4]{F_{#1}\left[\genfrac{}{}{0pt}{1}{#2}{#3};#4\right]}
\newcommand*\cFF[4]{\mathcal{F}_{#1}\left[\genfrac{}{}{0pt}{1}{#2}{#3};#4\right]}
\newcommand*\FL[5]{F_{#1}^{(#2)}\left[\genfrac{}{}{0pt}{1}{#3}{#4};#5\right]}
\newcommand*\cFL[5]{\mathcal{F}_{#1}^{(#2)}\left[\genfrac{}{}{0pt}{1}{#3}{#4};#5\right]}
\newcommand*\GG[5]{\,{}_{#1}F_{#2}\left[\genfrac{}{}{0pt}{1}{#3}{#4};#5\right]}
\newcommand*\cGG[5]{\,{}_{#1}\mathcal{F}_{#2}\left[\genfrac{}{}{0pt}{1}{#3}{#4};#5\right]}
\newcommand*\cGGt[5]{\,{}_{#1}\mathcal{F}^{\mathrm{2D}}_{#2}\left[\genfrac{}{}{0pt}{1}{#3}{#4};#5\right]}
\newcommand*\cGGtb[5]{\,{}_{#1}\overline{\mathcal{F}}^{\mathrm{2D}}_{#2}\left[\genfrac{}{}{0pt}{1}{#3}{#4};#5\right]}

\newcommand*\GH[4]{G_{#1}\left[\genfrac{}{}{0pt}{1}{#2}{#3};#4\right]}
\newcommand*\cGH[4]{\mathcal{G}_{#1}\left[\genfrac{}{}{0pt}{1}{#2}{#3};#4\right]}
\newcommand*\BB[4]{\mathcal{B}_{#1}\left[\genfrac{}{}{0pt}{1}{#2}{#3};#4\right]}
\newcommand*\CC[4]{\mathcal{C}_{#1}\left[\genfrac{}{}{0pt}{1}{#2}{#3};#4\right]}
\newcommand*\CCt[4]{\mathcal{C}^{\mathrm{2D}}_{#1}\left[\genfrac{}{}{0pt}{1}{#2}{#3};#4\right]}
\newcommand*\CCtb[4]{\overline{\mathcal{C}}^{\mathrm{2D}}_{#1}\left[\genfrac{}{}{0pt}{1}{#2}{#3};#4\right]}
\newcommand*\HH[4]{\mathcal{H}_{#1}\left[\genfrac{}{}{0pt}{1}{#2}{#3};#4\right]}
\newcommand*\PP[5]{\mathcal{P}^{(#1)}_{#2}\left[\genfrac{}{}{0pt}{1}{#3}{#4};#5\right]}
\newcommand*\TT[4]{\mathcal{T}_{#1}\left[\genfrac{}{}{0pt}{1}{#2}{#3};#4\right]}

\usepackage{color}

\newcommand{\ii}{\mathrm{i}}
\newcommand{\sgn}{\operatorname{s}}
\newcommand{\e}{\mathrm{e}}
\newcommand{\ep}{\kappa}
\newcommand{\bu}{\mathbf{u}}
\newcommand{\bbu}{\bar{\mathbf{u}}}
\newcommand{\ba}{\mathbf{a}}
\newcommand{\bba}{\mathbf{\bar{a}}}

\newcommand\doublequotient[3]{{_{\displaystyle #1}}\backslash{^{\displaystyle #2}}/{_{\displaystyle #3}}}

\begin{document}

\thispagestyle{empty}

\begin{flushright}\footnotesize
\texttt{BONN-TH-2025-29}
\end{flushright}
\vspace{.2cm}

\begin{center}%
{\LARGE\textbf{\mathversion{bold}%
Hypergeometry from $\mathrm{\widehat P}$-Symmetry: \\
Feynman Integrals in One and Two Dimensions}\par}

\vspace{1cm}

 \textsc{
 Gwena\"el Ferrando, 
 Florian Loebbert, \\
 Amelie Pitters, 
 Sven F. Stawinski
 }
\vspace{8mm} \\
\textit{%
Bethe Center for Theoretical Physics \\
Universit\"at Bonn, 53115, Germany
}
\vspace{.5cm}

\texttt{ \{gferrand, loebbert, apitters, sstawins\}@uni-bonn.de}

\par\vspace{15mm}

\textbf{Abstract} \vspace{5mm}

\begin{minipage}{13.2cm}

Feynman integrals with generic propagator powers in one and two spacetime dimensions are investigated from various perspectives. 
In particular, we argue that the class of track integrals at any loop order is fixed by the recently found $\mathrm{\widehat P}$-symmetries of Yangian type. All track integrals up to six external points (and four loops) are bootstrapped explicitly as well as the full family of one-loop integrals at any multiplicity. Moreover, the triangle tracks at generic loop order, which constitute the most generic family of track-type integrals, are bootstrapped in this way. The results are compared to the direct evaluation via a spectral transform from the integrability toolbox that turns out to be particularly efficient for position-space tree integrals in lower dimensions. We prove that all $\mathrm{\widehat P}$-symmetries of these integrals can be derived from the framework of Aomoto-Gelfand hypergeometric functions, which applies to integrals in one and two dimensions. 
Finally, we also demonstrate the method's applicability to conformal integrals by deriving the complete results for all comb-channel conformal partial waves as well as the conformal double-box integral. We explicitly go through all examples of the above integrals in 1D and then provide a straightforward recipe for how to read off their 2D counterparts.
\end{minipage}
\end{center}

\newpage 

\tableofcontents
\bigskip
\hrule


\section{Introduction}

Feynman integrals still constitute the most important building blocks for phenomenological predictions within the framework of quantum field theory. At the same time their computation poses a hard problem of mathematical physics and consequently has turned into a research area on its own. This difficulty is associated with 
how to define and represent Feynman integrals in an optimal way, and what it actually means to have computed a Feynman integral successfully.
\medskip

In many areas of physics,
it proved useful to address similar questions starting from an integrable toy model. In
quantum field theory, such a model is given by the planar $\superN=4$ super Yang--Mills (SYM) theory, whose integrability can be associated to
a Yangian symmetry \cite{Beisert:2010jq,Torrielli:2010kq,Loebbert:2016cdm}. This Yangian enhances the model's (super)conformal symmetry to an infinite-dimensional algebraic structure. Notably, the infinite set of symmetry operators is generated by the model's conformal Lie algebra symmetry and a single additional generator $\levo{P}^\mu$, the so-called \emph{level-one momentum}. In this sense, the operator $\levo{P}^\mu$ can be considered as the additional symmetry, which pushes the model across the boundary to integrability.%
\footnote{When mapped into a dual space, which is related to the original kinematical variables via a non-local transformation, this generator $\levo{P}^\mu$ transforms into a local representation of the special conformal generator. This duality of conformal symmetries is deeply connected to the integrability of the planar AdS/CFT correspondence, see e.g.\ \cite{Drummond:2010km}.}
\medskip

Similar to understanding $\superN=4$ SYM theory as the ``harmonic oscillator of quantum field theory'',
fishnet Feynman integrals provide a suitable integrable starting point for the systematic understanding of generic Feynman integrals. These fishnet integrals represent exact correlation functions of the planar fishnet conformal field theories that arise from particular double-scaling limits of the gamma-deformed $\superN=4$ SYM theory (or their generalizations) \cite{Gurdogan:2015csr,Caetano:2016ydc,Kazakov:2018qbr,Kazakov:2018gcy,Loebbert:2020tje,Kazakov:2022dbd,Alfimov:2023vev}. Already in 1981, Zamolodchikov argued that fishnet integrals feature integrable structures  \cite{Zamolodchikov:1980mb}, which more recently have been phrased in the form of a Yangian invariance that extends the integrals' (bosonic) conformal Lie algebra symmetry~\cite{Chicherin:2017cns,Chicherin:2017frs,Loebbert:2019vcj,Loebbert:2020hxk,Loebbert:2022nfu,Kazakov:2023nyu,Loebbert:2025abz}. In addition to being invariant under the differential representation of the (level-zero) conformal algebra $  \levz{J}^a \in \{\levz{P}^{\mu}, \levz{L}^{\mu \nu}, \levz{D}, \levz{K}^{\mu}\}$, these position space integrals are annihilated by the following non-local differential operator 
\begin{equation}
\levo{P}^\mu=\frac{1}{2} {f^{\gen{P}^\mu}}_{bc} \sum^{n}_{j=1} \sum^{n}_{k=j+1}  \levz{J}^c_j \levz{J}^b_k+ \sum_{j=1} \Eval_j \gen{P}^\mu_j.
\label{eq:Phat}
\end{equation}
Here $f^a{}_{bc}$ denotes the conformal Lie algebra structure constants, $\levz{J}^a_j$ the first-order differential operators representing the conformal generators on leg $j$, and $\Eval_j$ corresponds to the so-called evaluation parameters of the Yangian representation that are fixed by the integral's topology. For conformal Feynman integrals, this additional symmetry implies differential equations in terms of cross ratios for the conformal functions characterizing the integral \cite{Loebbert:2019vcj,Loebbert:2020glj,Duhr:2022pch,Duhr:2024hjf,Levkovich-Maslyuk:2024zdy}.
\medskip

After various classes of fishnet-type Feynman integrals had been identified as being conformal Yangian invariants, most recently it was shown that all ---~not necessarily conformal~--- scalar planar Feynman integrals are annihilated by $\levo{P}^\mu$  \cite{Loebbert:2024qbw,Loebbert:2025abz}. This holds in full generality for  position-space tree integrals and extends to loops provided the propagator powers within the loops ($k$-gons) sum up to $D(k-2)/2$, see also \cite{Levkovich-Maslyuk:2024zdy}. The above $\levo{P}$-symmetry can be shown to follow from `planarizing' the solutions of the momentum space conformal Ward identities of \cite{Bzowski:2019kwd,Bzowski:2020kfw} and mapping the result into dual $x$-space. If in this case level-zero conformal symmetry is not assumed, $\levo{P}$-invariance implies constraining differential equations in terms of Mandelstam-like ratios of the kinematic variables. This will be the situation considered in the present paper, see also \cite{Loebbert:2020glj,Loebbert:2020aos} for other non-conformal applications of the $\levo{P}$-symmetry.
\medskip

Notably, a distinguished role is played by (position-space) \emph{tree} integrals, which have no loops and thus the symmetry requires no constraints. Despite being of tree shape, we emphasize that these position space integrals are highly non trivial as they correspond to multi-loop integrals in the dual momentum space, including for instance the frequently discussed train track Feynman graphs \cite{Bourjaily:2018ycu,Vergu:2020uur,Duhr:2022pch,Cao:2023tpx,McLeod:2023doa,Duhr:2024hjf}. Notably, these tree integrals are always annihilated by $\levo{P}^\mu$, i.e. for any parametric values of propagator powers. Even more so, this class of tree-level Feynman graphs is annihilated by an enlarged set of $\levo{P}$ sub-symmetries, which correspond to partial sums over the densities $\levo{P}_{jk}$ of the level-one momentum \eqref{eq:Phat}, see  \cite{Loebbert:2024qbw}. It is one aim of the present paper to understand if the set of these differential operators is complete and can be used to fully fix or even define the respective integrals. In particular, we will demonstrate this completeness for the class of generic (non-conformal) \emph{track integrals}, which correspond to tree graphs defined by the property that every integration vertex is connected to at most two internal propagators:
\begin{equation}
\includegraphics{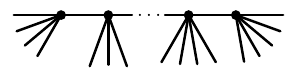}
\end{equation}
Here lines connecting points $j$ and $k$ correspond to position space propagators $\abs{x_{jk}}^{-2a_{jk}}$ with generic propagator powers~$a_{jk}$, while points indicate integrations over the respective coordinate $\int \dd x_j$. In particular, we do not impose any (e.g.\ conformal) constraints on the propagator powers. 

For various Feynman integrals it was shown that full Yangian plus permutation symmetry is sufficient to bootstrap their expressions in terms of polylogarithms or hypergeometric functions \cite{Loebbert:2019vcj,Loebbert:2020glj,Corcoran:2020epz}.
In particular, the corresponding analysis of fishnet integrals in two spacetime dimensions led to a new relation between integrability and Calabi--Yau geometry \cite{Duhr:2022pch,Duhr:2024hjf}. Here the set of differential equations arising from the Yangian plus the discrete permutation symmetries of a given graph can be identified with the Picard--Fuchs equations defining the period integrals of certain Calabi--Yau geometries. Importantly, a special property of two spacetime dimensions was used, namely that the linear combination of these Calabi--Yau periods
must be a single-valued double copy of (anti-)holomorphic periods, which is identified with the K\"ahler potential of the respective Calabi--Yau geometry. In the present paper, we perform a similar analysis of lower-dimensional Feynman integrals with the aim of generalizing the underlying assumptions of \cite{Duhr:2022pch,Duhr:2024hjf} as follows:
\begin{itemize}
\item conformal Yangian symmetry $\to$ only $\levo{P}$-symmetry,
\item fixed conformal propagator powers $\to $ generic parametric propagator powers,
\item two spacetime dimensions $\to$ one and two spacetime dimensions.
\end{itemize}
While  for the first two points the generalization is obvious, we should comment on the last item. A priori one can expect that lowering the spacetime dimension results in a simplification. There is, however, a little caveat here. As indicated above, Feynman integrals in two spacetime dimensions are computable via a well prescribed intersection pairing resulting in a single-valued double copy. In particular, in 2D this determines the linear combination of Yangian invariants up to an overall factor, cf.\ \cite{Duhr:2022pch,Duhr:2023bku,Duhr:2024hjf}. In one spacetime dimension no analogue of this double copy construction is known and hence the linear combination cannot be fixed in the same way. In this sense the analysis in 1D is closer to the general analysis in $D>2$ than the distinguished setup in 2D. Below we will go through a large number of 1D integrals explicitly (see \Tabref{tab:allints}) and then explain in \Secref{sec:1to2D} how to obtain the corresponding 2D integrals from these results.
\medskip

\begin{table}[t]
\begin{center}
  \renewcommand{\arraystretch}{1.3}
\begin{tabular}{|c||c|}\hline
Points|Loops& Track Feynman Graphs
\\\hline\hline
3|1&\includegraphicsbox[scale=.8]{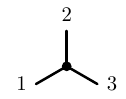}
\\\hline
4|1,2& \includegraphicsbox[scale=.8]{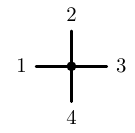} \quad  \includegraphicsbox[scale=.8]{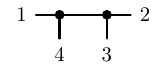}
\\\hline
5|1,2,3&\includegraphicsbox[scale=.8]{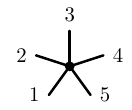}\quad \includegraphicsbox[scale=.8]{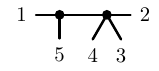}  \quad\includegraphicsbox[scale=.8]{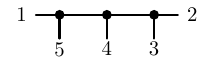}
\\\hline
6|1,2,3,4&\includegraphicsbox[scale=.8]{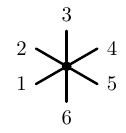} \quad
\begin{minipage}{2.3cm}
\includegraphicsbox[scale=.8]{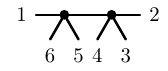}
\includegraphicsbox[scale=.8]{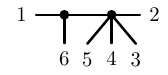}
\end{minipage}
\quad
\begin{minipage}{2.8cm}
\includegraphicsbox[scale=.8]{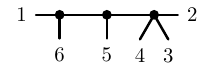}
\includegraphicsbox[scale=.8]{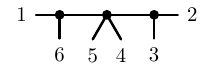}
\end{minipage}
\quad
\includegraphicsbox[scale=.8]{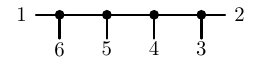}
\\\hline
$n|1$&\includegraphicsbox[scale=.8]{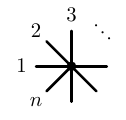}
\\\hline
$n|n-2$& \includegraphicsbox[scale=.8]{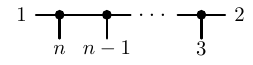}
\\\hline\hline
$6|2$ (conformal) & \includegraphicsbox[scale=.8]{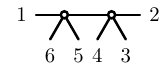}
\\\hline
$n|n-3$ (conformal) & \includegraphicsbox[scale=.8]{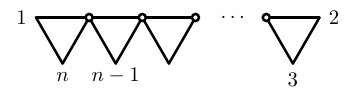}
\\\hline
\end{tabular}
\caption{Explicit Feynman graphs computed in this paper. White circles correspond to conformal integration points where propagator powers sum up to the spacetime dimension $D=1$ or $D=2$, respectively.}%
\label{tab:allints}
\end{center}
\end{table}

The approach to define Feynman integrals from their $\levo{P}$-symmetries is reminiscent of the systematic construction of Gelfand--Kapranov--Zelevinsky (GKZ) hypergeometric functions \cite{gkzToralManifolds,gkzEulerIntegrals,gkzTalk}. In fact, mapping the two approaches has recently been explored in \cite{Levkovich-Maslyuk:2024zdy} in the context of the full Yangian symmetry in $D$ spacetime dimensions. Here the generic spacetime dependence was used to argue for the independence of certain vectors in a conformal setup at sufficiently large~$D$, cf.\ also \cite{Loebbert:2019vcj}. In the present work we approach a similar question from the opposite direction by fixing the dimension to its smallest positive values $D=1$ and $D=2$. In particular, in one and two spacetime dimensions, Feynman integrals can be understood as Aomoto--Gelfand (AG) hypergeometric functions \cite{Duhr:2023bku}, which represent a subclass of the above GKZ systems \cite{aomotoHGFs,aomotoKita}. We will review this connection in \Secref{sec:AGfunctions} and demonstrate how the above $\levo{P}$-symmetries arise from the AG framework.
\medskip

Integrability is not known only for its symmetries, but also for the tools and representations these mathematical structures imply. One example is given by the \emph{separation of variables} (SoV) which represents a powerful concept applicable to a wide range of problems, ranging from the hydrogen atom in quantum mechanics to a sophisticated framework for non-compact spin chains and the above fishnet Feynman integrals \cite{Derkachov:2001yn,Derkachov:2002tf,Kirch:2004mk,Derkachov:2018rot,Derkachov:2019tzo,Derkachov:2021ufp}. In the present paper, we will demonstrate that rewriting single propagators via a \emph{spectral transform} provides a straightforward method for the evaluation of tree Feynman integrals in one dimension (see \Secref{sec:spectransf} for details). This trick from the integrability toolbox allows us to rewrite a propagator in such a way that all position space integrations for tree integrals can be straightforwardly performed using the so-called chain rule. The resulting expressions resemble, but differ from, a Mellin--Barnes representation and can be transformed into combinations of hypergeometric series. We will use this method as an efficient way to test and refine our bootstrap results.
\medskip

In the following sections we will discuss all cases of Feynman integrals in 1D explicitly; in \Secref{sec:1to2D} we will then explain in detail how to straightforwardly obtain the corresponding 2D integrals from the 1D expressions.
The paper is structured as follows.
In \Secref{sec:Phat} we review the $\levo{P}$-symmetries of position space Feynman integrals and we outline the bootstrap algorithm that is employed in the subsequent sections.
In \Secref{sec:spectransf} the spectral transform method is introduced and illustrated on a simple three-point example, as well as for the more elaborate case of the full family of comb-channel conformal blocks in one dimension.
In the following \Secref{sec:3pt} to \Secref{sec:6pt}, we explicitly bootstrap all examples of the above track integrals up to six external points and four loops, see \Tabref{tab:allints}. This $\levo{P}$-bootstrap is complemented by alternative derivations using the spectral transform, for details see \Appref{app:spectranscomps}.

In  \Secref{sec:polygons} we derive the full explicit results for the 1D family of $n$-point polygon (or star) integrals as linear combinations of hypergeometric functions:
\begin{equation}
I_n=\includegraphicsbox{FigNPointStar.pdf}\,.
\end{equation}
In dual momentum space defined via $p_j=x_j-x_{j+1}$ this graph maps to a (one-loop) polygon and we will refer to the respective integrals as polygon integrals in the following.

Extending the analysis, in \Secref{sec:triangletracks} we bootstrap the full result for all ($\ell$-loop) \emph{triangle track Feynman graphs}, which correspond to glued triangles in dual momentum space:
\begin{equation}
I_{2,1,\dots,1,2}=\includegraphicsbox{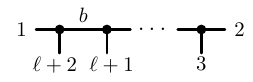}\,.
\end{equation}
Here the notation $I_{2,1,\dots,1,2}$ indicates how many external legs are attached to each integration vertex with a straightforward generalization to other graphs.
These triangle-track integrals can be considered the most general track-like integrals since we can obtain any other track diagram by taking coincidence limits of external points, as well as limits of propagator powers using the identity (later specified to $D=1$ or 2)
\begin{equation}\label{prop to delta}
	\lim_{b\to 0} \frac{b}{x^{2(D/2-b)}}=\frac{\pi^{D/2}}{\Gamma(\frac{D}{2})}\delta^{(D)}(x).
\end{equation}

In \Secref{sec:AGfunctions} we investigate the considered 1D Feynman integrals in the language of Aomoto--Gelfand hypergeometric functions and we show how the $\levo{P}$-symmetries can be derived from this perspective.

After describing the general recipe for how to read off expressions for 2D Feynman integrals from their 1D counterparts in \Secref{sec:1to2D},
 we illustrate it in  \Secref{sec:confdoubbox} on the generic conformal double box in 1D and 2D. We close with an outlook in \Secref{sec:outlook}.

\section{The $\levo{P}$-Bootstrap}
\label{sec:Phat}

In this section we introduce the non-local (spacetime) symmetries of tree-level Feynman integrals that we will employ in the subsequent sections for bootstrapping various examples of Feynman integrals.%
\footnote{For each position space Feynman graph, one of these symmetries, namely the full level-one momentum generator $\levo{P}^\mu$, is related to a momentum-space conformal symmetry $\gen{\bar K}^\mu$ in dual momentum space defined via $p^\mu=x_i^\mu-x_{i+1}^\mu$, cf.\ \cite{Loebbert:2020glj}.}
In particular, we will argue that the most general class of track-like integrals is fixed by the corresponding family of differential operators.
\subsection{$\levo{P}$-Symmetries of Tree Integrals}
\label{sec:PhatForTrees}
Let us briefly review the $\levo{P}$-symmetries of tree-level Feynman integrals that were identified in \cite{Loebbert:2024qbw}. These are constructed from the bilocal density of the level-one momentum generator which can be defined as
\begin{equation}
    \levo{P}^{\mu}_{jk} := \frac{i}{2} \left( \levz{P}^{\mu}_j \levz{D}_k + \levz{P}_{j \nu} \levz{L}^{\mu \nu}_k - i a_k \levz{P}^{\mu}_j - (j \leftrightarrow k) \right).
    \label{eq:Phatdens}
\end{equation}
Here the densities of the conformal generators in position space take the form
\begin{align}
    \levz{P}^{\mu}_j &= -i \partial^{\mu}_j, 
    &
    \levz{D}_j &= -i( x_{j \mu} \partial^{\mu}_j + \Delta_j ), 
    \nonumber\\
    \levz{L}^{\mu \nu}_j &= i( x^{\mu}_j \partial^{\nu}_j - x^{\nu}_j \partial^{\mu}_j ), 
    &
    \levz{K}^{\mu}_j &= -i(2 x^{\mu}_j x^{\nu}_j \partial_{j \nu} - x^2_j \partial^{\mu}_j + 2 \Delta_j x^{\mu}_j),
\end{align}
where the parameter $\Delta_j$ denotes a scaling dimension that will typically be set to the propagator power $a_j$ of external leg $j$ in the following. In particular, for $D=1$ dimension with $\Delta_j=a_j$ and $\Delta_k=a_k$ we simply have
\begin{align}\label{1D Pjk}
\levo{P}_{jk}
=\frac{i}{2}\brk[s]1{\levz{P}_j\levz{D}_k-\levz{P}_k\levz{D}_j-ia_k \levz{P}_j+ia_j \levz{P}_k}
=
\frac{i}{2}\brk[s]1{(x_{j}-x_k) \partial_j \partial_k -2a_k\partial_j+2a_j\partial_k}.
\end{align}
This nonlocal expression provides the main building block for the annihilating differential operators used below for bootstrapping Feynman integrals in one dimension (or the holomorphic parts of their 2D counterparts).
\paragraph{Two-Point Symmetries.}
The above bilocal density $ \levo{P}^{\mu}_{jk}$ given in \eqref{eq:Phatdens} is distinguished by the fact that it annihilates the product of two propagators with one overlapping leg $X$ as follows (see \cite{Loebbert:2020glj,Loebbert:2024qbw}):
\begin{equation}
    \levo{P}^{\mu}_{jk} \, \frac{1}{(x^2_{j X})^{a_j} (x^2_{k X})^{a_k}} = 0.
\end{equation}
In particular this implies that $\levo{P}^{\mu}_{jk}$ annihilates any integral for which the external points $j$ and $k$ are connected to the same integration vertex \cite{Loebbert:2024qbw}:
\begin{equation}
\includegraphicsbox{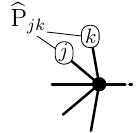}
\end{equation}
 We refer to these symmetries as \emph{two-point symmetries}.
\paragraph{(Generalized) End-Vertex Symmetries.}

When acting on external legs attached to different integration vertices, the above densities on their own do not annihilate the integrals. However, if we sum over all external legs attached to a vertex or subgraph, one can still identify bilocal symmetries, which differ from the full level-one momentum symmetry that is obtained by summing over all external legs of a given graph. In particular, the (generalized) \emph{end-vertex symmetries} are obtained when one of the legs of the bilocal operator acts on the external legs of a fixed vertex $X$, while the other leg is summed over all legs of the attached tree graph \cite{Loebbert:2024qbw}:
\begin{equation}\label{eq:endvertexsym}
\includegraphicsbox{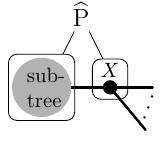}
\end{equation}
In the wording of \cite{Loebbert:2024qbw}, here the term `generalized end-vertex symmetry' refers to the fact that we  have a non-trivial sub-tree in the above figure, while the symmetry was simply denoted `end-vertex symmetry' if the sub-tree is given by a single integration vertex. In the following, we will not make this distinction and simply speak about `end-vertex symmetries'.
\paragraph{(Generalized) Bridge-Vertex Symmetries.}
Similar to the above situation, one finds a \emph{bridge-vertex symmetry} if the two legs of $\levo{P}$ act on different tree graphs attached to the same integration vertex \cite{Loebbert:2024qbw}:
\begin{equation}
\includegraphicsbox{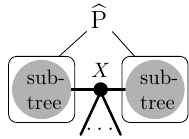}
\end{equation}
We note that the different types of above symmetries need not be independent. In particular, we will argue below that two-point and bridge-vertex symmetries are sufficient to fix all track integrals. Also here we will not distinguish between generalized and ordinary bridge-vertex symmetries in the following.
\paragraph{Example.}
As a concrete example consider the following triangle-track integral at six-loop order with internal propagator powers $b_j$ and $\bar b_k$ named with regard to the below example operator \eqref{eq:exampleop}:
\begin{equation*}
\includegraphics{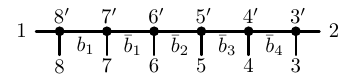}
\end{equation*}
This graph is invariant under the above two-point symmetries acting on the pairs $j,k$ of external points $8,1$ or $2,3$. 

Moreover, we have bridge-vertex symmetries with the two legs of $\levo{P}$ acting on all legs on the left or on the right of a bridge vertex $X=4',5',6'$ or $7'$, respectively. Choosing for instance the bridge vertex $X=7'$, the annihilating differential operator takes the form \cite{Loebbert:2024qbw}
\begin{equation}\label{eq:exampleop}
    \levo{P}_\text{bridge}=\sum^{N_1}_{p=1} \sum^{N_2}_{q=1} \levo{P}^{\mu}_{Y_{p} Z_{q}}  - \frac{1}{2} \left( N_2 D - 2 \sum^{N_2}_{i=1} \bar b_i \right) \sum^{N_1}_{p=1} \levz{P}^{\mu}_{Y_p}  + \frac{1}{2} \left( N_1 D - 2 \sum^{N_1}_{i=1} b_i \right) \sum^{N_2}_{q=1} \levz{P}^{\mu}_{Z_q} ,
\end{equation}
with $N_1=1$ integration vertex $Y_1=8'$ on the left of $7'$ and $N_2=4$ integration vertices $Z_{1,2,3,4}=6',5',4',3'$ on the right. Here $D=1$, $\mu=1$ and e.g.\ $ \levz{P}^{\mu}_{Y_p}$ denotes the sum of $\levz{P}_j^\mu$ over all legs $j$ attached to $Y_p$, and similarly for the bilocal  $\levo{P}^{\mu}_{Y_{p} Z_{q}}$. We will use these bridge-vertex symmetries to fix all triangle-track integrals in \Secref{sec:triangletracks}.

Finally, any of the primed integration points can serve as an end vertex for the end-vertex symmetries indicated in \eqref{eq:endvertexsym}.

\subsection{Bootstrap Algorithm}
\label{sec:Algorithm}

Below we will explore the extent to which the aforementioned $\levo{P}$-symmetries can be used to characterize the underlying Feynman integrals when combined with minimal additional assumptions such as boundary values or permutation symmetries. Such a definition  via an underlying set of differential operators is common practice in the context of related geometries, e.g.\ for Calabi--Yau motives (cf.\ \cite{Bonisch:2021yfw} and references therein). 
In case the integral also admits (dual/position-space) conformal symmetry, the $\levo{P}$-symmetry implies invariance under the full Yangian algebra, which makes connection to the Yangian bootstrap employed in \cite{Loebbert:2019vcj,Loebbert:2020glj,Corcoran:2020epz,Duhr:2022pch,Duhr:2024hjf}. In particular for the case of (position-space) tree integrals, the $\levo{P}$-symmetries split up into the bilocal symmetries reviewed in the above \Secref{sec:PhatForTrees}. Even in the absence of conformal symmetry, non-trivial integrals can be fully fixed by the $\levo{P}$-symmetries as we will see in the following.

To describe the bootstrap procedure, consider some planar position-space tree-level  Feynman integral $I(x_1,\dots ,x_n)$ in one dimension, which implicitly depends on the (unconstrained) external and internal propagator powers $a_i$ and $b_i$, respectively. Here we assume generic external kinematics with none of the external points being coincident. We can always write the integral in the form
\begin{equation}\label{kinematics}
	I(x_1,\dots ,x_n)=V(x_1,\dots ,x_n) \phi(\chi_1,\dots,\chi_{n-2}) \,,
\end{equation}
where $V(x_1,\dots ,x_n)$ is a product of the distances $|x_{ij}|=|x_i-x_j|$ raised to some powers, and carries the dimension of the integral such that the remaining function $\phi$ only depends on $n-2$ variables $\chi_i$, which are dimensionless rational functions of the differences. Our aim will be to bootstrap the function~$\phi$. Note that it is simple to write down an integral representation for $\phi$ which manifestly depends only on the variables $\chi_i$ by appropriately shifting and rescaling the integration variables. To bootstrap $\phi$, we will perform the following steps:
\begin{enumerate}
\item \textbf{Choose a set of variables $\chi_i$ and prefactor $V$.} This choice defines the function we want to bootstrap and hence sets up the problem. It can also crucially affect the following steps. In particular for our problem to be well-defined we need to ensure that we are expanding our integral around a normal-crossing singularity in kinematic space.\footnote{This means that the intersections of $k$ singular hypersurfaces always have codimension $k$. In the plane, for example, this excludes a situation where 3 lines cross at the same point.} Since in our setup all singularities arise from external points becoming coincident, the problem of choosing ``good'' variables can be mapped to the corresponding problem in the moduli space of punctured Riemann spheres $\mathcal{M}_{0,n}$, which has been fully solved by Brown in \cite{brownUVars}.\footnote{We are grateful to Claude Duhr for pointing this out to us.}
\item \textbf{Find a set of differential equations.} Using the $\levo{P}$-symmetries we derive differential equations for the function $\phi$. At this stage it is convenient to derive as many differential equations as possible as it simplifies the below step 4. After that we will restrict to a convenient subset of these.
\end{enumerate}
We are looking for solutions to these differential equations of the form
\begin{equation}
\label{eq:solForm}
		\phi(\vec{\chi})=\sum_{i=1}^m c_i\chi_1^{r_1^{(i)}}\dots \chi_{n-2}^{r_{n-2}^{(i)}}f_i(\vec{\chi}) \,,
\end{equation}
where $\vec{\chi}=(\chi_1,\dots ,\chi_{n-2})$, the $c_i$ are ratios of $\Gamma$-functions depending on the propagator powers, the $r_k^{(i)}$ are at most linear polynomials in the propagator powers and the $f_i$ are power series in the $\chi_i$ with the coefficients being ratios of $\Gamma$-functions depending linearly on the propagator powers. Hence this computes the Feynman integral as a linear combination of hypergeometric functions. The fact that expressing these Feynman integrals in terms of hypergeometric functions is always possible is clear from the integral representation of the one-dimensional Feynman integrals considered here \cite{Duhr:2023bku,aomotoKita,aomotoHGFs}, and is in fact true for all Feynman integrals in any dimension \cite{feynmanIntsAndGKZDModules,Vanhove:2018mto,delaCruz:2019skx,Klausen:2019hrg}, see also the discussion in \Secref{sec:AGfunctions}.
In the following steps we will determine each of the ingredients in the solution \eqref{eq:solForm}, thereby making the solution fully explicit. In one variable a solution of the form \eqref{eq:solForm} can be algorithmically found using the \emph{Frobenius method} (see e.g., \cite{inceBook}) and our algorithm will be a straightforward extension to multiple variables. Note that this is essentially a special case of the more general Saito--Sturmfels--Takayama algorithm \cite{sstBook}, see also \cite{Henn:2023tbo} for an application of this algorithm to physics.
\begin{enumerate}[resume]
\item \textbf{Compute the basis size $m$.} The basis size $m$, also called \emph{holonomic rank} can in principle be algorithmically computed using algebraic methods \cite{sstBook}. An often faster but less rigorous way of determining $m$ is the following. Set the $\dd\log$ of the integrand of $\phi$ with respect to all integration variables to zero and view it as a system of equations for the integration variables. The number of solutions to this system should then equal $m$ (cf.\ \Secref{sec:3pt} for a simple example).%
\footnote{In principle, this computes not the holonomic rank but a closely related quantity, namely the dimension of the underlying \emph{twisted (co)homology group} \cite{aomotoKita,Lee:2013hzt,Frellesvig:2019uqt}. These two numbers have been shown to coincide for wide classes of integrals \cite{Agostini:2022cgv} and this is expected to be true more generally, see e.g., the discussion in \cite{Chestnov:2025whi}. To our knowledge this is however not a proven fact for all integrals considered here.}

\item \textbf{Compute the indicials.} We plug an ansatz of the form
\begin{equation}
	\chi_1^{r_1}\dots \chi_{n-2}^{r_{n-2}}f(\vec{\chi}) \,,
\end{equation}
into the differential equations obtained in step 2, where $f(\vec{\chi})$ is a power series in the $\chi_i$. Requiring the lowest non-trivial orders in the $\chi_i$ of the resulting expression to vanish yields polynomial equations for the $r_i$, the \emph{indicial equations}. The solution set to these equations should be discrete and finite with the corresponding solutions $(r_1^{(i)},\dots ,r_{n-2}^{(i)})$, $i=1,\dots ,m$ being called the \emph{indicials}. Note that in many cases the first non-trivial order already admits a finite set of solutions, however one sometimes has to consider higher orders as well.%
\footnote{Note that it can also happen that even when one has found a finite set of solutions, higher orders might still impose further restrictions.}
If one finds a finite set of indicials, which are however more complicated, i.e., rational or even algebraic in the propagator powers, this is a sign of a bad choice of variables. In our experience, it then often suffices to simply invert one or multiple variables $\chi_i$. Further, it can happen that one finds a finite set of linear indicials that is too small, which can e.g., be due to a degeneracy. While this can in principle be dealt with by making a more general ansatz, we will choose to also change variables in order to find a non-degenerate set of $m$ distinct indicials.%
Conventionally we will choose the prefactor $V(x_1,\dots ,x_n)$ such that $r_i^{(1)}=0$ for all $i$, which means that the first solution is analytic around the origin. We will refer to this analytic solution as the \emph{fundamental solution}.
\item \textbf{Find the fundamental solution $f_1$.} By plugging a series ansatz into the differential equations, we turn them into recurrence relations for the series coefficients. At this stage we pick a convenient subset (including possible linear combinations) of the recurrence equations, such that ideally every equation only contains a shift in one direction in the summation variable space and all directions in this space are covered by some equation. These recurrence equations can then be solved in closed form yielding the fundamental solution $f_1$. Note that the choice of variables and prefactor can influence how difficult it is to decouple the recurrence equations into the various directions. In case the decoupling cannot be obtained in an obvious way, the algorithm proposed in \cite{Liu:2025udl} might be helpful. The fundamental solution typically takes the form
\begin{equation}
	f_1(\vec{\chi})=\sum_{m_1,\dots ,m_{n-2}=0}^{\infty}\frac{1}{\prod_{i=1}^{k}(\alpha_i(\vec{a},\vec{b}))_{\nu_i(\vec{m})}}\frac{\prod_{i=1}^{n-2}(\eta_i\chi_i)^{m_i}}{\prod_{i=1}^{n-2}\Gamma(1+\mu_i(\vec{m}))} \,,
\end{equation}  
in terms of Pochhammer symbols $(a)_n=\Gamma(a+n)/\Gamma(a)$ and we conventionally moved all Pochhammer symbols to the denominator using $(a)_n(1-a)_{-n}=(-1)^n$. Here the $\alpha_i$ are at most linear polynomials, $\nu_i,\mu_i$ are linear forms and the $\eta_i=\pm$ are signs. The $\Gamma(1+\mu_i(\vec{m}))$ determine the effective summation ranges and hence, by shifting summation variables, the arguments of the corresponding hypergeometric function as products of the $\eta_i\chi_i$. We call the fundamental solution \emph{minimal} if $k$ equals the number of propagators. We always aim to find such a minimal solution, possibly by changing the variables. In most cases however we find that if the steps so far have worked out we already land on a minimal fundamental solution.
\item \textbf{Generate the other series $f_i$.} The other solutions can now be generated by shifting the summation variables $m_1,\dots ,m_{n-2}$ in $f_1$ by the respective indicials $m_j\rightarrow m_j+r_j^{(i)}$ for all $i=2,\dots,m$. This naturally yields the powers of the $\chi_i$ as a prefactor times another power series which we identify as $f_i$
\begin{equation}
		f_1(\vec{\chi})\rightarrow \chi_1^{r_1^{(i)}}\dots \chi_{n-2}^{r_{n-2}^{(i)}}f_i(\vec{\chi})  \,.
	\end{equation}
\item \textbf{Computing the coefficients $c_i$.}  We are now left with the task of finding the explicit forms of the coefficients $c_i$. This can be achieved by evaluating the original Feynman integral in appropriate limits. Explicitly to compute $c_i$ we consider
	\begin{equation}
		\lim_{\chi_1\rightarrow 0}\dots \lim_{\chi_{n-2}\rightarrow 0} \chi_1^{-r_1^{(i)}}\dots \chi_{n-2}^{-r_{n-2}^{(i)}} \phi(\chi_1,\dots ,\chi_{n-2})=c_i \,.
	\end{equation}
	Note that here we have restricted the $a_i,b_i$ to some open set in which all other terms go to zero in this limit (which we assume to exist). Since the result will be some ratio of $\Gamma$-functions it is trivial to then analytically continue $c_i$ to arbitrary values of the propagator powers. The left-hand side can be explicitly computed as follows. First we rescale the integration variables of the integral representation of $\phi$ to explicitly cancel the prefactor and make the integral manifestly finite in the limit. The resulting integral can then typically be explicitly computed by known formulas. For all examples considered here, we could compute the coefficients through iterated application of the chain relation, cf.\ \eqref{chain}:
	\begin{equation}
		\int\frac{\dd x_0}{\sqrt{\pi}}\frac{1}{|x_{01}|^{2a_1}|x_{02}|^{2a_2}}=\frac{\Gamma\left( \frac{1}{2}-a_1\right)\Gamma\left( \frac{1}{2}-a_2\right)\Gamma\left(a_1+a_2-\frac{1}{2}\right)}{\Gamma(a_1)\Gamma(a_2)\Gamma(1-a_1-a_2)}\frac{1}{|x_{12}|^{2(a_1+a_2)-1}} \,.
	\end{equation}
\end{enumerate}

In principle, the above algorithm should also be applicable to graphs including position-space loops as well as to integrals in higher spacetime dimensions. 
In the presence of loops, we expect step 6 to become less straightforward.%
\footnote{It is also less clear which symmetries are present for integrals with position-space loops. While the two-point symmetries still hold, this is currently not clear for the other partial symmetries reviewed in  \Secref{sec:PhatForTrees}. It is hence an interesting question for future work to see if (some of) the partial symmetries generalize in some way to loop graphs or if maybe even new symmetries emerge and if the symmetries are still sufficient to fix the integrals. Note that it has recently been proven that the full $\levo{P}$-symmetry generalizes to loop graphs as long as the propagator powers satisfy certain linear constraints \cite{Loebbert:2025abz}.} For spacetime dimensions $D>2$, step 2 is more involved since extracting a set of differential equations in the scale invariant variables requires to identify independent (and still tractable) equations from the vector differential operators $\levo{P}^\mu$. Also the construction of \cite{brownUVars} does not help us in constructing ``good`` variables anymore since the singularity structure becomes more general than that of $\mathcal{M}_{0,n}$. There is however a more general construction that should work in higher dimensions as described in \cite{sstBook}.

\section{1D Trees and Spectral Transform}
\label{sec:spectransf}

In this section we introduce a particular `spectral transform' inspired by the separation of variables method coming from integrability. In particular, this method allows to compute tree-level Feynman integrals in position space with generic propagator powers in one and two spacetime dimensions straightforwardly.

\subsection{Spectral Transform Method}
\label{sec:ThreePointSpectral}

\begin{figure}
\begin{center}
	\includegraphicsbox{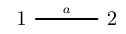}
	$=
	\sum \int \dd u$\includegraphicsbox{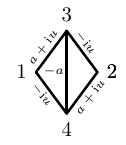}
\caption{Spectral transform of a single propagator. We introduce auxiliary spacetime points $3$ and $4$ as well as a spectral parameter $u$ into the propagator powers.}
\label{fig:spectrans}
\end{center}
\end{figure}

For our direct computations of one-dimensional, tree integrals, we mostly rely on the following spectral representation of a single propagator%
\footnote{We are not aware of any article in which this exact equation, or its higher-dimensional analogues presented in \secref{sec:1to2D} and \appref{app:higher d}, appears. However, we stress that they are the simplest application of the completeness relation for the SoV bases constructed in \cite{Belitsky:2014rba,Derkachov:2014gya,Basso:2019xay} using techniques introduced in \cite{Gaudin:1992ci,Derkachov:1999pz,Derkachov:2001yn}.},
cf.\ \Figref{fig:spectrans},
\begin{equation}\label{SoV prop}
	\frac{1}{|x_{12}|^{2 a}} = \sum_{\ep = 0}^1\! \int_{\mathbb{R} + \ii \eta} \! \frac{|x_{34}|^{2a}\sgn^\ep(x_{13} x_{24} x_{14} x_{23}) }{(|x_{13}| |x_{24}|)^{2(a+\ii u)} (|x_{14}| |x_{23}|)^{-2\ii u}} \frac{A_0(a)}{A_\ep(-\ii u) A_\ep(a+\ii u)} \frac{\dd u}{2\sqrt{\pi}}\, ,
\end{equation}
where $(x_1,x_2,x_3,x_4) \in\mathbb{R}^4$, the integration contour runs parallel to the real axis, the constant $\eta\in ]0;\Re(a)[$ is arbitrary, and we introduced the functions
\begin{equation}\label{eq:defsign}
	\sgn(x) = \sign(x)\quad \text{and}\quad A_{\ep}(\alpha) = \frac{\Gamma\!\left(\frac{1+\ep}{2}-\alpha\right)}{\Gamma\!\left(\frac{\ep}{2}+\alpha\right)} = \frac{1}{A_{\ep}\!\left(\frac{1}{2} -\alpha\!\right)}\, .
\end{equation}
Equation \eqref{SoV prop} is easily verified by deforming the contour in the lower/upper half-plane and picking the residues of the simple poles coming from $A^{-1}_\ep(-\ii u)$ or $A^{-1}_\ep(a+\ii u)$ respectively. It is particularly useful to specify \eqref{SoV prop} in the limit $|x_4|\to\infty$, we then have
\begin{equation}\label{SoV prop inf}
	\frac{1}{|x_{12}|^{2 a}} = \sum_{\ep = 0}^1\! \int_{\mathbb{R} + \ii \eta} \! \frac{\sgn^\ep(x_{13} x_{23}) }{|x_{13}|^{2(a+\ii u)} |x_{23}|^{-2\ii u}} \frac{A_0(a)}{A_\ep(-\ii u) A_\ep(a+\ii u)} \frac{\dd u}{2\sqrt{\pi}}\, .
\end{equation}

The integration contour in \eqref{SoV prop} and \eqref{SoV prop inf} need not be horizontal; it can be deformed arbitrarily as long as it does not cross any pole and the integral remains convergent at infinity. For instance, if $\Re(a)$ becomes negative, then we cannot even choose a horizontal contour anymore. In practice, such deformations might be required for some intermediate steps in the computations. Indeed, when using the spectral transform to compute a Feynman integral we have to first rewrite the integrand using \eqref{SoV prop} or \eqref{SoV prop inf}, and then change the order of integration to perform the integrals over the spacetime points. This usually requires further, integral-specific restrictions on the range of allowed propagator exponents for which the contours cannot be horizontal. We will make this precise for a simple three-point example, see the paragraph at the end of this subsection and  \Appref{app:triangle}, but we will not expand on this point for any of the other integrals. In all considered cases, once all spacetime integrals have been performed, and before any of the spectral integrals is evaluated, there exists an infinite vertical strip of allowed complex values for each propagator powers such that the contours may be taken to be horizontal. Moreover, this ensures exactly that each (half-infinite) series of poles should lie entirely on one side of the horizontal integration contours. Other values of the exponents may then be reached by analytic continuation through contour deformation. 

In \Appref{app:higherd} we give a generalization of the spectral representation \eqref{SoV prop}, namely \eqref{SoV prop sign}, that works for both scalar and `spinning' propagators $\sgn^\ep(x_{12})/|x_{12}|^{2a}$ for $\ep\in\{0,1\}$. This relation is needed (and sufficient) to evaluate tree integrals beyond track integrals for which \eqref{SoV prop} is enough. A similar relation for spinning propagators exists in two dimensions, see \eqref{SoV prop 2d}, and it is equally useful for the computation of tree integrals. In higher dimensions, the angular dependence is more complicated, see \eqref{SoV prop d} for scalar propagators.

\paragraph{Star-Triangle Identity.}
We will also need the so-called star-triangle relation, which we give in its generalized form using the sign-factors of \eqref{eq:defsign}: 
\begin{equation}\label{star-triangle}
	\includegraphicsbox[scale=.8]{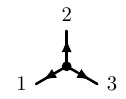}
	\hspace{-4mm}
	=\!\int\! \prod_{i=1}^3 \frac{\sgn^{\ep_i}(x_{i0})}{|x_{i0}|^{2a_i}} \frac{\dd x_0}{\sqrt{\pi}} = (-1)^{\ep_1 + \ep_2\ep_3} \prod_{i=1}^3 A_{\ep_i}(a_i) \frac{\sgn^{\ep_3}(x_{12}) \sgn^{\ep_1}(x_{23}) \sgn^{\ep_2}(x_{31})}{|x_{12}|^{1-2a_3} |x_{23}|^{1-2a_1} |x_{31}|^{1-2a_2}}
	\propto\hspace{-4mm}\includegraphicsbox[scale=.8]{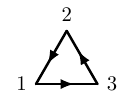}.
\end{equation}
Here the parameters $(a_i,\ep_i)\in\mathbb{C}\times \{0,1\}$ satisfy the constraints  $a_1 + a_2 + a_3 = 1$ and 
$\ep_1 + \ep_2 + \ep_3 \equiv 0 \pmod2$.

\paragraph{Two-Point Integral.}
Taking the limit $|x_3|\to\infty$ of the star-triangle identity yields the so-called \emph{chain relation}
\begin{equation}\label{chain}
	\int \frac{\sgn^{\ep_1}(x_{10}) \sgn^{\ep_2}(x_{02})}{|x_{10}|^{2a_1} |x_{20}|^{2a_2}} \frac{\dd x_0}{\sqrt{\pi}} = (-1)^{\ep_1\ep_2} A_{\ep_1}(a_1) A_{\ep_2}(a_2) A_{[\ep_1+\ep_2]}(1-a_1-a_2) \frac{\sgn^{\ep_1 + \ep_2}(x_{12})}{|x_{12}|^{2a_1 + 2a_2 - 1}}\, ,
\end{equation}
where the parameters $(a_1,\ep_1)$ and $(a_2,\ep_2)$ are arbitrary and we introduced the notation $[\ep_1+\ep_2]$ defined by
\begin{equation}\label{kappasum}
	[\ep_1+\ep_2] \in \{0,1\}\, ,\qquad [\ep_1+\ep_2] \equiv \ep_1+\ep_2 \pmod 2\, .
\end{equation}

\paragraph{Three-Point Integral.}
Consider the following three-point integral, which corresponds to a triangle in dual momentum space:
\begin{equation}\label{star 3}
	I_3=\includegraphicsbox{FigThreePointStar.pdf} = \int \prod_{i=1}^3 \frac{1}{|x_{i0}|^{2a_i}} \frac{\dd x_0}{\sqrt{\pi}}\, .
\end{equation}
We explain in detail in \appref{app:triangle} how to carefully derive the following spectral representation of this integral:
\begin{equation}\label{SoV star}
	I_3 = \frac{A_0(a_2) A_0(a_3)}{|x_{12}|^{2(a_1+a_2+a_3)-1}} \sum_{\ep = 0}^1 \sgn^\ep(\chi) \int_{\mathbb{R} + \ii \eta}\!\!\!\! \frac{A_{\ep}(a_1+a_3+\ii u) A_{\ep}(1-a_1-a_2-a_3-\ii u)}{A_{\ep}(-\ii u) A_{\ep}(a_3+\ii u)} \frac{|\chi|^{2\ii u} \dd u}{2\sqrt{\pi}}\, ,
\end{equation}
where the various parameters are such that the poles of $A_\ep(...\pm\ii u)$ are all below/above the integration contour, and the ratio is $\chi=x_{13}/x_{12}$. Assuming that $|\chi|<1$, we may close the contour in the lower half-plane. There are two series of simple poles: one in $-\ii(\ep/2+\mathbb{N})$ and one in $-\ii((1+\ep)/2-a_1-a_3+\mathbb{N})$. Summing over their residues yields the final result
\begin{equation}\label{3-pt result}
	I_3 = \frac{A_0(a_2)A_0(a_3)}{|x_{12}|^{2\scriptstyle\sum_ia_i-1}}\left( \cGG{2}{1}{2a_3,2{\scriptstyle \sum_i} a_i-1}{2(a_1+a_3)}{\chi} + |\chi|^{1-2a_1-2a_3}\cGG{2}{1}{2a_2,1-2a_1}{2(1-a_1-a_3)}{\chi} \right)\,,
\end{equation}
where we introduced the rescaled Gauss hypergeometric function
\begin{equation}\label{2F1 rescaled}
	\cGG{2}{1}{a,b}{c}{x}=\frac{A_0(\sfrac{c}{2})}{A_0(\sfrac{a}{2})A_0(\sfrac{b}{2})}\GG{2}{1}{a,b}{c}{x} \,.
\end{equation}

Note that the spectral representation \eqref{SoV star} holds for all values of $\chi$, but equation \eqref{3-pt result} was obtained assuming that $-1<\chi<1$. To compute the integral for $|\chi|>1$, we \textit{a priori} need to start from \eqref{SoV star} again and close the contour in the upper half-plane. This gives\footnote{In this simple example, we could have arrived at this result using the invariance of the integral under the exchange $(x_2,a_2)\leftrightarrow (x_3,a_3)$, but we will not have this possibility for more complicated integrals.}
\begin{equation}\label{3-pt result 2}
	I_3 = \frac{A_0(a_2)A_0(a_3)}{|x_{13}|^{2\scriptstyle\sum_ia_i-1}}\left( \cGG{2}{1}{2a_2,2{\scriptstyle \sum_i} a_i-1}{2(a_1+a_2)}{\frac{1}{\chi}} + |\chi|^{1-2a_1-2a_2}\cGG{2}{1}{2a_3,1-2a_1}{2(1-a_1-a_2)}{\frac{1}{\chi}} \right)\,.
\end{equation}
However, we remark that the analytic continuation of \eqref{3-pt result} for $\chi<-1$ gives exactly \eqref{3-pt result 2}, if we use known properties of the Gauss hypergeometric function.

\subsection{Comb-Channel Conformal Partial Waves from Spectral Transform}
\label{sec:comb}
One advantage of the above `spectral transform method'  as compared to the $\levo{P}$ differential equations is that it also applies to integrals with external coincidence limits, without requiring any constraints on the propgatator powers.
Let us therefore apply the above spectral transform to such an example, namely the full family of $n$-point conformal partial waves  in the comb channel as considered by Rosenhaus in \cite{Rosenhaus:2018zqn}. For this purpose, we first introduce the scalar conformal three-point structure
\begin{equation}
	\Phi_{h_1,h_2,h_3}(x_1,x_2,x_3) 
	=\includegraphicsbox{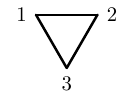}
	= |x_{12}|^{h_3-h_1-h_2} |x_{23}|^{h_1-h_2-h_3} |x_{13}|^{h_2-h_1-h_3}\, ,
\end{equation}
where $h_i$ is typically the conformal dimension of an operator located at position~$x_i$. Then, the $n$-point conformal partial wave in the comb channel is given by the conformal integral
\begin{equation}
	\Psi_{n} 
	=\includegraphicsbox{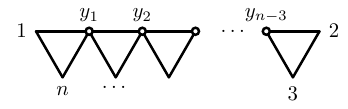}
	= \int \prod_{i=1}^{n-2} \Phi_{h_{n+1-i},g_{i-1},1-g_{i}}(x_{n+1-i},y_{i-1},y_{i}) \prod_{i=1}^{n-3} \frac{\dd y_i}{\sqrt{\pi}}\, ,
\end{equation}
where we set $(y_0,g_0) = (x_1,h_1)$ and $(y_{n-2},g_{n-2}) = (x_2,1-h_2)$. Here conformal integration vertices are indicated by white circles and the parameters $g_1,\dots,g_{n-3}$ correspond to the conformal dimensions of the exchanged operators, i.e.\ those whose positions we integrate over.

As explained in \cite{Rosenhaus:2018zqn}, the partial wave can be expanded into a sum of $2^{n-3}$ so-called conformal blocks, each one of them being associated with a choice of one element from each of the sets $\{g_1,1-g_1\},\dots,\{g_{n-3},1-g_{n-3}\}$. We show here how the spectral transform can be used to give an elementary, rigorous derivation of this expansion, including the explicit coefficient multiplying each of the blocks.

We first apply the spectral representation \eqref{SoV prop} with the replacement
\begin{equation}
	(x_1,x_2,x_3,x_4,a,u) \to (y_{i-1},y_{i},x_{n-i},x_{n+1-i},(1+g_{i-1,i}-h_{n+1-i})/2,u_{n-2-i})
\end{equation}
for each $i\in\{1,\dots,n-3\}$. After this step, the integrations over the internal vertices can all be performed using the star-triangle relation \eqref{star-triangle}. As mentioned before, this requires some restriction on the values of the propagator exponents for the integrals to be convergent and the change in the order of integration to be permitted. Assuming these restrictions to hold, we arrive at
\begin{multline}\label{spectral comb 1}
	\Psi_{n} = V_n \prod_{i=1}^{n-2} A_0\!\left(\frac{1+g_{i-1,i}-h_{n+1-i}}{2}\right) \prod_{j=1}^{n-3} \sum_{\ep_j} \int_{\mathbb{R}+\ii \eta_j} \frac{\dd u_j}{2\sqrt{\pi}} \frac{\sgn^{\ep_j}(\chi_j)  |\chi_j|^{1-g_{n-2-j}+2\ii u_j}}{A_{\ep_j}(-\ii u_j) A_{\ep_j}(g_{n-2-j}-1/2 -\ii u_j)}\\
	\times \prod_{j=2}^{n-3} (-1)^{\ep_{j-1}\ep_j} A_{[\ep_{j-1}+\ep_j]}\!\left(\frac{h_{j+2}+g_{n-2-j}+g_{n-1-j}-1}{2}-\ii(u_{j-1}+u_j)\right)\\
	\times A_{\ep_1}\!\!\left(\frac{h_{32}+g_{n-3}}{2}-\ii u_{1}\right)  A_{\ep_{n-3}}\!\!\left(\frac{h_{n1}+g_1}{2}-\ii u_{n-3}\right)\, ,
\end{multline}
where each of the (half-infinite) series of poles should fully lie on one side of the integration contours. The kinematical prefactor and the cross ratios are the same as in \cite{Rosenhaus:2018zqn}, up to our relabelling of the external points, namely
\begin{equation}
	\chi_j = \frac{x_{j+1,j+2}\, x_{j+3,j+4}}{x_{j+1,j+3}\, x_{j+2,j+4}}\quad\text{for}\quad j\in\{1,\dots,n-3\}\, ,
\end{equation}
and
\begin{equation}
	V_n = \left(\frac{|x_{n-1,n}|}{|x_{n-1,1}| |x_{1,n}|}\right)^{h_1} \left(\frac{|x_{34}|}{|x_{32}| |x_{24}|}\right)^{h_2} \prod_{i=3}^{n} \left(\frac{|x_{i-1,i+1}|}{|x_{i-1,i}| |x_{i,i+1}|}\right)^{h_i}\, ,
\end{equation}
where $x_{n+1} = x_1$.
In order to make the formula appear more symmetric we choose $\eta_j = (g_{n-2-j}-1)/2$ and make the change of variables $u_j\mapsto u_j+\ii\eta_j$, we thus obtain\footnote{Requiring the series of poles to lie on a definite side of the integration contours imposes different restrictions on the parameters $h_j$ and $g_j$ for \eqref{spectral comb 1} and \eqref{spectral comb 2}. However, these have an overlap for which the contour manipulation is permitted and thus \eqref{spectral comb 2} is indeed the analytic continuation of \eqref{spectral comb 1}.}
\begin{multline}\label{spectral comb 2}
	\Psi_{n} = V_n \prod_{i=1}^{n-2} A_0\!\left(\sfrac{1+g_{i-1,i}-h_{n+1-i}}{2}\right) \prod_{j=1}^{n-3} \sum_{\ep_j} \int_{\mathbb{R}} \frac{\dd u_j}{2\sqrt{\pi}} \sgn^{\ep_j}(\chi_j)  |\chi_j|^{2\ii u_j} \frac{A_{\ep_j}(g_{n-2-j}/2 +\ii u_j)}{A_{\ep_j}(g_{n-2-j}/2 -\ii u_j)}\\
	\prod_{j=2}^{n-3} (-1)^{\ep_{j-1}\ep_j}\! A_{[\ep_{j-1}+\ep_j]}\!\left(\sfrac{h_{j+2}+1}{2}-\ii(u_{j-1}+u_j)\!\right) A_{\ep_1}\!\!\left(\sfrac{h_{32}+1}{2}-\ii u_1\right) A_{\ep_{n-3}}\!\!\left(\sfrac{h_{n1}+1}{2}-\ii u_{n-3}\right).
\end{multline}

Now, assuming that the cross ratios are all smaller than $1$ in absolute value, for instance when $x_2<\dots<x_n<x_1$, we can compute the integrals by summing over the residues of the simple poles situated in the lower half-plane. Independently of the order in which we compute the integrals, there are always the same two series of poles for each of the integration variables $u_j$: in $-\ii ((g_{n-2-j}+\ep_j)/2+\mathbb{N})$ and in $-\ii ((1-g_{n-2-j}+\ep_j)/2+\mathbb{N})$. Choosing one of these series for each of these $n-3$ variables will generate exactly one conformal block, and there are $2^{n-3}$ such choices. The full result is then
\begin{multline}
	\Psi_{n} = V_n \prod_{i=1}^{n-2} A_0\!\left(\frac{1+g_{i-1,i}-h_{n+1-i}}{2}\right) \prod_{j=1}^{n-3} \sum_{\delta_j\in\{g_{n-2-j},1-g_{n-2-j}\}} |\chi_j|^{\delta_j} \\
	\times \mathcal{F}_K\!\left(\substack{\delta_1+h_{23};\delta_1+\delta_2-h_4,\dots,\delta_{n-4}+\delta_{n-3}-h_{n-1};\delta_{n-3}+h_{1n} \\ 2\delta_1,\dots,2\delta_{n-3}};\chi_1,\dots,\chi_{n-3}\right)\, ,
\end{multline}
where we use (a rescaled version of) the hypergeometric function introduced in \cite{Rosenhaus:2018zqn}:
\begin{multline}
	\mathcal{F}_K\!\left[\substack{a_1;b_1,\dots,b_{n-1};a_2 \\ c_1,\dots,c_n};x_1,\dots,x_n\right] = \frac{\prod_{i=1}^{n} A_0\!\left(\frac{c_i}{2}\right)}{A_0\!\left(\frac{a_1}{2}\right) A_0\!\left(\frac{a_2}{2}\right)\prod_{i=1}^{n-1} A_0\!\left(\frac{b_i}{2}\right)}\\
	\times \sum_{m_1,\dots,m_n=0}^{+\infty}\!\!\! \frac{(a_1)_{m_1} \prod_{i=1}^{n-1}(b_i)_{m_{i}+m_{i+1}} (a_2)_{m_n}}{\prod_{i=1}^n (c_i)_n} \prod_{i=1}^n \frac{x_i^{m_i}}{m_i!}\, .
\end{multline}

The above spectral transform method has been very useful for understanding the below (families of) Feynman integrals in one and two dimensions, see \Appref{app:spectranscomps} for further details. In the following, however, we will demonstrate how to obtain those integrals from the $\levo{P}$-bootstrap outlined in \Secref{sec:Phat}.

\section{Three-Point Integral}
\label{sec:3pt}
We begin by bootstrapping the generic three-point integral \eqref{star 3} that already served as an illustrative example for the spectral transform method in \Secref{sec:ThreePointSpectral}. Let us decompose the integral as
\begin{equation}\label{eq_32_decomposition}
	I_3=
	\includegraphicsbox{FigThreePointStar.pdf}
	=
	\frac{\phi_3(\chi)}{|x_{12}|^{2(a_1+a_2+a_3)-1}} \,, \qquad \chi=\frac{x_{13}}{x_{12}}.
\end{equation}
We will now bootstrap the function $\phi_3(\chi)$, which admits the integral representation
\begin{equation}
\label{eq:intrepfphi3}
	\phi_3(\chi)=\int\frac{\dd y}{\sqrt{\pi}}\frac{1}{|y|^{2a_1} |1-y|^{2a_2} |y-\chi|^{2a_3}} \,.
\end{equation}
We can obtain differential equations from two-point symmetries with respect to any pair of points. All of these yield the same differential equation that can be identified with the Gauss hypergeometric differential equation, thus allowing us to immediately write down a basis of the solution space. Depending on the chosen decomposition in \eqref{eq_32_decomposition}, the precise form of the differential equation changes, but not the fact that it is a Gauss hypergeometric differential equation. For the above choice of decomposition for instance, we find
\begin{equation}
	\chi (\chi-1) \phi_3''(\chi) + 2((a_1+a_2+2a_3)\chi-a_1-a_3) \phi_3'(\chi) + 2 a_3 (2{\textstyle \sum_i a_i}-1) \phi_3(\chi) = 0\,.
\end{equation}
This is solved by an arbitrary linear combination of two $_2F_1$ hypergeometric functions,
\begin{multline}
	\phi_3(\chi) = c_1\, {}_2F_1 \left(2a_3,2{\textstyle \sum_i} a_i-1;2(a_1+a_3);\chi\right)\\
	+ c_2 \chi^{1-2(a_1+a_3)} {}_2F_1 \left(2a_2,1-2a_1;2(1-a_1-a_3);\chi\right)\, .
\end{multline}
Knowing that the integral is symmetric under the exchange of two legs, we can fix the ratio $c_1/c_2$. Then, considering a coincidence limit, e.g. $\chi\to 0$, allows us to fix the integral completely, obtaining in the end the same result as from the spectral transform in \Secref{sec:ThreePointSpectral}.

While this procedure works for simple differential operators --- especially when the underlying functions are understood well enough to know relations amongst them involving variable transformations --- we do not have this option in more complicated cases. Hence, let us redo the calculation of the solution basis without recognizing the Gauss hypergeometric function right away, but following the bootstrap algorithm layed out in \Secref{sec:Algorithm}. Computing the indicials we find
\begin{equation}
	r^{(1)}=0,\qquad r^{(2)}=1-2(a_1+a_3) \,.
\end{equation}
This suggests that we can expect a $2$-dimensional solution space, which is indeed correct. We get to the same result by counting the zeroes of
\begin{equation}
	\dd\log\left( \frac{1}{y^{2a_1} (1-y)^{2a_2} (y-\chi)^{2a_3}} \right)=\frac{-2(a_1(y-1)(y-\chi)+a_2y(y-\chi)+y(y-1)a_3)}{y(y-1)(y-\chi)}\dd y \,.
\end{equation}
The zeros correspond to critical points of the associated \emph{Morse function} which by Morse theory \cite{milnorBook} correspond to the independent cycles for the integral and hence yield the number of expected basis functions \cite{Lee:2013hzt, Frellesvig:2019uqt}. Here we see that the numerator is a univariate polynomial of degree two (with the zeros not cancelled by the denominator) and hence there are two critical points, i.e., we find a two-dimensional solution space.

We can now compute the fundamental solution by making a series ansatz and solving the resulting recursion relation for the coefficients. This yields
\begin{equation}
	f_1(a,\chi) = \sum_{k=0}^{\infty}\frac{(2a_3)_k (2{\textstyle \sum_i} a_i-1)_k}{(2(a_1+a_3))_k}\frac{\chi^k}{k!} = {}_2F_1 \left(2a_3,2{\textstyle \sum_i} a_i-1;2(a_1+a_3);\chi\right)\,,
\end{equation}
where $(a)_k$ is the Pochhammer symbol, which we could identify as the Gauss hypergeometric function, as expected. Shifting the summation variable $k\rightarrow k+1-2(a_1+a_3)$ we can obtain the second basis function
\begin{equation}
	f_2(a,\chi) = \sum_{k=0}^{\infty}\frac{(2a_2)_k (1-2a_1)_k}{(2(1-a_1-a_3))_k}\frac{\chi^k}{k!} = {}_2F_1 \left(2a_2,1-2a_1;2(1-a_1-a_3);\chi\right) \,.
\end{equation}
We are left with the task of computing the coefficients $c_i$. Taking appropriate limits of the integral representation of $\phi_3(\chi)$ in \eqref{eq:intrepfphi3}, as explained in \secref{sec:Algorithm}, we compute the coefficients to be
\begin{align}
	c_1 &= \int\frac{\pi^{-1/2} \dd y}{|y|^{2(a_1+a_3)} |1-y|^{2a_2}} = \frac{A_0(a_2) A_0(a_1+a_3)}{A_0({\textstyle \sum_i} a_i-1/2)}\,, \nonumber \\
	c_2 &= \int\frac{\pi^{-1/2} \dd y}{|y|^{2a_1} |1-y|^{2a_3}} = \frac{A_0(a_3) A_0(1-a_1-a_3)}{A_0(1/2-a_1)} \,.
\end{align}
Putting everything together, we find the result\footnote{While this result is strictly only valid for $0<\chi<1$, we saw in the previous section using the spectral transform how to extend it to $\chi<0$: it suffices to replace the power of $\chi$ in front of the second basis element by the same power of $|\chi|$ and to use the approriate analytic continuation of the Gauss hypergeometric function for $\chi<-1$.} 
\begin{equation}
	I_3=\frac{A_0(a_2)A_0(a_3)}{|x_{12}|^{2\scriptstyle\sum_ia_i-1}}\left( \cGG{2}{1}{2a_3,2{\scriptstyle \sum_i} a_i-1}{2(a_1+a_3)}{\chi}  +\chi^{1-2a_1-2a_3}\cGG{2}{1}{2a_2,1-2a_1}{2(1-a_1-a_3)}{\chi} \right)\,,
\end{equation}
where we used the rescaled Gauss hypergeometric function defined in \eqref{2F1 rescaled}.

\section{Four-Point Integrals}
\label{sec:4pt}

There are two four-point tree integrals (excluding integrals with coincidence limits of external points), namely the box integral 
which forms part of an infinite family of $n$-point polygon (or star integrals, see \Secref{sec:polygons}, as well as
the H-integral
which belongs to the class of triangle tracks treated in \Secref{sec:triangletracks}.
We will compute both of them using the $\levo{P}$-bootstrap.

\subsection{Box Integral}
\label{sec:Box}

Consider first the box integral that we decompose as 
\begin{equation}
	I_4= \includegraphicsbox{FigFourPointStar.pdf}=\frac{\phi_{4}(\chi_1,\chi_2)}{|x_{12}|^{2a_1+2a_2-1}|x_{13}|^{2a_3}|x_{14}|^{2a_4}}\,,
\end{equation}
with
\begin{equation}
	\chi_1=\frac{x_{12}}{x_{13}},\qquad \chi_2=\frac{x_{13}}{x_{14}} \,.
\end{equation}
In order to bootstrap the function $\phi_{4}$, we derive a set of two differential equations for it. For example the two-point symmetries acting on the points $x_2,x_4$ and $x_3,x_4$, respectively, yield the following PDEs for $\phi_{4}$
\begin{align}
	&\Big[2a_4(2a_1-1)\chi_1-2a_4\chi_1^2\partial_{\chi_1}+(1-2a_1-2a_2+(2a_1-1)\chi_1\chi_2)\partial_{\chi_2} \nonumber\\
	&\qquad\qquad +\chi_1(1-\chi_1\chi_2)\partial_{\chi_1}\partial_{\chi_2}\Big]\phi_4=0 \,, \\
	&\Big[2a_4\chi_1\partial_{\chi_1}+(1-2a_3-(1+2a_4)\chi_2)\partial_{\chi_2}+\chi_1(\chi_2-1)\partial_{\chi_1}\partial_{\chi_2}+\chi_2(1-\chi_2)\partial_{\chi_2}^2\Big]\phi_4=0 \,.
\end{align}
Computing the indicials of these differential operators yields the list
\begin{align}
	&(0,0) \,, \nonumber \\
	&(2a_1+2a_2-1,0) \,, \\
	&(2a_1+2a_2-1,2a_1+2a_2+2a_3-1) \,. \nonumber 
\end{align}
We hence expect a solution of the form
\begin{equation}
\label{eq:genBoostrapSolS4}
	\phi_{4}(\chi_1,\chi_2)=c_1 f_1(\chi_1,\chi_2)+c_2 \chi_1^{2a_1+2a_2-1} f_2(\chi_1,\chi_2)+c_3 \chi_1^{2a_1+2a_2-1}\chi_2^{2a_1+2a_2+2a_3-1} f_3(\chi_1,\chi_2) \,,
\end{equation}
for some constant coefficients $c_i$ and some power series $f_i$. By making a series ansatz we can compute the fundamental series $f_1$ to be
\begin{equation}
	f_1(\chi_1,\chi_2)=\FF{1}{1-2a_1;2a_3,2a_4}{2-2a_1-2a_2}{\chi_1,\chi_1\chi_2} \,,
\end{equation}
in terms of the Appell $F_1$ function defined in \appref{app:defs}. The other series can be found by shifting the summation variables of the fundamental series by the indicials. We find
\begin{align}
	f_2(\chi_1,\chi_2)&=\GH{2}{2a_2,2a_4}{1-2a_1-2a_2,2(a_1+a_2+a_3)-1}{-\chi_1,-\chi_2} \,, \\
	f_3(\chi_1,\chi_2)&=\FF{1}{2(a_1+a_2+a_3+a_4)-1;2a_2,2a_3}{2(a_1+a_2+a_3)}{\chi_1\chi_2,\chi_2} \,,
\end{align}
where $G_2$ refers to a Horn function defined in \appref{app:defs}. Finally we fix the coefficients $c_i$ by evaluating the integral representation
\begin{equation}
	\phi_{4}(\chi_1,\chi_2)=\int_{\mathbb{R}}\frac{\dd y}{\sqrt{\pi}}\frac{1}{|y|^{2a_1}|1-y|^{2a_2}|1-\chi_1 y|^{2a_3}|1-\chi_1\chi_2 y|^{2a_4}} \,,
\end{equation}
in certain limits. Explicitly we obtain
\begin{align}
	c_1&=\lim_{\chi_1,\chi_2\rightarrow 0}\phi_{4}(\chi_1,\chi_2) = A_0(a_1) A_0(a_2) A_0(1-a_1-a_2)\,, \\
	c_2&=\lim_{\chi_1,\chi_2\rightarrow 0}\chi_1^{1-2a_1-2a_2}\phi_{4}(\chi_1,\chi_2)=\frac{A_0(a_1+a_2)A_0(a_3)}{A_0( a_1+a_2+a_3-1/2)} \,, \\
	c_3&=\lim_{\chi_1,\chi_2\rightarrow 0}\chi_1^{1-2a_1-2a_2}\chi_2^{1-2(a_1+a_2+a_3)}\phi_{4}(\chi_1,\chi_2)=\frac{A_0(a_1+a_2+a_3)A_0(a_4)}{A_0( a_1+a_2+a_3+a_4-1/2)} \,.
\end{align}
Plugging everything into \eqref{eq:genBoostrapSolS4}, we find
\begin{align}\label{star4Result}
	I_4 &= \frac{A_0(a_2) A_0(a_3) A_0(a_4)}{|x_{12}|^{2a_1+2a_2-1}|x_{13}|^{2a_3}|x_{14}|^{2a_4}} \bigg[ \cFF{1}{1-2a_1,2a_3,2a_4}{2-2a_1-2a_2}{\chi_1,\chi_1\chi_2}  \nonumber\\
	& \qquad\qquad\qquad\quad+\chi_1^{2a_1+2a_2-1}\cGH{2}{2a_2,2a_4}{1-2a_1-2a_2,2(a_1+a_2+a_3)-1}{-\chi_1,-\chi_2} \nonumber
	\\
	&\qquad \qquad\qquad\quad+\chi_1^{2a_1+2a_2-1}\chi_2^{2a_1+2a_2+2a_3-1}\cFF{1}{2(a_1+a_2+a_3+a_4)-1;2a_2,2a_3}{2(a_1+a_2+a_3)}{\chi_1\chi_2,\chi_2}  \bigg] \, . 
\end{align}
Here we introduced the rescaled Appell $F_1$ and Horn $G_2$ functions 
\begin{align}
	\cFF{1}{a;b_1,b_2}{c}{x_1,x_2} &=\frac{A_0(\sfrac{c}{2})}{A_0(\sfrac{a}{2})A_0(\sfrac{b_1}{2})A_0(\sfrac{b_2}{2})}\FF{1}{a;b_1,b_2}{c}{x_1,x_2} \,, \\
	\cGH{2}{a_1,a_2}{b_1,b_2}{x_1,x_2} &=\frac{1}{A_0(\sfrac{a_1}{2})A_0(\sfrac{a_2}{2})A_0(\sfrac{b_1}{2})A_0(\sfrac{b_2}{2})}\GH{2}{a_1,a_2}{b_1,b_2}{x_1,x_2}  \,.
\end{align}
If, moreover, 
we have $|\chi_1|,|\chi_2|<1$ and $\chi_1>0,\chi_2<0$,
we may use formula \eqref{G2 to F1}, which relates the Horn function to the analytic continuation of the Appell function, and rewrite the result as 
\begin{align}
	I_4&=\frac{1}{\sqrt{\pi}|x_{12}|^{2a_1+2a_2-1}|x_{13}|^{2a_3}|x_{14}|^{2a_4}} \nonumber\\
	&\quad \times \Big[-\frac{\Gamma(2\tilde{a}_1)\Gamma(2a_{12}-1)\sin(2\pi a_{12})\sin(\pi a_{23})}{\Gamma(2a_2)\cos(\pi a_2)\cos(\pi a_3)} \FF{1}{2\tilde{a}_1;2a_4,2a_3}{2-2a_{12}}{\chi_1\chi_2,\chi_1}  \\
	&\qquad +\frac{2\Gamma(2\tilde{a}_1)\Gamma(2\tilde{a}_4)\sin(\pi a_{12})\sin(\pi a_{13})}{\Gamma(2-2a_{41})\cos(\pi a_3)}\chi_1^{2a_1-1}(-\chi_2)^{2a_1-1} \FF{1}{2\tilde{a}_1;2a_2,2a_3}{2-2a_{41}}{\frac{1}{\chi_1\chi_2},\frac{1}{\chi_2}} \nonumber\\
	&\qquad -\frac{2\Gamma(1-2a_{13})\Gamma(2\tilde{a}_4)\sin(\pi a_{13})}{\Gamma(2-2a_{14})}\left( \frac{\sin(\pi a_{12})}{\cos(\pi a_3)}+\frac{\sin(\pi a_4)}{\cos(\pi a_{14})}\right) \nonumber \\
	&\qquad\qquad\qquad \times \chi_1^{2a_{12}-1} (-\chi_2)^{2a_{13}-1}\FF{1}{2a_{14}-1;2a_2,2a_3}{2a_{13}}{\chi_1\chi_2,\chi_2}\Big] \,. \nonumber
\end{align}
Here we used the abbreviations $\tilde{a}=1/2-a$ and $a_{ij}=a_i+\dots +a_j$ for $i<j$ with cyclic identifications. In particular $a_{41}=a_4+a_1$. Note that the above expression only involves the Appell $F_1$ function, there is however no region in $(\chi_1,\chi_2)$ space where all terms converge as series (c.f., \eqref{eq:defF1}). Indeed there is no known local solution to the $F_1$ differential equation system which is expressed solely in terms of $F_1$  \cite{Goto:2022uo}. 

\subsection{H-Integral  (Two-Loop Triangle Track)}

Let us now turn to the $H$-integral, which we decompose as
\begin{equation}
	I_{2,2}	= \includegraphicsbox{FigHIntegral.pdf}
	= \frac{\phi_{2,2}(\chi_1,\chi_2)}{|x_{12}|^{2(a_1+a_2+a_3+a_4+b-1)}} \,, 
\end{equation}
with 
\begin{equation}
\label{eq:chisH}
	\chi_1=\frac{x_{23}}{x_{21}},\qquad\chi_2=\frac{x_{14}}{x_{12}}\,.
\end{equation}
We will proceed to bootstrap the function $\phi_{2,2}(\chi_1,\chi_2)$. 
We can find a complete set of two differential equations from the two-point symmetries acting on $x_2,x_3$ and $x_1,x_4$, respectively. We find 
\begin{align}
	&\Big[ 4a_3(a_{14}+b-1)+\brk1{(2a_{14}+2a_3+2b-1)\chi_1-2a_2-2a_3}\partial_{\chi_1} +2a_3\chi_2\partial_{\chi_2}  \nonumber\\
	& \qquad +\chi_1\chi_2\partial_{\chi_1}\partial_{\chi_2}+\chi_1(\chi_1-1)\partial_{\chi_1}^2 \Big]\phi_{2,2}=0 \,, \\
	&\Big[ 4a_4(1-a_{14}-b)-2a_4\chi_1\partial_{\chi_1}+\brk1{2a_1+2a_4+(1-2a_{14}-2a_4-2b)\chi_2}\partial_{\chi_2} \nonumber \\
	& \qquad -\chi_1\chi_2\partial_{\chi_1}\partial_{\chi_2}+\chi_2(1-\chi_2)\partial_{\chi_2}^2\Big]\phi_{2,2}=0 \,,
\end{align}
with $a_{14}=a_1+a_2+a_3+a_4$. These equations can be identified with the Appell $F_2$ system to immediately write down a basis for the solution space, but we will not do so to test and showcase the bootstrap. For these differential operators we find the indicials
\begin{align}
	&(0,0) ,  \nonumber\\
	&(1-2a_2-2a_3,0),  \\
	&(0,1-2a_1-2a_4), \nonumber \\
	&(1-2a_2-2a_3,1-2a_1-2a_4)  \,. \nonumber
\end{align}
We hence expect a solution of the form
\begin{equation}
	\phi_{2,2}=c_1 f_1+c_2\chi_1^{1-2a_2-2a_3}f_2+c_3\chi_2^{1-2a_1-2a_4}f_3+c_4\chi_1^{1-2a_2-2a_3}\chi_2^{1-2a_1-2a_4}f_4 \,,
\end{equation}
for coefficients $c_i$ and power series $f_i$. We can find the fundamental solution $f_1$ by making a series ansatz. The resulting series can, as expected, be identified as an Appell $F_2$ function
\begin{equation}
	f_1(\chi_1,\chi_2)=\FF{2}{2a_{14}+2b-1;2a_3,2a_4}{2(a_2+a_3),2(a_1+a_4)}{\chi_1,\chi_2} \,.
\end{equation}
Shifting the summation variables by the indicials yields the other series
\begin{align}
	f_2(\chi_1,\chi_2)&= \FF{2}{2(a_1+a_4+b)-1;1-2a_2,2a_4}{2(1-a_2-a_3), 2(a_1+a_4)}{\chi_1,\chi_2} \,, \nonumber \\
	f_3(\chi_1,\chi_2)&=\FF{2}{2(a_2+a_3+b)-1;2a_3,1-2a_1}{2(a_2+a_3),2(1-a_1-a_4)}{\chi_1,\chi_2} \,, \\
	f_4(\chi_1,\chi_2)&=\FF{2}{2b;1-2a_2,1-2a_1}{2(1-a_2-a_3),2(1-a_1-a_4)}{\chi_1,\chi_2} \,. \nonumber
\end{align}
Finally we need to compute the coefficients $c_i$ as limits of the integral representation of $\phi_{2,2}$:
\begin{equation}
	\phi_{2,2}(\chi_1,\chi_2)=\int_{\mathbb{R}^2}\frac{\dd y_1\dd y_2}{\pi}\frac{1}{|y_1|^{2a_1}|y_2|^{2a_2}|y_2-\chi_1|^{2a_3}|y_1-\chi_2|^{2a_4}|1-y_1-y_2|^{2b}} \,.
\end{equation}
We obtain
\begin{align}
	c_1&=\lim_{\chi_1,\chi_2\rightarrow 0}\phi_{2,2}(\chi_1,\chi_2)=\frac{A_0(a_2+a_3)A_0(a_1+a_4)A_0(b)}{A_0\left(a_{14}+b-1 \right)} \,, \\
	c_2&=\lim_{\chi_1,\chi_2\rightarrow 0}\chi_1^{2a_2+2a_3-1}\phi_{2,2}(\chi_1,\chi_2)=\frac{A_0(a_2)A_0(a_3)A_0(b)A_0(a_1+a_4)A_0(1-a_2-a_3)}{A_0(a_1+a_4+b-1/2)} \,, \\
	c_3&=\lim_{\chi_1,\chi_2\rightarrow 0}\chi_2^{2a_1+2a_4-1}\phi_{2,2}(\chi_1,\chi_2)=\frac{A_0(a_1)A_0(a_4)A_0(b) A_0(a_2+a_3) A_0(1-a_1-a_4)}{A_0(a_2+a_3+b-1/2)} \,, \\
	c_4&=\lim_{\chi_1,\chi_2\rightarrow 0}\chi_1^{2a_2+2a_3-1}\chi_2^{2a_1+2a_4-1}\phi_{2,2}(\chi_1,\chi_2)=\frac{A_0(a_1)A_0(a_2)A_0(a_3)A_0(a_4)}{A_0(a_1+a_4-1/2)A_0(a_2+a_3-1/2)} \,.
\end{align}
Putting everything together, we find
\begin{align}\label{H explicit}
	I_{2,2} = \frac{A_0(a_3)A_0(a_4)A_0(b)}{|x_{12}|^{2a_{14}+2b-2}}\bigg[&\mathcal{F}_2\brk[s]*{\substack{2a_{14}+2b-2;2a_3,2a_4\\  2(a_2+a_3),2(a_1+a_4)};\chi_1,\chi_2}
	\nonumber\\
	&+ \chi_1^{1-2a_2-2a_3} \mathcal{F}_2\brk[s]*{\substack{2(a_1+a_4-\tilde{b});2\tilde{a}_2,2a_4 \\ 2(\tilde{a}_2+\tilde{a}_3), 2(a_1+a_4)};\chi_1,\chi_2}  
	 \nonumber \\
	&+ \chi_2^{1-2a_1-2a_4}\mathcal{F}_2\brk[s]*{\substack{2(a_2+a_3-\tilde{b});2a_3,2\tilde{a}_1 \\ 2(a_2+a_3),2(\tilde{a}_1+\tilde{a}_4)};\chi_1,\chi_2}
	\nonumber\\
	&+\chi_1^{1-2a_2-2a_3}  \chi_2^{1-2a_1-2a_4} \mathcal{F}_2\brk[s]*{\substack{2b;2\tilde{a}_2,2\tilde{a}_1 \\ 2(\tilde{a}_2+\tilde{a}_3),2(\tilde{a}_1+\tilde{a}_4)};\chi_1,\chi_2} \bigg]\,,
\end{align}
where we used $\tilde{a} = 1/2 - a$ (also recall $a_{14}=a_1+\dots +a_4$), and introduced the rescaled Appell $F_2$ function
\begin{equation}
	\cFF{2}{a;b_1,b_2}{c_1,c_2}{x_1,x_2}=\frac{A_0(\sfrac{c_1}{2})A_0(\sfrac{c_2}{2})}{A_0(\sfrac{a}{2})A_0(\sfrac{b_1}{2})A_0(\sfrac{b_2}{2})}\FF{2}{a;b_1,b_2}{c_1,c_2}{x_1,x_2} \,.
\end{equation}

\paragraph{Star Limit.}

We remark that we can use the identity \eqref{prop to delta}
 (here specified to $D=1$)
to recover the box integral as a limit of the integral $I_{2,2}$ just computed:
\begin{equation}
	I_4 
	=
	\includegraphicsbox{FigFourPointStar.pdf}= \lim_{b\to 1/2} {A_0^{-1}(b)}\includegraphicsbox{FigHIntegral.pdf}
	= \lim_{b\to 1/2} \frac{I_{2,2}}{A_0(b)}\, .
\end{equation}
Taking the limit directly in the final result \eqref{H explicit} yields
\begin{align}
	I_4 = \frac{A_0(a_3)A_0(a_4)}{|x_{12}|^{2\scriptstyle\sum a_i-1}}\bigg[
	&\mathcal{F}_2\brk[s]*{\substack{2\sum_i a_i-1;2a_3,2a_4\\ 2(a_2+a_3),2(a_1+a_4)};\chi_1,\chi_2}\nonumber\\
	&+ \chi_1^{1-2a_2-2a_3} \mathcal{F}_2\brk[s]*{\substack{2(a_1+a_4);2\tilde{a}_2,2a_4\\  2(\tilde{a}_2+\tilde{a}_3), 2(a_1+a_4)};\chi_1,\chi_2} \nonumber\\
	&+ \chi_2^{1-2a_1-2a_4}\mathcal{F}_2\brk[s]*{\substack{2(a_2+a_3);2a_3,2\tilde{a}_1 \\ 2(a_2+a_3),2(\tilde{a}_1+\tilde{a}_4)};\chi_1,\chi_2}\bigg]\,,
\end{align}
Using the properties \cite{bailey1935generalized}
\begin{equation}
	\FF{2}{\alpha;\beta,\beta'}{\gamma,\alpha}{x,y} = \frac{1}{(1-y)^{\beta'}} \FF{1}{\beta;\alpha-\beta',\beta' }{\gamma}{x,\frac{x}{1-y}}
\end{equation}
and \cite{srivastavaBook}\footnote{We are grateful to Souvik Bera and Tanay Pathak for pointing us to this reference.}
\begin{equation}
	\FF{2}{\gamma+\gamma'-1;\beta,\beta'}{\gamma,\gamma'}{x,y} = \frac{1}{(1-x)^{\beta}(1-y)^{\beta'}} \GH{2}{\beta,\beta'}{1-\gamma,1-\gamma'}{\frac{x}{1-x},\frac{y}{1-y}}
\end{equation}
of the Appell and Horn functions, which can e.g. be derived using \cite{Ananthanarayan:2021yar}, we can rewrite our result as
\begin{align}\label{S explicit}
	I_4 =& \frac{A_0(1-\sum_i a_i) A_0(a_1+a_4) A_0(a_2+a_3)}{|x_{12}|^{2(a_1+a_2)-1} |x_{13}|^{2a_3} |x_{24}|^{2a_4}} \GH{2}{2a_3,2a_4}{1-2(a_2+a_3),1-2(a_1+a_4)}{\frac{\chi_1}{1-\chi_1},\frac{\chi_2}{1-\chi_2}} \nonumber\\
	&+ \frac{A_0(a_2) A_0(a_3) A_0(\tilde{a}_2+\tilde{a}_3)}{|x_{12}|^{2a_1} |x_{23}|^{2(a_2+a_3)-1} |x_{24}|^{2a_4}} \FF{1}{2\tilde{a}_2;2a_1,2a_4}{2(\tilde{a}_2+\tilde{a}_3)}{\chi_1,\frac{\chi_1}{1-\chi_2}} \nonumber\\
	&+ \frac{A_0(a_1) A_0(a_4) A_0(\tilde{a}_1+\tilde{a}_4)}{|x_{12}|^{2a_2} |x_{13}|^{2a_3} |x_{14}|^{2(a_1+a_4)-1}} \FF{1}{2\tilde{a}_1;2a_3,2a_2}{2(\tilde{a}_1+\tilde{a}_4)}{\frac{\chi_2}{1-\chi_1},\chi_2} \, .
\end{align}
To compare this result to our earlier result \eqref{star4Result} for the star integral we need to consider a permutation of the result \eqref{star4Result} such that the resulting series expression has a common domain of convergence with the expression found here. Indeed the permutation $x_2\leftrightarrow x_4, \, a_2\leftrightarrow a_4$ achieves precisely this. The common domain of convergence is given by  $\chi_1\in (0,\sfrac{1}{2}), \chi_2\in (-1+\chi_1,\sfrac{1}{2})$ with $\chi_1,\chi_2$ as in \eqref{eq:chisH}. We hence have two different expressions for the star integral in this domain which are not obviously the same. They only agree if the following hypergeometric identity holds, which we have not found in the literature but checked numerically:
\begin{align}
\label{eq:G2F1Identity}
	&\frac{A_0(a_2+a_3)A_0(a_1+a_4)}{A_0(a_{14}-\sfrac{1}{2})}(1-\chi_1)^{-2a_3}(1-\chi_2)^{-2a_4} \GH{2}{2a_3,2a_4}{1-2a_2-2a_3,1-2a_1-2a_4}{\frac{\chi_1}{1-\chi_1},\frac{\chi_2}{1-\chi_2}}  \nonumber\\
	&\quad - \frac{A_0(a_3)A_0(a_1+a_4)}{A_0(a_{31}-\sfrac{1}{2})}(1-\chi_1)^{1-2a_1-2a_3-2a_4}\GH{2}{2a_4,2a_2}{1-2a_1-2a_4,2a_{31}-1}{\frac{\chi_2}{\chi_1-1},\chi_1-1} \nonumber\\
	&=\frac{A_0(a_2)A_0(a_{31})}{A_0(a_{14}-\sfrac{1}{2})} \FF{1}{2a_{14}-1,2a_4,2a_3}{2a_{31}}{\chi_2,1-\chi_1} \\
	&\quad -\frac{A_0(a_2)A_0(a_3)}{A_0(a_2+a_3-\sfrac{1}{2})}\chi_1^{1-2a_2-2a_3}(1-\chi_2)^{-2a_4}\FF{1}{2\tilde{a}_2,2a_1,2a_4}{2(\tilde{a}_2+\tilde{a}_3)}{\chi_1,\frac{\chi_1}{1-\chi_2}} 	 \,. \nonumber
\end{align}
Here $a_{31}=a_3+a_4+a_1$.

\section{Five-Point Integrals}
\label{sec:5pt}

We will now move on to five-point integrals. We will only consider the triangle-box integral, as the other five-point integrals (the pentagon and the three-loop triangle-track) are members of integral families to be considered in full generality in \Secref{sec:polygons} and \Secref{sec:triangletracks}.


\subsection{Triangle-Box Integral}

Let us consider
\begin{equation}
	I_{2,3} =
	\includegraphicsbox{FigTriangleBox.pdf}
	=
	 \int \frac{\pi^{-1} \dd x_0\, \dd x_{0'}}{|x_{10}|^{2a_1} |x_{50}|^{2a_5} |x_{00'}|^{2b} |x_{40'}|^{2a_4} |x_{30'}|^{2a_3} |x_{20'}|^{2a_2}}\, .
\end{equation}
We decompose the integral as
\begin{equation}
	I_{2,3}=\frac{\phi_{2,3}(\chi_1,\chi_2,\chi_3)}{|x_{12}|^{2(a_1+a_5+b)-1}|x_{24}|^{2(a_2+a_3+a_4)-1}} \,,
\end{equation}
with
\begin{equation}
	\chi_1=\frac{x_{23}}{x_{24}},\qquad \chi_2=\frac{x_{24}}{x_{21}},\qquad \chi_3=\frac{x_{15}}{x_{12}} \,.
\end{equation}
From the two-point symmetries acting on the external points $(x_1,x_5)$, $(x_2,x_3)$ and $(x_3,x_4)$, respectively, we find the following differential equations for $\phi_{2,3}$:
\begin{align}
	&\Big[ 2a_5(1\!-\!2a_{51}-2b)-2a_5\chi_2\partial_{\chi_2}+2(a_1\!+\!a_5-(a_{51}+b\!+\!a_5)\chi_3)\partial_{\chi_3} \nonumber\\
	&\qquad\qquad -\chi_2\chi_3\partial_{\chi_2}\partial_{\chi_3}+\chi_3(1\!-\!\chi_3)\partial_{\chi_3}^2\Big]\phi_{2,3}=0 \,, \\
	&\Big[  2a_3(2a_{24}-1+(2a_{51}+2b-1)\chi_2)+2a_3\chi_2(\chi_2-1)\partial_{\chi_2}  +2a_3\chi_2\chi_3\partial_{\chi_3} \nonumber \\
	&\qquad\qquad+((2a_{24}+2a_3+(2a_{51}+2b-1)\chi_2)\chi_1-2a_2-2a_3)\partial_{\chi_1}    \\
	&\qquad\qquad +\chi_1\chi_2(\chi_2-1)\partial_{\chi_1}\partial_{\chi_2}+\chi_1\chi_2\chi_3\partial_{\chi_1}\partial_{\chi_3} +\chi_1(\chi_1-1)\partial_{\chi_1}^2 \Big] \phi_{2,3}=0 \,, \nonumber \\
	&\Big[ 2a_3(2a_{24}\!-\!1)+2((a_{24}\!+\!a_3)\chi_1\!-\!a_2\!-\!a_3)\partial_{\chi_1}-2a_3\chi_2\partial_{\chi_2}+ \nonumber\\
	&\qquad\qquad (1\!-\!\chi_1)(\chi_2\partial_{\chi_2} \!-\! \chi_1\partial_{\chi_1})\partial_{\chi_1} \Big]\phi_{2,3}=0 \,,  
\end{align}
with the shorthands
\begin{equation}
	a_{51}=a_5+a_1,\qquad a_{24}=a_2+a_3+a_4 \,.
\end{equation}
From these differential equations we find the indicials $(r_1^{(i)},r_2^{(i)},r_3^{(i)})$ for $i=1,\dots 6$
\begin{align}
	&(0,0,0), \nonumber \\
	&(1-2a_2-2a_3,0,0), \nonumber\\
	&(0,2(a_2+a_3+a_4)-1, 0) , \nonumber\\
	&(0,0,1-2a_1-2a_5),  \\
	&(1-2a_2-2a_3,0,1-2a_1-2a_5), \nonumber \\
	&(0,2(a_2+a_3+a_4)-1,1-2a_1-2a_5) \,.\nonumber
\end{align}
The solution will then take the general form
\begin{equation}
	\phi_{2,3}(\chi_1,\chi_2,\chi_3)=\sum_{i=1}^6 c_i \, \chi_1^{r_1^{(i)}}\chi_2^{r_2^{(i)}}\chi_3^{r_3^{(i)}} f_i(\chi_1,\chi_2,\chi_3) \,,
\end{equation}
for some coefficients $c_i$ and power series $f_i$. The fundamental series is given by
\begin{equation}
	f_1(\chi_1,\chi_2,\chi_3)=\HH{1}{2a_3,2a_5,2(a_1+a_5+b)-1}{2(a_2+a_3+a_4)-1,2(a_2+a_3),2(a_1+a_5)}{\chi_1,\chi_2,\chi_3} \,,
\end{equation}
with the $\mathcal{H}_1$ series defined in \Appref{app:defs}. Shifting the summation indices by the indicials we can then find the other solutions.
By taking appropriate limits we can finally compute the coefficients $c_i$, as before. This yields the result 
\begin{align}
\label{eq:resultTriangleBox}
	I_{2,3} = &\frac{A_0(a_3)A_0(a_4)A_0(a_5)A_0(b)}{|x_{12}|^{2(a_1+a_5+b)-1}|x_{24}|^{2(a_2+a_3+a_4)-1}} 
	\nonumber\\
	\times&\bigg[
	\HH{1}{2a_3,2a_5;2b+2a_{51}-1}{2a_{24}-1,2a_{23};2a_{51}}{\chi_1,\chi_2,\chi_3} \nonumber \\
	&+ \chi_1^{1-2a_{23}}\HH{2}{2a_4,2a_5;2\tilde{a}_2;2b+2a_{51}-1}{2-2a_{23};2a_{51}}{\chi_1,\chi_1\chi_2,\chi_3} \\
	&+\chi_2^{2a_{24}-1} \HH{3}{2a_3,2a_4,2a_5;2(b+a_{15}-1)}{2a_{24},2a_{51}}{\chi_1\chi_2,\chi_2,\chi_3} \nonumber\\
	&+ \chi_3^{1-2a_{51}}  \HH{1}{2a_3,2\tilde{a}_1;2b}{2a_{24}-1,2a_{23};2-2a_{51}}{\chi_1,\chi_2,\chi_3} \nonumber\\	
	&+ \chi_1^{1-2a_{23}}\chi_3^{1-2a_{51}} \HH{2}{2a_4,2\tilde{a}_1;2\tilde{a}_2;2b}{2-2a_{23};2-2a_{51}}{\chi_1,\chi_1\chi_2,\chi_3}\nonumber \\
	&+\chi_2^{2a_{24}-1}\chi_3^{1-2a_{51}}  \HH{3}{2a_3,2a_4,2\tilde{a}_1;2(b+a_{24})-1}{2a_{24},2-2a_{51}}{\chi_1\chi_2,\chi_2,\chi_3} \bigg] \,, \nonumber
\end{align}
with the abbreviations $\tilde{a}=1/2-a$ and $a_{ij}=a_i+\dots +a_j$ with $i<j$ understood cyclically. In particular $a_{51}=a_5+a_1$. The hypergeometric series in the above expression are defined in \appref{app:defs}.

\section{Six-Point Integrals}
\label{sec:6pt}

In this section we will consider six-point integrals. We will compute the double-box integral as well as the triangle-triangle-box and triangle-box-triangle integrals explicitly from the bootstrap. The remaining six-point integrals, namely the four-loop triangle-track integral and the hexagon are omitted here as they are members of integral families considered in full generality in \Secref{sec:polygons} and \Secref{sec:triangletracks}.

\subsection{Double-Box Integral (Two-Loop Train Track)}
\label{sec:DoubleBoxNonConf}

Consider the following non-conformal, double-box integral:
\begin{equation}
	I_{3,3} 
	=\includegraphicsbox{FigDoubleBox.pdf}
	= \int \frac{\pi^{-1} \dd x_0\, \dd x_{0'}}{|x_{10}|^{2a_1} |x_{60}|^{2a_6} |x_{50}|^{2a_5} |x_{00'}|^{2b} |x_{40'}|^{2a_4} |x_{30'}|^{2a_3} |x_{20'}|^{2a_2}}\, .
\end{equation}
We will decompose the integral as
\begin{equation}
	I_{3,3}=\frac{\phi_{3,3}(\chi_1,\chi_2,\chi_3,\chi_4)}{|x_{12}|^{2b}|x_{24}|^{2(a_2+a_3+a_4)-1}|x_{16}|^{2(a_1+a_5+a_6)-1}} \,,
\end{equation}
with the ratios
\begin{equation}
	\chi_1=\frac{x_{23}}{x_{24}},\qquad \chi_2=\frac{x_{24}}{x_{21}}, \qquad \chi_3=\frac{x_{15}}{x_{16}}, \qquad \chi_4=\frac{x_{16}}{x_{12}} \,.
\end{equation}
From the two-point symmetries acting on the pairs of external points $(x_1,x_6)$, $(x_5,x_6)$, $(x_3,x_4)$,  $(x_2,x_4)$ we find the differential equations
\begin{align}
	&\Big[ (1-2a_1-2a_5)\chi_2\chi_4\partial_{\chi_2}+2(a_1+a_5-(a_{51}+a_5+b\chi_4)\chi_3)\partial_{\chi_3}-\chi_2\chi_3\chi_4\partial_{\chi_2}\partial_{\chi_3} \nonumber\\
	&\qquad -2(a_5(2a_{51}-1)+b(2a_1+2a_5-1)\chi_4)+2\chi_4(a_{51}+a_5-1+(1-a_1-a_5+b)\chi_4)\partial_{\chi_4}  \nonumber\\
	&\qquad -\chi_4(1+(\chi_4-2)\chi_3)\partial_{\chi_3}\partial_{\chi_4}+\chi_2\chi_4^2\partial_{\chi_2}\partial_{\chi_4}+\chi_3(1-\chi_3)\partial_{\chi_3}^2+\chi_4^2(\chi_4-1)\partial_{\chi_4}^2\Big]\phi_{3,3}=0 \,,  \nonumber\\
	&\Big[2a_5(1-2a_{51})+(2a_1+2a_5-2(a_{51}+a_5)\chi_3)\partial_{\chi_3}+2a_5\chi_4\partial_{\chi_4}  \nonumber\\
	&\qquad +\chi_4(\chi_3-1)\partial_{\chi_3}\partial_{\chi_4}+\chi_3(1-\chi_3)\partial_{\chi_3}^2\Big]\phi_{3,3}=0 \,,  \nonumber \\
	&\Big[2a_3(2a_{24}-1)+2((a_{24}+a_3)\chi_1-a_2-a_3)\partial_{\chi_1}-2a_3\chi_2\partial_{\chi_2}\\
	&\qquad +\chi_2(1-\chi_1)\partial_{\chi_1}\partial_{\chi_2}+\chi_1(\chi_1-1)\partial_{\chi_1}^2\Big]\phi_{3,3}=0 \,,  \nonumber\\
	&\Big[2(a_3(1-2a_{24})+b(1-2a_2-2a_3)\chi_2)+2(a_2+a_3-(a_{24}+a_3+b\chi_2)\chi_1)\partial_{\chi_1}  \nonumber\\
	&\qquad +2\chi_2(a_{24}+a_3-1+(1+b-a_2-a_3)\chi_2)\partial_{\chi_2}+(1-2a_2-2a_3)\chi_2\chi_4\partial_{\chi_4}  \nonumber\\
	&\qquad -(1+(\chi_2-2)\chi_1)\chi_2\partial_{\chi_1}\partial_{\chi_2}-\chi_1\chi_2\chi_4\partial_{\chi_1}\partial_{\chi_4}  \nonumber\\
	&\qquad +\chi_2^2\chi_4\partial_{\chi_2}\partial_{\chi_4}+\chi_1(1-\chi_1)\partial_{\chi_1}^2+\chi_2^2(\chi_2-1)\partial_{\chi_2}^2\Big]\phi_{3,3}=0 \,,\nonumber
\end{align}
with the abbreviations $a_{51}=a_5+a_6+a_1$ and $a_{24}=a_2+a_3+a_4$. From the differential equations we find the following indicials $(r_1^{(i)},\dots r_4^{(i)})$ for $i=1,\dots ,9$
 \begin{align}
 	&(0,0,0,0), \nonumber\\
	&(1-2a_2-2a_3,0,0,0), \nonumber\\
	&(0,2(a_2+a_3+a_4)-1,0,0),\nonumber \\
	&(0,0,1-2a_1-2a_5,0),\nonumber \\
	&(0,0,0,2(a_1+a_5+a_6)-1), \\
	&(1-2a_2-2a_3,0,1-2a_1-2a_5,0), \nonumber\\
	&(1-2a_2-2a_3,0,0,2(a_1+a_5+a_6)-1) , \nonumber \\
	&(0,2(a_2+a_3+a_4)-1,1-2a_1-2a_5,0) , \nonumber\\
	&(0,2(a_2+a_3+a_4)-1,0,2(a_1+a_5+a_6)-1)  \,. \nonumber
 \end{align}
The solution hence takes the general form
\begin{equation}
	\phi_{3,3}(\chi_1,\dots \chi_4)=\sum_{i=1}^9 c_i \chi_1^{r_1^{(i)}}\chi_2^{r_2^{(i)}}\chi_3^{r_3^{(i)}}\chi_4^{r_4^{(i)}} f_i(\chi_1,\dots ,\chi_4) \,,
\end{equation}
for some coefficients $c_i$ and power series $f_i$. The fundamental solution $f_1$ is found to be
\begin{equation}
	f_1(\chi_1,\chi_2,\chi_3,\chi_4)=\BB{1}{2a_3,2a_5;2b}{2a_{24}-1,2(a_2+a_3),2a_{51}-1,2(a_1+a_5)}{\chi_1,\chi_2,\chi_3,\chi_4} \,,
\end{equation}
defined in \appref{app:defs}. Shifting the summation variables by the indicials yields the other series.
Computing the coefficients as before we find the full result 
\begin{align}\label{eq:resultDoubleBox}
	I_{3,3}=&\frac{A_0(a_3)A_0(a_4)A_0(a_5)A_0(a_6)A_0(b)}{|x_{12}|^{2b}|x_{24}|^{2a_{24}-1}|x_{16}|^{2a_{51}-1}}
	\nonumber \\
	&\times \bigg[ \BB{1}{2a_3,2a_5;2b}{2a_{24}-1,2(a_2+a_3),2a_{51}-1,2(a_1+a_5)}{\chi_1,\chi_2,\chi_3,\chi_4} \nonumber\\
	&\quad +\chi_1^{1-2a_2-2a_3}  \BB{2}{2a_4,2a_5;2b;2\tilde{a}_2}{2(\tilde{a}_2+\tilde{a}_3),2a_{51}-1,2(a_1+a_5)}{\chi_1,\chi_1\chi_2,\chi_3,\chi_4}\nonumber \\
	&\quad + \chi_2^{2a_{24}-1}\BB{3}{2a_3,2a_4,2a_5;2b+2a_{24}-1}{2a_{51}-1,2(a_1+a_5);2a_{24}}{\chi_1\chi_2,\chi_2,\chi_3,\chi_4}\nonumber\\
	&\quad + \chi_3^{1-2a_1-2a_5} \BB{2}{2a_6,2a_3;2b;2\tilde{a}_1}{2(\tilde{a}_1+\tilde{a}_5),2a_{24}-1,2(a_2+a_3)}{\chi_3,\chi_3\chi_4,\chi_1,\chi_2} \\
	&\quad +\chi_4^{2a_{51}-1}\BB{3}{2a_5,2a_6,2a_3;2b+2a_{51}-1}{2a_{24}-1,2(a_2+a_3);2a_{51}}{\chi_3\chi_4,\chi_4,\chi_1,\chi_2} \nonumber\\
	&\quad + \chi_1^{1-2a_2-2a_3}\chi_3^{1-2a_1-2a_5}\BB{4}{2a_4,2a_6;2b;2\tilde{a}_2,2\tilde{a}_1}{2(\tilde{a}_2+\tilde{a}_3),2(\tilde{a}_1+\tilde{a}_5)}{\chi_1,\chi_1\chi_2,\chi_3,\chi_3\chi_4} \nonumber \\
	&\quad +\chi_1^{1-2a_2-2a_3}\chi_4^{2a_{51}-1}\BB{5}{2a_5,2a_6,2a_4;2b+2a_{51}-1;2\tilde{a}_2}{2a_{51},2(\tilde{a}_2+\tilde{a}_3)}{\chi_3\chi_4,\chi_4,\chi_1,\chi_1\chi_2} \nonumber\\
	&\quad +\chi_2^{2a_{24}-1}\chi_3^{1-2a_1-2a_5}\BB{5}{2a_3,2a_4,2a_6;2b+2a_{24}-1;2\tilde{a}_1}{2a_{24},2(\tilde{a}_1+\tilde{a}_5)}{\chi_1\chi_2,\chi_2,\chi_3,\chi_3\chi_4}\nonumber\\
	&\quad + \chi_2^{2a_{24}-1}\chi_4^{2a_{51}-1} \BB{6}{2a_3,2a_4,2a_5,2a_6;2b+2a_{16}-2}{2a_{24},2a_{51}}{\chi_1\chi_2,\chi_2,\chi_3\chi_4,\chi_4} \bigg] \,, \nonumber
\end{align}
with the series defined in \appref{app:defs}. Here we used the abbreviations $\tilde{a}=1/2-a$ and $a_{16}=a_1+\dots +a_6$. Also recall $a_{51}=a_5+a_6+a_1,\, a_{24}=a_2+a_3+a_4$.


\subsection{Triangle-Pentagon Integral}

Let us now consider the integral
\begin{equation}
	I_{2,4}=\includegraphicsbox{FigTrianglePentagon.pdf}
	=\int\frac{\pi^{-1}\dd x_0\dd x_{0'}}{|x_{10}|^{2a_1}|x_{60}|^{2a_6}|x_{20'}|^{2a_2}|x_{30'}|^{2a_3}|x_{40'}|^{2a_4}|x_{50'}|^{2a_5}|x_{00'}|^{2b}} \,.
\end{equation}
We will decompose it as
\begin{equation}
	I_{2,4}=\frac{\phi_{2,4}(\chi_1,\chi_2,\chi_3,\chi_4)}{|x_{12}|^{2(a_1+a_6+b)-1}|x_{25}|^{2(a_2+a_3+a_4+a_5)-1}} \,,
\end{equation}
with
\begin{equation}
	\chi_1=\frac{x_{23}}{x_{24}},\qquad \chi_2=\frac{x_{24}}{x_{25}}, \qquad \chi_3=\frac{x_{25}}{x_{21}}, \qquad \chi_4=\frac{x_{16}}{x_{12}} \,.
\end{equation}
We can obtain a of complete set of differential equations from the two-point symmetries acting on the pairs of external points $\{(x_1,x_6),\, (x_2,x_3), \, (x_3,x_4), \, (x_4,x_5) \}$. The differential equations read 
\begin{align}
	&\Big[2a_6(1-2(a_1+a_6+b))-2a_6\chi_3\partial_{\chi_3}+(2a_1+2a_6-2(a_1+2a_6+b)\chi_4)\partial_{\chi_4} \nonumber\\
	&\qquad -\chi_3\chi_4\partial_{\chi_3}\partial_{\chi_4}+\chi_4(1-\chi_4)\partial_{\chi_4}^2\Big]\phi_{2,4}=0 \,,\nonumber \\
	&\Big[(-2(a_2+a_3)+(1+2a_3+(2a_{25}-1+(2(a_1+a_6+b)-1)\chi_3)\chi_2)\chi_1)\partial_{\chi_1} \nonumber\\
	&\qquad +2a_3\chi_2(2a_{25}-1+(2(a_1+a_6+b)-1)\chi_3)+2a_3\chi_2(\chi_2-1)\partial_{\chi_2}+2a_3\chi_2\chi_3(\chi_3-1)\partial_{\chi_3}\nonumber \\
	&\qquad +2a_3\chi_2\chi_3\chi_4\partial_{\chi_4}+\chi_1\chi_2(\chi_2-1)\partial_{\chi_1}\partial_{\chi_2}+\chi_1\chi_2\chi_3(\chi_3-1)\partial_{\chi_1}\partial_{\chi_3} \nonumber\\
	&\qquad +\chi_1\chi_2\chi_3\chi_4\partial_{\chi_1}\partial_{\chi_4}+\chi_1(\chi_1-1)\partial_{\chi_1}^2\Big]\phi_{2,4}=0 \,, \\
	&\Big[(2a_4-1+(1+2a_3)\chi_1)\partial_{\chi_1}-2a_3\chi_2\partial_{\chi_2}+\chi_2(1-\chi_1)\partial_{\chi_1}\partial_{\chi_2}+\chi_1(\chi_1-1)\partial_{\chi_1}^2\Big]\phi_{2,4}=0 \,, \nonumber\\
	&\Big[2a_4(2a_{25}-1)\chi_2+\chi_1(2a_{24}-1+(1-2a_{25})\chi_2)\partial_{\chi_1}+2\chi_2((a_{25}+a_4)\chi_2-a_{24})\partial_{\chi_2}\nonumber \\
	&\qquad -2a_4\chi_2\chi_3\partial_{\chi_3}+\chi_1\chi_2(1-\chi_2)\partial_{\chi_1}\partial_{\chi_2}+\chi_1\chi_3(\chi_2-1)\partial_{\chi_1}\partial_{\chi_3} \nonumber\\
	&\qquad +(1-\chi_2)\chi_2\chi_3\partial_{\chi_2}\partial_{\chi_3}+\chi_2^2(\chi_2-1)\partial_{\chi_2}^2\Big]\phi_{2,4}=0 \,, \nonumber
\end{align}
with the abbreviations $a_{25}=a_2+a_3+a_4+a_5$ and $a_{24}=a_2+a_3+a_4$. From these differential equations we find the indicials
\begin{align}
	&(0,0,0,0), \notag\\
	&(0,1-2a_2-2a_3-2a_4,0,0), \notag\\
	&(0,0,2(a_2+a_3+a_4+a_5)-1,0), \notag\\
	&(0,0,0,1-2a_1-2a_6), \\
	&(1-2a_2-2a_3,1-2a_2-2a_3-2a_4,0,0), \notag\\
	&(0,1-2a_2-2a_3-2a_4,0,1-2a_1-2a_6),\notag \\
	&(0,0,2(a_2+a_3+a_4+a_5)-1,1-2a_1-2a_6), \notag\\
	&(1-2a_2-2a_3,1-2a_2-2a_3-2a_4,0,1-2a_1-2a_6) \,. \notag
\end{align}
As before we compute the basis solutions and coefficients, which leads to the final result
\begin{align}
	&I_{2,3}=\frac{A_0(a_3)A_0(a_4)A_0(a_5)A_0(a_6)A_0(b)}{|x_{12}|^{2(a_1+a_6+b)-1}|x_{25}|^{2a_{25}-1}} \notag\\
	&\times \bigg[
	\HH{4}{2a_3,2a_4,2a_6;2(b+a_1+a_6)-1}{2a_{25}-1,2a_{24};2(a_1+a_6)}{\chi_1\chi_2,\chi_2,\chi_3,\chi_4}\notag\\
	&\quad +\chi_2^{1-2a_{24}}\HH{5}{2a_3,2a_5,2a_6;2(a_1+a_6+b)-1}{2a_{24}-1,2(a_2+a_3);2(a_1+a_6)}{\chi_1,\chi_2,\chi_2\chi_3,\chi_4} \notag\\
	&\quad +\chi_3^{2a_{25}-1}\HH{6}{2a_3,2a_4,2a_5,2a_6;2a_{16}+2b-2}{2a_{25};2(a_1+a_6)}{\chi_1\chi_2\chi_3,\chi_2\chi_3,\chi_3,\chi_4} \notag\\
	&\quad + \chi_4^{1-2a_1-2a_6}\HH{4}{2a_3,2a_4,2\tilde{a}_1;2b}{2a_{25}-1,2a_{24};2(\tilde{a}_1+\tilde{a}_6)}{\chi_1\chi_2,\chi_2,\chi_3,\chi_4} \\
	&\quad +\chi_1^{1-2a_2-2a_3}\chi_2^{1-2a_{24}}\HH{7}{2a_4,2a_5,2a_6;2(a_1+a_6+b)-1}{2\tilde{a}_2,2(\tilde{a}_2+\tilde{a}_3);2(a_1+a_6)}{\chi_1,\chi_1\chi_2,\chi_1\chi_2\chi_3,\chi_4} \notag\\
	&\quad+\chi_2^{1-2a_{24}}\chi_4^{1-2a_1-2a_6}\HH{5}{2a_3,2a_5,2\tilde{a}_1;2b}{2a_{24}-1,2(a_2+a_3);2(\tilde{a}_1+\tilde{a}_6)}{\chi_1,\chi_2,\chi_2\chi_3,\chi_4}\notag \\
	&\quad +\chi_3^{2a_{25}-1}\chi_4^{1-2a_1-2a_6}\HH{6}{2a_3,2a_4,2a_5,2\tilde{a}_1;2a_{25}+2b-1}{2a_{25};2(\tilde{a}_1+\tilde{a}_6)}{\chi_1\chi_2\chi_3,\chi_2\chi_3,\chi_3,\chi_4}\notag \\
	&\quad +\chi_1^{1-2a_2-2a_3}\chi_2^{1-2a_{24}}\chi_4^{1-2a_1-2a_6}\HH{7}{2a_4,2a_5,2\tilde{a}_1;2b}{2\tilde{a}_2,2(\tilde{a}_2+\tilde{a}_3);2(\tilde{a}_1+\tilde{a}_6)}{\chi_1,\chi_1\chi_2,\chi_1\chi_2\chi_3,\chi_4}\bigg]\,, \notag
\end{align}
with the hypergeometric series defined in \appref{app:defs}. Here we used the abbreviations $\tilde{a}=1/2-a$ and $a_{16}=a_1+\dots +a_6$. Also recall $a_{25}=a_2+a_3+a_4+a_5, \, a_{24}=a_2+a_3+a_4$.
\subsection{Triangle-Triangle-Box Integral}
Let us consider the integral
\begin{align}
	I_{2,2,3}&=\includegraphicsbox{FigTriangleTriangleBox.pdf}
	\nonumber\\
	&=\int\frac{\pi^{-\sfrac{3}{2}}\dd x_0\dd x_{0'} \dd x_{0''}}{|x_{01}|^{2a_1}|x_{06}|^{2a_6}|x_{00'}|^{2b_1}|x_{0'5}|^{2a_5}|x_{0'0''}|^{2b_2}|x_{0''2}|^{2a_2}|x_{0''3}|^{2a_3}|x_{0''4}|^{2a_4}} \,,
\end{align}
which we decompose as
\begin{equation}
	I_{2,2,3}=\frac{\phi_{2,2,3}(\chi_1,\chi_2,\chi_3,\chi_4)}{|x_{16}|^{2a_1}|x_{34}|^{2(a_2+a_3+a_4)-1}|x_{45}|^{2a_5+2b_2-1}|x_{56}|^{2a_6+2b_1-1}} \,,
\end{equation}
with
\begin{equation}
	\chi_1=\frac{x_{56}}{x_{16}}, \qquad  \chi_2=\frac{x_{45}}{x_{65}}, \qquad \chi_3=\frac{x_{34}}{x_{54}}, \qquad \chi_4=\frac{x_{24}}{x_{34}} \,.
\end{equation}
From the two-point symmetries acting on the pairs of points $(x_1,x_6),\,(x_2,x_3),\,(x_2,x_4)$ we find the differential equations
\begin{align}
	&\Big[2a_1(2a_6-1)+2(1-a_6-b_1-(1+a_1-a_6)\chi_1)\partial_{\chi_1}-\chi_2\partial_{\chi_1}\partial_{\chi_2}+\chi_1(1-\chi_1)\partial_{\chi_1}^2\Big]\phi_{2,2,3}=0 \,, \nonumber \\
	&\Big[2a_2(1-2a_{24})+2a_2\chi_3\partial_{\chi_3}+2(a_2+a_4-(a_{24}+a_2)\chi_4)\partial_{\chi_4} \nonumber \\
	&\qquad +\chi_3(\chi_4-1)\partial_{\chi_3}\partial_{\chi_4}+\chi_4(1-\chi_4)\partial_{\chi_4}^2\Big]\phi_{2,2,3}=0 \,,  \\
	&\Big[2a_2(2a_{24}-1+(2a_5+2b_2-1)\chi_3)-2a_2\chi_2\chi_3\partial_{\chi_2}+2a_2\chi_3(\chi_3-1)\partial_{\chi_3} \nonumber \\
	&\qquad +((2(a_{24}+a_2)+(2a_5+2b_2-1)\chi_3)\chi_4-2(a_2+a_4))\partial_{\chi_4}-\chi_2\chi_3\chi_4\partial_{\chi_2}\partial_{\chi_4} \nonumber\\
	&\qquad +(\chi_3-1)\chi_3\chi_4\partial_{\chi_3}\partial_{\chi_4}+\chi_4(\chi_4-1)\partial_{\chi_4}^2\Big]\phi_{2,2,3}=0 \,, \nonumber
\end{align}
with the abbreviation $a_{24}=a_2+a_3+a_4$. From the bridge-vertex symmetry with respect to $x_{0'}$ we furthermore find the differential equation
\begin{align}
	&\Big[(2a_5-1)(2a_6+2b_1-1)+(1-2a_5)\chi_1\partial_{\chi_1}+(2(1-a_5-b_2)+(2(a_5-a_6-b_1)-1)\chi_2)\partial_{\chi_2} \nonumber \\
	&\qquad +\chi_1\chi_2\partial_{\chi_1}\partial_{\chi_2}-\chi_3\partial_{\chi_2}\partial_{\chi_3}+\chi_2(1-\chi_2)\partial_{\chi_2}^2\Big]\phi_{2,2,3}=0 \,.
\end{align}
From these we find the indicials
\begin{align}
	&(0,0,0,0) \,, \nonumber \\
	&(2a_6+2b_1-1,0,0,0) \,, \nonumber\\
	&(0,2a_5+2b_2-1,0,0) \,, \nonumber\\
	&(0,0,2(a_2+a_3+a_4)-1,0) \,, \nonumber\\
	&(0,0,0,1-2a_2-2a_4) \,,\nonumber \\
	&(0,2a_5+2b_2-1,0,1-2a_2-2a_4) \,, \\
	&(0,2(a_{25}+b_2-1),2a_{24}-1,0) \,, \nonumber\\
	&(2a_6+2b_1-1,0,0,1-2a_2-2a_4) \,, \nonumber\\
	&(2a_6+2b_1-1,0,2a_{24}-1,0) \,, \nonumber\\
	&(2(a_5+a_6+b_1+b_2-1),2a_5+2b_2-1,0,0) \,,\nonumber \\
	&(2(a_5+a_6+b_1+b_2-1),2a_5+2b_2-1,0,1-2a_2-2a_4) \,, \nonumber\\
	&(2(a_{26}+b_1+b_2)-3,2(a_{25}+b_2-1),2a_{24}-1,0) \,,\nonumber
\end{align}
with $a_{2k}=a_2+\dots +a_k$ for $k>2$. Following the same steps as before we obtain the result 
\begin{align}
	&I_{2,2,3}=\frac{A_0(a_1)A_0(a_2)A_0(a_3)A_0(b_1)A_0(b_2)}{|x_{16}|^{2a_1}|x_{34}|^{2a_{24}-1}|x_{45}|^{2a_5+2b_2-1}|x_{56}|^{2a_6+2b_1-1}} \nonumber\\
	&\times\bigg[\HH{8}{2a_1,2\tilde{a}_6,2\tilde{a}_5,2a_2}{1-2a_2-2a_4,2-2a_{24};2(\tilde{a}_6+\tilde{b}_1);2(\tilde{a}_5+\tilde{b}_2)}{\chi_1,-\chi_2,-\chi_3,\chi_4}\nonumber \\
	&+\chi_1^{2a_6+2b_1-1}\HH{9}{2\tilde{a}_5,2a_2;2b_1,2(a_1+a_6+b_1)-1,2a_6+2b_1}{1-2a_2-2a_4,2-2a_{24};2(\tilde{a}_5+\tilde{b}_2)}{\chi_1,-\chi_1\chi_2,-\chi_3,\chi_4}\nonumber \\
	& +\chi_2^{2a_5+2b_2-1}\HH{10}{2a_1,2\tilde{a}_6,2a_2;2b_2,2(a_5+b_2)}{1-2a_2-2a_4,2-2a_{24};3-2(a_5+a_6+b_{12})}{\chi_1,-\chi_2,\chi_2\chi_3,\chi_4} \nonumber\\
	& +\chi_3^{2a_{24}-1}\HH{11}{2a_1,2\tilde{a}_6,2\tilde{a}_5,2a_3,2a_2}{2(\tilde{a}_6+\tilde{b}_1);2a_{24};3-2a_{25}-2b_2}{\chi_1,-\chi_2,-\chi_3,-\chi_3\chi_4} \nonumber\\
	& +\chi_4^{1-2a_2-2a_4}\HH{12}{2a_1,2\tilde{a}_6,2\tilde{a}_5,2a_3}{2\tilde{a}_4,2(\tilde{a}_2+\tilde{a}_4);2(\tilde{a}_6+\tilde{b}_1);2(\tilde{a}_5+\tilde{b}_2)}{\chi_1,-\chi_2,-\chi_3\chi_4,\chi_4}\nonumber \\
	& +\chi_2^{2a_5+2b_2-1}\chi_4^{1-2a_2-2a_4}\HH{13}{2a_1,2\tilde{a}_6,2a_3;2b_2,2(a_5+b_2)}{2\tilde{a}_4,2(\tilde{a}_2+\tilde{a}_4);3-2(a_5+a_6+b_{12})}{\chi_1,-\chi_2,\chi_2\chi_3\chi_4,\chi_4} \\
	& +\chi_2^{2(a_{25}+b_2-1)}\chi_3^{2a_{24}-1}\HH{14}{2a_1,2\tilde{a}_6,2a_3,2a_2}{2a_{24}+2b_2-1;2a_{25}+2b_2-1;2a_{24},4-2a_{26}-2b_{12}}{\chi_1,-\chi_2,\chi_2\chi_3,\chi_2\chi_3\chi_4}\nonumber \\
	& +\chi_1^{2a_6+2b_1-1}\chi_4^{1-2a_2-2a_4}\HH{15}{2\tilde{a}_5,2a_3;2b_1,2(a_1+a_6+b_1)-1,2a_6+2b_1}{2\tilde{a}_4,2(\tilde{a}_2+\tilde{a}_4);2(\tilde{a}_5+\tilde{b}_2)}{\chi_1,-\chi_1\chi_2,-\chi_3\chi_4,\chi_4} \nonumber\\
	&+ \chi_1^{2a_6+2b_1-1}\chi_3^{2a_{24}-1}\HH{16}{2\tilde{a}_5,2a_3,2a_2;2b_1,2(a_1+a_6+b_1)-1,2a_6+2b_1}{2a_{24};3-2a_{25}-2b_2}{\chi_1,-\chi_1\chi_2,-\chi_3,-\chi_3\chi_4} \nonumber\\
	&+ \chi_1^{2a_{56}+2b_{12}-2}\chi_2^{2a_5+2b_2-1}\HH{17}{2a_2;2b_2,2a_5+2b_2;1-2a_2-a_4,2-2a_{24}}{2a_{51}+2b_{12}-2,2a_5+2b_{12}-1,2a_{56}+2b_{12}-1}{\chi_1,-\chi_1\chi_2,\chi_1\chi_2\chi_3,\chi_4} \nonumber\\
	&+\chi_1^{2a_{56}+2b_{12}-2}\chi_2^{2a_5+2b_2-1}\chi_4^{1-2a_{24}}\HH{18}{2a_3;2b_2,2a_5+2b_2;2\tilde{a}_4,2\tilde{a}_2+2\tilde{a}_4}{2a_{51}+2b_{12}-2,2a_5+2b_{12}-1,2a_{56}+2b_{12}-1}{\chi_1,-\chi_1\chi_2,\chi_{1234},\chi_4}\nonumber \\
	&+\chi_1^{2a_{26}+2b_{12}-3}\chi_2^{2a_{25}+2b_2-2}\chi_3^{2a_{24}-1} \nonumber\\
	&\qquad \times\HH{19}{2a_3,2a_2;2a_{24}+2b_2-1,2a_{25}+2b_2-1}{2a_{25}+2b_{12}-2,2a_{16}+2b_{12}-3,2a_{26}+2b_{12}-2;2a_{24}}{\chi_1,-\chi_1\chi_2,\chi_1\chi_2\chi_3,\chi_{1234}}\bigg]\,, \nonumber
\end{align}
where the hypergeometric series are defined in \appref{app:defs}. Here we made use of the abbreviations $\tilde{a}=1/2-a$, $b_{12}=b_1+b_2$, $\chi_{1234}=\chi_1\chi_2\chi_3\chi_4$, as well as $a_{ij}=a_i+\dots+ a_j$ for $i<j$ understood cyclically. In particular $a_{51}=a_5+a_6+a_1$.

\subsection{Triangle-Box-Triangle Integral}
\label{sec:triBoxTri}
Finally let us consider the integral 
\begin{align}
	I_{2,3,2}
	&=\includegraphicsbox{FigTriangleBoxTriangle.pdf}\nonumber\\
	&=\int\frac{\pi^{-\sfrac{3}{2}}\dd x_0 \dd x_{0'} \dd x_{0''}}{|x_{01}|^{2a_1}|x_{06}|^{2a_6}|x_{00'}|^{2b_1}|x_{0'4}|^{2a_4}|x_{0'5}|^{2a_5}|x_{0'0''}|^{2b_2}|x_{0''2}|^{2a_2}|x_{0''3}|^{2a_3}} \,,
\end{align}
which we decompose as
\begin{equation}
	I_{2,3,2}=\frac{\phi_{2,3,2}(\chi_1,\chi_2,\chi_3,\chi_4)}{|x_{45}|^{2(a_2+a_3+a_4+a_5+b_2-1)}|x_{56}|^{2(a_1+a_6+b_1)-1}} \,,
\end{equation}
with
\begin{equation}
	\chi_1=\frac{x_{16}}{x_{56}},\qquad \chi_2=\frac{x_{45}}{x_{65}}, \qquad \chi_3=\frac{x_{35}}{x_{45}}, \qquad \chi_4=-\frac{x_{23}}{x_{53}} \,.
\end{equation}
From the two-point symmetries with respect to the pairs of external points $(x_1,x_6)$, $(x_2,x_3)$ and $(x_4,x_5)$ we find
\begin{align}
	&\Big[ 2a_1(1-2a_{61}-2b_1)+2(a_{61}-(2a_1+a_6+b_1)\chi_1)\partial_{\chi_1}-2a_1\chi_2\partial_{\chi_2}-\chi_1\chi_2\partial_{\chi_1}\partial_{\chi_2} \nonumber \\
	&\quad +\chi_1(1-\chi_1)\partial_{\chi_1}^2\Big]\phi_{2,3,2}=0\,, \\
	&\Big[-2a_2\chi_3\partial_{\chi_3}+(2a_{23}+(2a_2+1)\chi_4)\partial_{\chi_4}-\chi_3\chi_4\partial_{\chi_3}\partial_{\chi_4}+\chi_4(\chi_4+1)\partial_{\chi_4}^2\Big]\phi_{2,3,2}=0 \,, \\
	&\Big[2(a_{25}+b_2-1)(2a_{23}+2b_2-1)\chi_3+2(2a_{61}+2b_1-1)(a_{23}+a_5+b_2-1)\chi_2\chi_3 \nonumber \\
	&\quad +2(a_{23}+a_5+b_2-1)\chi_1\chi_2\chi_3\partial_{\chi_1} +2(a_{23}+a_5+b_2-1)\chi_4\partial_{\chi_4} \nonumber \\
	&\quad  -2(a_{25}+a_{23}+2b_2-2+(1+a_{61}-a_{23}-a_5+b_1-b_2)\chi_2)\chi_2\chi_3\partial_{\chi_2}\nonumber \\
	&\quad +(1-2a_{23}-2a_5-2b_2+(2(a_{25}+a_{23}+2b_2-1)+(2a_{61}+2b_1-1)\chi_1)\chi_3)\chi_3\partial_{\chi_3}\nonumber \\
	&\quad -\chi_1\chi_2^2\chi_3\partial_{\chi_1}\partial_{\chi_2}+\chi_1\chi_2\chi_3^2\partial_{\chi_1}\partial_{\chi_3}+(1-2\chi_3+\chi_2\chi_3)\chi_2\chi_3\partial_{\chi_2}\partial_{\chi_3}-\chi_2\chi_4\partial_{\chi_2}\partial_{\chi_4}\nonumber\\
	&\quad +\chi_3\chi_4\partial_{\chi_3}\partial_{\chi_4} +(1-\chi_2)\chi_2^2\chi_3\partial_{\chi_2}^2+(\chi_3-1)\chi_3^2\partial_{\chi_3}^2\Big]\phi_{2,3,2}=0 \,.
\end{align}
Furthermore we find the following partial differential equation from the bridge-vertex symmetry with respect to $x_{0'}$:
\begin{align}
	&\Big[(1-2a_{61}-2b_1)(2a_{23}+2b_2-1)\chi_3+(1-2a_{23}-2b_2)\chi_1\chi_3\partial_{\chi_1} +(1-2a_{23}-2b_2)\chi_2\chi_3\partial_{\chi_2} \nonumber \\
	&\quad +(1-2a_{61}-2b_1)\chi_3^2\partial_{\chi_3}-\chi_1\chi_3^2\partial_{\chi_1}\partial_{\chi_3} +(1-\chi_2\chi_3)\chi_3\partial_{\chi_2}\partial_{\chi_3}-\chi_4\partial_{\chi_2}\partial_{\chi_4}\Big]\phi_{2,3,2}=0  \,.
\end{align}
Here and below we are making use of the abbreviations $b_{12}=b_1+b_2$ and $a_{ij}=a_i+\dots +a_j$ with $i<j$, understood cyclically. In particular $a_{61}=a_6+a_1$. From the differential equations we obtain the indicials
\begin{align}
	&(0,0,0,0) \,, \nonumber\\
	&(1-2a_1-2a_6,0,0,0) \,,  \nonumber \\
	&(0, 2a_{25}+2b_2-2,0,0) \,, \nonumber \\
	&(0,0,2(1-a_2-a_3-a_5-b_2),0) \,,\nonumber \\
	&(1-2a_1-2a_6,2a_{25}+2b_2-2,0,0) \,,\nonumber \\
	&(1-2a_1-2a_6,0,2(1-a_2-a_3-a_5-b_2),0) \,, \\
	&(0,0,1-2a_2-2a_3,1-2a_2-2a_3) \,, \nonumber\\
	&(0,0,2(1-a_2-a_3-a_5-b_2),1-2a_2-2a_3) \,, \nonumber \\
	&(1-2a_1-2a_6,0, 1-2a_2-2a_3,1-2a_2-2a_3) \,, \nonumber \\
	&(1-2a_1-2a_6,0,2(1-a_2-a_3-a_5-b_2),1-2a_2-2a_3) \,, \nonumber \\
	&(0,2a_4+2a_5+2b_2-1,1-2a_2-2a_3,1-2a_2-2a_3) \,,\nonumber \\
	&(1-2a_1-2a_6,2a_4+2a_5+2b_2-1,1-2a_2-2a_3,1-2a_2-2a_3) \,.\nonumber
\end{align}
Proceeding as before we find the final result
\begin{align}
	&I_{2,3,2}=\frac{A_0(a_1)A_0(a_2)A_0(a_4)A_0(b_1)A_0(b_2)}{|x_{45}|^{2(a_2+a_3+a_4+a_5+b_2-1)}|x_{56}|^{2(a_1+a_6+b_1)-1}}  \nonumber \\
	&\quad\times\bigg[\HH{20}{2a_1,2a_2;2a_{61}+2b_1-1,2a_{23}+2b_2-1}{2-2a_{23}-2a_5-2b_2;3-2a_{25}-2b_2,2a_{61},2a_{23}}{\chi_1,\chi_2,\chi_3,\chi_3\chi_4} \nonumber\\
	&\quad + \chi_1^{1-2a_{61}}\HH{20}{2\tilde{a}_6,2a_2;2b_1,2a_{23}+2b_2-1}{2-2a_{23}-2a_5-2b_2,3-2a_{25}-2b_2;2-2a_{61},2a_{23}}{\chi_1,\chi_2,\chi_3,\chi_3\chi_4}\nonumber \\
	&\quad + \chi_2^{2a_{25}+2b_2-2}\HH{21}{2a_1,2a_4,2a_2;2a_{23}+2b_2-1}{2a_{16}+2b_{12}-3;2a_{61},2a_{23};2a_{25}+2b_2-1}{\chi_1,\chi_2,\chi_2\chi_3,\chi_2\chi_3\chi_4} \nonumber\\
	&\quad +\chi_3^{2-2a_{23}-2a_5-2b_2}\HH{22}{2a_1,2a_4,2a_2;2a_{61}+2b_1-1,2\tilde{a}_5}{2a_{61},2a_{23};3-2a_{23}-2a_5-2b_2}{\chi_1,\chi_2\chi_3,\chi_3,\chi_4} \nonumber\\
	&\quad +\chi_1^{1-2a_{61}}\chi_2^{2a_{25}+2b_2-2}\HH{21}{2\tilde{a}_6,2a_4,2a_2;2a_{23}+2b_2-1}{2a_{25}+2b_{12}-2;2-2a_{61},2a_{23};2a_{25}+2b_2-1}{\chi_1,\chi_2,\chi_2\chi_3,\chi_2\chi_3\chi_4} \nonumber\\
	&\quad + \chi_1^{1-2a_{61}}\chi_3^{2-2a_{23}-2a_5-2b_2}\HH{22}{2\tilde{a}_6,2a_4,2a_2;2b_1,2\tilde{a}_5}{2-2a_{61},2a_{23};3-2a_{23}-2a_5-2b_2}{\chi_1,\chi_2\chi_3,\chi_3,\chi_4} \\
	&\quad +\chi_3^{1-2a_{23}}\chi_4^{1-2a_{23}}\HH{20}{2a_1,2\tilde{a}_3;2a_{61}+2b_1-1,2b_2}{1-2a_5-2b_2,2-2a_{45}-2b_2;2a_{61},2-2a_{23}}{\chi_1,\chi_2,\chi_3,\chi_3\chi_4} \nonumber\\
	&\quad + \chi_3^{2-2a_{23}-2a_5-2b_2}\chi_4^{1-2a_{23}}\HH{22}{2a_1,2a_4,2\tilde{a}_3;2a_{61}+2b_1-1,2\tilde{a}_5}{2a_{61},2-2a_{23};2-2a_5-2b_2}{\chi_1,\chi_2\chi_3,\chi_3,\chi_4}\nonumber \\
	&\quad + \chi_1^{1-2a_{61}}\chi_3^{1-2a_{23}}\chi_4^{1-2a_{23}}\HH{20}{2\tilde{a}_6,2\tilde{a}_3;2b_1,2b_2}{1-2a_5-2b_2,2-2a_{45}-2b_2;2-2a_{61},2-2a_{23}}{\chi_1,\chi_2,\chi_3,\chi_3\chi_4}\nonumber \\
	&\quad + \chi_1^{1-2a_{61}}\chi_3^{2-2a_{23}-2a_5-2b_2}\chi_4^{1-2a_{23}}\HH{22}{2\tilde{a}_6,2a_4,2\tilde{a}_3;2b_1,2\tilde{a}_5}{2-2a_{61},2-2a_{23};2-2a_5-2b_2}{\chi_1,\chi_2\chi_3,\chi_3,\chi_4}\nonumber \\
	&\quad +\chi_2^{2a_{45}+2b_2-1}\chi_3^{1-2a_{23}}\chi_4^{1-2a_{23}}\HH{21}{2a_1,2a_4,2\tilde{a}_3;2b_2}{2a_{41}+2b_{12}-2;2a_{61},2-2a_{23};2a_{45}+2b_2}{\chi_1,\chi_2,\chi_2\chi_3,\chi_2\chi_3\chi_4}\nonumber \\
	&\quad +\chi_1^{1-2a_{61}}\chi_2^{2a_{45}+2b_2-1}\chi_3^{1-2a_{23}}\chi_4^{1-2a_{23}} \nonumber\\
	&\qquad \times\HH{21}{2\tilde{a}_6,2a_4,2\tilde{a}_3;2b_2}{2a_{45}+2b_{12}-1;2-2a_{61},2-2a_{23};2a_{45}+2b_2}{\chi_1,\chi_2,\chi_2\chi_3,\chi_2\chi_3\chi_4}\bigg] \,, \nonumber
\end{align}
with the hypergeometric series defined in \appref{app:defs}. Also recall the abbreviations $\tilde{a}=\sfrac{1}{2}-a$, $b_{12}=b_1+b_2$ and $a_{ij}=a_i+\dots a_j$ for $i<j$, understood cyclically. In particular, $a_{61}=a_6+a_1$ and $a_{41}=a_4+a_5+a_6+a_1$.

\section{Generic Polygon Integrals}
\label{sec:polygons}

In this section we will study polygon integrals for general $n$:
\begin{equation}
	I_n=\includegraphicsbox{FigNPointStar.pdf}=\int\frac{\dd x_0}{\sqrt{\pi}} \frac{1}{\prod_{i=1}^n|x_{0i}|^{2a_i}}\,.
\end{equation}
We will decompose these as\footnote{Note that these conventions differ from those chosen for the triangle integral in \secref{sec:3pt}. It is however simple to transform the result to the conventions used here using known analytic continuations of the Gauss hypergeometric function.} 
\begin{equation}
	I_n=|x_{12}|^{1-2a_1-2a_2}\prod_{i=3}^n|x_{1i}|^{-2a_i}\phi_n(\chi_1,\dots ,\chi_{n-2}) \,,
\end{equation}
with the ratios\footnote{Note that a fully cyclic choice of variables (cf. \eqref{eq:varsTriangleTrack}) does not lead to minimal hypergeometric series.} 
\begin{equation}
	\chi_i=\frac{x_{1,i+1}}{x_{1,i+2}} ,\qquad i=1,\dots,n-2 \,.
\end{equation}
Note that these are precisely a subset of the $u$-variables described in \cite{brownUVars}, and hence ensure that we are always expanding around normal-crossing singularities. An integral representation for $\phi_n$ is given by
\begin{equation}
	\phi_n(\chi_1,\dots ,\chi_{n-2})=\int\frac{\dd y}{\sqrt{\pi}}\frac{1}{|y|^{2a_1}|1-y|^{2a_2}\prod_{i=1}^{n-2}|1-y\prod_{j=1}^i z_j| ^{2a_{i+2}}} \,.
\end{equation}
We will now proceed to bootstrap the function $\phi_n$ and hence the integral $I_n$.

\paragraph{$\levo{P}$-Symmetries and Recurrences.}
The polygon integrals admit two-point symmetries with respect to any pair of external points. A convenient subset that fixes the integrals turns out to be given by the $\levo{P}_{jk}$-equations with $(j,k)\in\{ (2,n),\,(3,n),\dots, (n-1,n)\}$. The resulting differential equations are given by
\begin{align}
	&\Big[2(2a_1-1)a_n \textstyle\prod_{i=1}^{n-3}\chi_i+(1-2a_1-2a_2+(2a_1-1)\textstyle\prod_{i=1}^{n-2}\chi_i)\partial_{\chi_{n-2}}-2a_n \chi_1\textstyle\prod_{i=1}^{n-3}\chi_i\partial_{\chi_1} \notag\\
	&\qquad +\chi_1(1-\textstyle\prod_{i=1}^{n-2}\chi_i)\partial_{\chi_1}\partial_{\chi_{n-2}}\Big]\phi_n=0 \,, \notag\\
	&\Big[2a_n\textstyle\prod_{i=k+1}^{n-3}\chi_i(\chi_k\partial_{\chi_k}-\chi_{k+1}\partial_{\chi_{k+1}})-2a_{k+2}\partial_{\chi_{n-2}} \\
	&\qquad+(1-\textstyle\prod_{i=k+1}^{n-2}\chi_i)(\chi_{k+1}\partial_{\chi_{k+1}}\partial_{\chi_{n-2}}-\chi_k\partial_{\chi_k}\partial_{\chi_{n-2}})\Big]\phi_n=0 \,, \quad k=1,\dots,n-4 \notag \\
	&\Big[ (1-2a_{n-1}-\chi_{n-2}-2a_n \chi_{n-2})\partial_{\chi_{n-2}}-\chi_{n-2}^2\partial_{\chi_{n-2}}^2+\chi_{n-3}(2a_n\partial_{\chi_{n-3}}-\partial_{\chi_{n-3}}\partial_{\chi_{n-2}}) \notag\\
	&\qquad + \chi_{n-2}(\partial_{\chi_{n-2}}^2+\chi_{n-3}\partial_{\chi_{n-3}}\partial_{\chi_{n-2}})\Big]\phi_n=0 \,. \notag
\end{align}
In principle we can now proceed to find the indicials as before. It however turns out that this requires us to go to higher orders in the expansion or to make use of all two-point symmetries. While this can be done for general $n$ we have a simpler argument using the spectral transform presented in \appref{app:spectralpolygon}. This results in the following $n-1$ sets of indicials $(r_1^{(j)},\dots ,r_{n-2}^{(j)})$ given by
\begin{equation}\label{indicials polygon}
	r_i^{(j)}=\left\{\begin{array}{ll} 2\sum_{k=1}^{i+1}a_k-1 &  i\leqslant j-1 \,, \\  0 & \, \textrm{otherwise}\,, \end{array} \right.
\end{equation}
for $j=1,\dots ,n-1$. The solution will hence take the form
\begin{equation}
	\phi_n(\chi_1,\dots ,\chi_{n-2})=\sum_{k=1}^{n-1}c_k\left(\prod_{j=1}^{k-1}\chi_j^{2\sum_{i=1}^{j+1}a_i-1} \right)f_k(\chi_1,\dots ,\chi_{n-2}) \,,
\end{equation}
for some coefficients $c_k$ and power series $f_k$. By making a series ansatz
\begin{equation}
	f_1(\chi_1,\dots ,\chi_{n-2})=\sum_{m_1,\dots ,m_{n-2}}c_{m_1,\dots ,m_{n-2}}\chi_1^{m_1}\dots \chi_{n-2}^{m_{n-2}}\,,
\end{equation}
we can translate the differential equations into recurrence equations
\begin{align}
	&\Big[(2a_1-m_1)(2a_n+m_{n-2})\prod_{i=1}^{n-3}\hat{S}_i^{-1}+(1-2a_1-2a_2+m_1)(1+m_{n-2})\hat{S}_{n-2}\Big]c_{m_1,\dots m_{n-2}}=0 \,,\nonumber\\
	&\Big[(1+m_k-m_{k+1})(2a_n+m_{n-2})\prod_{i=k+1}^{n-3}\hat{S}_i^{-1}\\
	&\qquad -(2a_{k+2}+m_k-m_{k+1})(1+m_{n-2}) \hat{S}_{n-2}\Big]c_{m_1,\dots m_{n-2}}=0\,, \qquad k=1,\dots ,n-4\,, \nonumber\\
	&\Big[(1+m_{n-2})(1-2a_{n-1}-m_{n-3}+m_{n-2})\hat{S}_{n-2}+(m_{n-3}-m_{n-2})(2a_n+m_{n-2})\Big]c_{m_1,\dots m_{n-2}}=0 \,. \nonumber 
\end{align}
Here the $\hat{S}_i^k$ are shift operators acting on the series coefficients as $\hat{S}_i^k c_{m_1,\dots ,m_i,\dots ,m_{n-2}}=c_{m_1,\dots ,m_i+k,\dots ,m_{n-2}}$. 

\paragraph{Bootstrapping the Integral.}

We can solve the recurrence equations iteratively to find the fundamental solution
\begin{align}
	f_1(\vec{\chi})&=\sum_{m_1,\dots ,m_{n-2}=0}^{\infty}\left[\Gamma(2a_1-m_1)\Gamma(2-2a_{12}+m_1)\left(\textstyle\prod_{i=1}^{n-3}\Gamma(1-2a_{i+2}-m_i+m_{i+1})\right)\right.\nonumber \\
	&\qquad\times \left. \Gamma(1-2a_n-m_{n-2})\left(\textstyle\prod_{i=1}^{n-3}\Gamma(1+m_i-m_{i+1})\right)\Gamma(1+m_{n-2})\right]^{-1} \prod_{j=1}^{n-2}\chi_j^{m_j} \,,
\end{align}
with $a_{12}=a_1+a_2$ and where we moved all $\Gamma$-functions to the denominator for convenience and introduced the shorthand $\vec{\chi}=(\chi_1,\dots ,\chi_{n-2})$. Note that redefining the summation variables by $m_i\rightarrow m_i+\dots + m_{n-2}$ the series can be identified with the Lauricella $F_D$ function defined in \appref{app:defs}. To find the other basis solutions we shift the summation variables by the indicials $m_i\rightarrow m_i+r_i^{(k)}$ for $1<k<n-1$ which yields
\begin{align}
	&f_k(\vec{\chi})=\sum_{m_1,\dots ,m_{n-2}=0}^{\infty}\left[\Gamma(1-2a_2-m_1)\left(\textstyle\prod_{i=1}^{k-2}\Gamma(1-2a_{i+2}+m_i-m_{i+1})\right)\right.\nonumber  \\
	&\quad\times\Gamma\left(2a_{1k}+m_{k-1}-m_k\right) \Gamma\left(2-2a_{1,k+1}-m_{k-1}+m_k \right) \\
	&\quad \times \left(\textstyle\prod_{i=k}^{n-3}\Gamma(1-2a_{i+2}-m_i+m_{i+1})\right)\Gamma(1-2a_n-m_{n-2}) \nonumber \\
	&\quad\left.\times \Gamma(1+m_1)\left( \textstyle\prod_{i=1}^{k-2}\Gamma(1-m_i+m_{i+1}) \textstyle\prod_{i=k}^{n-3}\Gamma(1+m_i-m_{i+1})\right) \Gamma(1+m_{n-2})\right]^{-1} \prod_{j=1}^{n-2}\chi_j^{m_j}\,, \nonumber
\end{align}
where we used the abbreviation $a_{ij}=a_i+\dots +a_j$ for $i<j$. Shifting by the indicial $r^{(n-1)}$ we find the final series
\begin{align}
	f_{n-1}(\vec{\chi})&=\sum_{m_1,\dots ,m_{n-2}=0}^{\infty}\left[\Gamma(1-2a_2-m_1)\left( \textstyle\prod_{i=1}^{n-3}\Gamma(1-2a_{i+2}+m_i-m_{i+1})\right) \Gamma(1+m_1)\right.\nonumber  \\
	&\left.\times \Gamma\left(2-2a_{1n}-m_{n-2}\right)\Gamma\left(2a_{1,n-1}+m_{n-2} \right)\left( \textstyle\prod_{i=1}^{n-3}\Gamma(1-m_i+m_{i+1})\right) \right]^{-1}\prod_{j=1}^{n-2}\chi_j^{m_j} \,.
\end{align}
Also this series can be identified with Lauricella $F_D$ after shifting the summation variables as $m_i\rightarrow m_1+\dots +m_i$.
Finally to compute the coefficients we need to evaluate the original integral in various limits. First of all we compute
\begin{equation}
	\lim_{\chi \rightarrow 0}\phi_n(\vec{\chi})=\int\frac{\dd y}{\sqrt{\pi}}\frac{1}{|y|^{2a_1}|1-y|^{2a_2}}= A_0(a_1) A_0(a_2) A_0(1-a_1-a_2) \,,
\end{equation}
where we use the shorthand $\lim_{\chi\rightarrow 0}=\lim_{\chi_1,\dots ,\chi_{n-2}\rightarrow0}$. Furthermore for $2\leqslant k\leqslant n-1$ we compute
\begin{align}
	&\lim_{\chi \rightarrow 0} \textstyle\prod_{j=1}^{k-1}\chi_j^{1-2\sum_{i=1}^{j+1}a_i} \phi_n(\vec{\chi}) \nonumber\\
	&= \lim_{\chi\rightarrow 0}\int\frac{\dd y}{\sqrt{\pi}}|y|^{-2a_1}\left(\prod_{i=1}^{k-1}\left|\textstyle\prod_{j=i}^{k-1}\chi_j-y\right|^{-2a_{i+1}}\right)|1-y|^{-2a_{k+1}}\left(\prod_{i=k}^{n-2}\left|1-y\textstyle\prod_{j=k}^i \chi_k\right|^{-2a_{i+2}} \right) \nonumber\\
	&=\int\frac{\dd y}{\sqrt{\pi}}\frac{1}{|y|^{2a_{1k}}|1-y|^{2a_{k+1}}} = A_0\left(a_{1k}\right) A_0(a_{k+1}) A_0(1-a_{1,k+1}) \,.
\end{align}
Putting everything together we find the final result 
\begin{align}
	I_n&=\frac{\prod_{i=2}^nA_0(a_i)}{|x_{12}|^{2a_1+2a_2-1}\prod_{i=3}^n|x_{1i}|^{2a_i}}\Big[\cFL{D}{n-2}{2\tilde{a}_1;2a_3,\dots,2a_n}{2(\tilde{a}_1+\tilde{a}_2)}{z_1^{(n,1)},\dots ,z_{n-2}^{(n,1)}} \notag \\
	&\qquad +\sum_{k=2}^{n-2}\left(\prod_{j=1}^{k-1}\chi_j^{2a_{1,j+1}-1} \right)\PP{n}{k}{2a_2,\dots,2\check{a}_{k+1},\dots,2a_n}{2a_{1,k+1}-1,2a_{1,k}}{z_1^{(n,k)},\dots ,z_{n-2}^{(n,k)}} \\
	&\qquad +\left(\prod_{j=1}^{n-2}\chi_j^{2a_{1,j+1}-1} \right)  \cFL{D}{n-2}{2a_{1,n}-1;2a_2,\dots ,2a_{n-1}}{2a_{1,n-1}}{z_1^{(n,n-1)},\dots ,z_{n-2}^{(n,n-1)}} \notag \,,
 \end{align}
where the $2\check{a}_{k+1}$ means that this term is left out. Furthermore we used the abbrevation $\tilde{a}=1/2-a$ and we recall that $a_{ij}=a_i+\dots +a_j$ for $i<j$. The hypergeometric functions $\mathcal{P}^{(n)}_{k}$ are defined in \Appref{app:defs} and their arguments read
\begin{align}
	z_j^{(n,1)}&=\prod_{i=1}^j\chi_i , \qquad j=1,\dots ,n-2 \,, \\
	z_j^{(n,k)}&=\left\{\begin{array}{ll} \prod_{i=j}^{k-1}\chi_i &  j\in\{1,\dots, k-1\} \,, \\  \prod_{i=k}^{j}\chi_i &  j\in\{k,\dots, n-2\} \,. \end{array} \right.
\end{align}
Furthermore we introduced the rescaled Lauricella $F_D$ function by
\begin{equation}
	\cFL{D}{n}{a;b_1,\dots ,b_n}{c}{x_1,\dots ,x_n}= \frac{A_0(\sfrac{c}{2})}{A_0(\sfrac{a}{2})A_0(\sfrac{b_1}{2})\dots A_0(\sfrac{b_n}{2})}\FL{D}{n}{a;b_1,\dots ,b_n}{c}{x_1,\dots ,x_n} \,.
\end{equation}
We note that it is possible to express the generic polygon integral fully in terms of Lauricella $F_D$ functions, as implied e.g., by the results of \cite{Duhr:2023bku}. However the different terms will then not converge in the same region as series but only make sense as analytic continuations thereof. Indeed there is no known local solution space of the Lauricella $F_D$ differential equation system that is fully expressible in terms of Lauricella $F_D$ \cite{Goto:2022uo}, c.f., the discussion in \secref{sec:Box}.

\section{Generic Triangle-Track Integrals}
\label{sec:triangletracks}

In this section we will bootstrap the family of generic (non-conformal) triangle-track integrals at any loop order $\ell$. This is the most general family of track integrals since any other track can be obtained by taking coincidence limits of external points and pinching internal propagators using the identity \eqref{prop to delta}. In particular, our results imply that every track integral is fixed by the $\levo{P}$-symmetries.

The $\ell+2$-point triangle-track integral is defined as the following $\ell$-loop integral (cf.\ \cite{Duhr:2024hjf}): 
\begin{align}
	I_{2,1^{\ell-2},2} 
		&=\includegraphicsbox{FigTriangleTracksNobs.pdf}
		\\
	&= \int  \frac{\prod_{j=1}^{\ell} \pi^{-1/2} \dd y_j}{|x_1 - y_1|^{2a_1} \prod_{j=3}^{\ell+2}|x_j-y_{\ell+3-j}|^{2a_{j}} \prod_{j=1}^{\ell-1}|y_{j,j+1}|^{2b_j} |y_{\ell} - x_2|^{2a_2}}.
	\nonumber
\end{align}
We note that these integrals have a natural relation to the train-track integrals that have been discussed in various papers \cite{Bourjaily:2018ycu,Vergu:2020uur,Duhr:2022pch,Cao:2023tpx,McLeod:2023doa,Duhr:2024hjf}. The $\ell$-loop train-track integral is the $2\ell+2$-point integral defined by 
\begin{align}
	&I_{3,2^{\ell-2},3} 
	= \includegraphicsbox{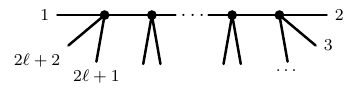}\\
	&= \int  \frac{\prod_{j=1}^{\ell} \pi^{-1/2} \dd y_j}{|x_1 - y_1|^{2a_1} \prod_{j=2}^{\ell+1} |x_{2j}-y_{\ell+2-j}|^{2a_{2j}} |x_{2j-1}-y_{\ell+2-j}|^{2a_{2j-1}} \prod_{j=1}^{\ell-1}|y_{j,j+1}|^{2b_j} |y_\ell - x_2|^{2a_2}}\, .
	\nonumber
\end{align}
The train-track integrals can be obtained from the triangle tracks via the limit
\begin{equation}
	I_{3,2^{\ell-2},3}= \lim_{b_1,b_3,\dots,b_{2\ell-1}\to 1/2} \frac{I_{2,1^{2\ell-2},2}}{\prod_{j=1}^\ell A_0(b_{2j-1})}
\end{equation}
and the renaming $b_{2j}\to b_j$ on the right-hand side after the limit.

Let us decompose the triangle-track integrals as
\begin{equation}\label{eq:DecompositionTriangleTracks}
	I_{2,1^{\ell-2},2}=|x_{23}|^{1-2a_2-2a_3}\prod_{j=4}^{\ell+2}|x_{j-1,j}|^{1-2a_j-2b_{\ell+3-j}}|x_{1,\ell+2}|^{-2a_1}\phi_{2,1^{\ell-2},2}(\chi_1,\dots,\chi_{\ell}) \,,
\end{equation}
with
\begin{equation}\label{eq:varsTriangleTrack}
	\chi_{j} = \frac{x_{j+1,j+2}}{x_{j+3,j+2}}\quad \text{for}\quad 1\leqslant j\leqslant \ell\, ,
\end{equation}
where we identify $x_{\ell+3} \equiv x_1$. We can derive the integral representation
\begin{align}
	\phi_{2,1^{\ell-2},2}(\chi_1,\dots ,\chi_{\ell})&=|Z_{1,\ell}|^{\ell-2(a_1+a_2)+\sum_{i=1}^{\ell-1}b_i}\int\prod_{i=1}^{\ell}\frac{\dd y_i}{\sqrt{\pi}}\,|1-\chi_\ell y_1|^{-2a_1}|y_{\ell}|^{-2a_2} \nonumber\\
	&\qquad\times\prod_{j=1}^{\ell-1}|y_j-\chi_{\ell-j}y_{j+1}|^{-2b_j}\prod_{j=1}^{\ell}|Z_{j+1,\ell}-(-1)^j Z_{1,\ell}y_j|^{-2a_{\ell+3-j}} \,,
\end{align}
where $Z_{i,j}$ is defined recursively via
\begin{equation}
	Z_{i,j}=1-\chi_{\ell+1-i} Z_{i+1,j},\qquad Z_{i,i}=1-\chi_{\ell+1-i},\qquad Z_{i,j}=1 \text{ for } i>j \,.
\end{equation}
We will now proceed to bootstrap the function $\phi_{2,1^{\ell-2},2}(\chi_1,\dots ,\chi_{\ell})$.

\paragraph{$\levo{P}$-symmetries and Recurrences.}

From the two end-vertex symmetries acting on the pairs $(x_2,x_3)$ and $(x_1,x_n)$ we find the differential equations
\begin{align}
	&\Big[(2a_3-1)(2a_4+2b_{\ell-1}-1)- (2(a_2+a_3-1)+(1-2a_3+2a_4+2b_{\ell-1})\chi_1)\partial_{\chi_1} \notag\\
	&\quad +(1-2a_3)\chi_2\partial_{\chi_2} +\chi_1\chi_2\partial_{\chi_1}\partial_{\chi_2}+\chi_1(1-\chi_1)\partial_{\chi_1}^2\Big]\phi_{2,1^{\ell-2},2} =0 \,, \notag \\
	&\Big[2a_1(1-2a_{\ell+2})+2(b_1+a_{\ell+2}-1+(1+a_1-a_{\ell+2})\chi_\ell)\partial_{\chi_\ell}\\
	&\quad + \chi_{\ell-1}\partial_{\chi_{\ell-1}}\partial_{\chi_\ell} + \chi_\ell(\chi_\ell-1)\partial_{\chi_\ell}^2\Big]\phi_{2,1^{\ell-2},2} =0 \,. \notag
\end{align}
From the bridge-vertex symmetries with respect to the internal vertices $y_{\ell+1-k}$ with $k=2,\dots ,\ell-1$ we furthermore find the differential equations
\begin{align}\label{eq:TriangleTrackDiffEq}
	&\Big[(1-2a_{k+2})(2a_{k+3}+2b_{\ell-k}-1)+ (2a_{k+2}-1)\chi_{k+1}\partial_{\chi_{k+1}}  \notag \\
	&\quad + (2(a_{k+2}+b_{\ell+1-k}-1)+(1-2a_{k+2}+2a_{k+3}+2b_{\ell-k})\chi_k)\partial_{\chi_k} \\
	&\quad -\chi_k\chi_{k+1}\partial_{\chi_k}\partial_{\chi_{k+1}}+\chi_{k-1}\partial_{\chi_{k-1}}\partial_{\chi_k}+\chi_k(\chi_k-1)\partial_{\chi_k}^2\Big]\phi_{2,1^{\ell-2},2}	=0 \,. \notag
\end{align}
The differential equations for the end-vertex symmetries are straightforward to obtain. The proof of \eqref{eq:TriangleTrackDiffEq} is presented in \Appref{sec:ProofTriangleTrackEq}. 

\paragraph{Bootstrapping the Integral.}

From these differential equations we find that the indicial equations take the particularly simple form 
\begin{equation}
	r_1(r_1+1-2a_2-2a_3)=0, \qquad  r_j(r_j-r_{j-1}+1-2a_{j+2}-2b_{\ell+1-j})=0 \,,
\end{equation}
where $2\leqslant j\leqslant\ell$. Hence the indicials of the $\ell$-loop triangle track are in one-to-one correspondence with binary words $w\in\{0,1\}^\ell$. In particular there are precisely $2^{\ell}$ indicials and hence $2^{\ell}$ basis functions. To translate a binary word $w$ into the corresponding indicial $r^w=(r_1^w,\dots ,r_{\ell}^w)$ we simply go through the word from left to right and recursively set
\begin{align}\label{indicials triangle-track}
	r_1^{w}&=\left\{ \begin{array}{ll} 0 & \text{if } w_1 = 0 \\ 2a_2+2a_3-1 & \text{if } w_1 = 1  \end{array} \right. \,, \\
	r_{j}^{w}&=\left\{ \begin{array}{ll} 0 & \text{if } w_{j} = 0 \\ 2a_{j+2}+2b_{\ell+1-j}+r_{j-1}^w-1 & \text{if } w_j= 1  \end{array} \right. \,,
\end{align}
where $w_i$ denotes the letter of $w$ in position $i$. This can be compactly written as
\begin{equation}
	r^w_j = \sum_{i=1}^{j} \alpha^w_{i,j} (2a_{i+2}+2b_{\ell+1-i}-1)\, ,
\end{equation}
where we identify $b_\ell \equiv a_2$, and for $i\leqslant j$ we define
\begin{equation}\label{alphajkw}
	\alpha^w_{i,j} = \prod_{k=i}^{j} \delta_{w_k,1} = \prod_{k=i}^{j} w_k\, .
\end{equation}

To find the fundamental series solution to the above differential equations we make an ansatz of the form
\begin{equation}
	f_1(\chi_1,\dots ,\chi_{\ell})=\sum_{m_1,\dots ,m_{\ell}} c_{m_1,\dots ,m_{\ell}}\chi_1^{m_1}\dots \chi_{\ell}^{m_{\ell}} \,,
\end{equation}
which translates the differential equations into recurrence relations. We find
\begin{align}
	&\Big[(2a_3-1-m_1)(2a_4+2b_{\ell-1}-1+m_1-m_2) \notag\\
	&\qquad -(1+m_1)(2a_2+2a_3-2-m_1)\hat{S}_1\Big]c_{m_1\dots m_{\ell}} =0 \,, \notag\\
	&\Big[(2a_1+m_\ell)(2a_{\ell+2}-1-m_\ell)-(1+m_\ell)(2a_{\ell+2}+2b_1-2+m_{\ell-1}-m_\ell)\hat{S}_\ell\Big]c_{m_1\dots m_{\ell}} =0 \,, \notag \\ 
	&\Big[  (2a_{k+2}-1-m_k)(2a_{k+3}+2b_{\ell-k}-1+m_k-m_{k+1})   \\
	&\qquad - (1+m_k)(2a_{k+2}+2b_{\ell+1-k}-2+m_{k-1}-m_k)\hat{S}_k\Big]c_{m_1\dots m_{\ell}} =0 \,, \notag 
\end{align}
for $k=2,\dots ,\ell-1$. Here, the $\hat{S}_j$ are shift operators $\hat{S}_j c_{m_1,\dots ,m_j,\dots, m_{\ell}}=c_{m_1,\dots m_{j}+1,\dots m_{\ell}}$. Remarkably, the recurrence equations are completely factorized and can be solved in closed form. We find the following fundamental solution:
\begin{align}
	f_1(\vec{\chi})&=\sum_{m_1,\dots,m_{\ell} = 0}^{\infty}\Big[\Gamma(1-2a_1-m_\ell) \prod_{j=1}^{\ell}\Gamma(2a_{j+2}-m_j) \nonumber \\
	&\qquad\qquad\times\prod_{j=1}^{\ell} \Gamma(2(1-a_{j+2}-b_{\ell+1-j})+m_j-m_{j-1}) \Big]^{-1}\prod_{j=1}^{\ell}\frac{\chi_j^{m_j}}{m_j!}\, ,
\end{align}
where for convenience, we moved all $\Gamma$-functions to the denominator, introduced the shorthand $\vec{\chi}=(\chi_1,\dots ,\chi_{\ell})$, and defined $m_0=0$. To generate the other basis functions we shift the summation variables according to 
\begin{equation}
	m_j \rightarrow m_j+r_j^w+\sum_{i=1}^j \alpha^w_{i,j} m_{i-1}\, .
\end{equation}
We obtain the shifted solution
\begin{multline}
	f_w(\vec{\chi}) = \sum_{m_1,\dots,m_{\ell} = 0}^{\infty} \Big[\Gamma(1-2a_1-m_\ell-r_\ell^w-{\textstyle{\sum}}_{i=1}^\ell \alpha^w_{i,\ell} m_{i-1})\\
	\times\prod_{j:\,w_j=0}\! \Gamma(2a_{j+2}-m_j) \Gamma(2(1-a_{j+2}-b_{\ell+1-j})+m_j-m_{j-1} - r^w_{j-1} - {\textstyle{\sum}}_{i=1}^{j-1} \alpha^w_{i,j-1} m_{i-1})\\
	\times\prod_{j:\,w_j=1}\! \Gamma(2a_{j+2}-m_j-r_j^w-{\textstyle{\sum}}_{i=1}^j \alpha^w_{i,j} m_{i-1}) \Gamma(1+m_j+r_j^w+{\textstyle{\sum}}_{i=1}^j \alpha^w_{i,j} m_{i-1}) \Big]^{-1}\prod_{j=1}^{\ell}\frac{(z_j^w)^{m_j}}{m_j!}\, ,
\end{multline}
where we defined the variables
\begin{equation}\label{variables triangle-track}
	z_\ell^w=\chi_\ell,\qquad z_j^w = \chi_{j} \prod_{k=j+1}^{\ell} \chi_i^{\alpha^w_{j+1,k}} \quad \text{for} \quad j\leqslant \ell-1\,.
\end{equation}
We observe that the structure of the shifted solution, i.e., the appearing combinations of summation variables in the $\Gamma$-functions as well as the variables do not depend on the first entry of the word $w$ but only the concrete combinations of propagator powers appearing in the $\Gamma$-functions. In particular, the variables $z_i^w$ only depend on $w'$ with $w=w_1w'$ for some $w_1\in \{0,1\}$.

Finally to find the coefficients of the basis functions we need to compute
\begin{equation}
	c_w = \lim_{\chi_1,\dots ,\chi_{\ell}\rightarrow 0} \textstyle\prod_{i=1}^{\ell}\chi_i^{-r_i^w} \phi_{2,1^{\ell-2},2}(\chi_1,\dots ,\chi_{\ell}) \,,
\end{equation}
for every binary word $w$ of length $\ell$. By renaming and rescaling the integration variables according to
\begin{equation}
	y_{\ell+1-j}\rightarrow y_j \prod_{k=j}^\ell \chi_i^{-\alpha^w_{j,k}}\, ,
\end{equation}
we can derive 
\begin{equation}
	c_w = \int\prod_{i=1}^{\ell}\frac{\dd y_i}{\sqrt{\pi}}|y_1|^{-2a_2}|1-y_\ell|^{-2a_1w_\ell} \prod_{j=1}^{\ell}|1-w_j+(-1)^{\ell-j} y_j|^{-2a_{j+2}} \prod_{j=1}^{\ell-1}|y_{j+1}-w_jy_j|^{-2b_{\ell-j}} \,.
\end{equation}
For any word $w$, this last integral yields
\begin{equation}
	c_w = \prod_{j=1}^{\ell} A(w_j b_{\ell-j} \! +\! (1\!-w_j)a_{j+2},\, b_{\ell+1-j} + w_j a_{j+2} +r_{j-1}^w/2) \,,
\end{equation} 
with the identifications $b_0 \equiv a_1$, $b_\ell \equiv a_2$ and the abbreviation
\begin{equation}
	A(a,b) = A_0(a) A_0(b) A_0(1-a-b) \,.
\end{equation}
Putting everything together we find the result
\begin{multline}\label{eq:triangleTrackResult}
	I_{2,1^{\ell-2},2} = \frac{A_0(a_1)A_0(a_2)\prod_{i=1}^{\ell-1}A_0(b_i)}{|x_{23}|^{2a_2+2a_3-1}\prod_{j=4}^{\ell+2}|x_{j-1,j}|^{2a_j+2b_{\ell+3-j}-1}|x_{1,\ell+2}|^{2a_{\ell+2}}} \\
	\times\sum_{w=w_1w'} \prod_{j=1}^{\ell}\chi_j^{r_j^w} \TT{w'}{a^w;\vec{b}^w}{\vec{c}^w;\vec{d}^w;e^w}{(-1)^{\eta_1^{w'}} z_1^{w'},\dots,(-1)^{\eta_{\ell-2}^{w'}}z_{\ell-2}^{w'},-z_{\ell-1}^{w'},z_{\ell}^{w'}} \,,
\end{multline}
where $w'=w_2\dots w_\ell$ is a word of length $\ell-1$, the defintion of the hypergeometric functions can be found in \Appref{app:defs}, and the parameters are
\begin{align}
	a^w &= 2a_1 + r_\ell^w\, ,\\
	b^w_i &= 2\tilde{a}_{i+2} + r_i^w \quad \text{for} \quad 1\leqslant i\leqslant \ell\, ,\\
	c_i^w &= 2(\tilde{a}_{i+2}+\tilde{b}_{\ell+1-i}) - r_{i-1}^w \quad \text{for}\,\ i\geqslant 2\,\ \text{such that}\quad w'_i = 0\, ,\\
	d_i^w &= 1+r_i^w \quad \text{for}\,\ i\geqslant 2\,\ \text{such that}\quad w'_i = 1\, ,\\
	e^w &= 2w_1(a_2+a_3) + 2(1-w_1)(\tilde{a}_2 + \tilde{a}_3)\\
	\eta_i^{w'} &= 1 + {\textstyle{\sum}}_{j=i+1}^{\ell-1} \alpha^{w'}_{i+1,j} \quad \text{for} \quad 1\leqslant i\leqslant \ell-2\, .
\end{align}
We recall that the indicials $r_i^w$ are given by \eqref{indicials triangle-track}, the variables $z_i^w$ are defined in \eqref{variables triangle-track}, and the notation $\alpha_{i,j}^w$ was introduced in \eqref{alphajkw}. Equation \eqref{eq:triangleTrackResult} can also be derived using the spectral transform, this computation is outlined in \Appref{app:spectraltriangle}.

\section{Aomoto--Gelfand Hypergeometric Functions}
\label{sec:AGfunctions}

Throughout this paper we have studied one-dimensional Feynman integrals of the general form
\begin{equation}
	I=\int_{\mathbb{R}^\ell}\prod_{k=1}^\ell\frac{\dd y_k}{\sqrt{\pi}}\prod_{(ij)\in \mathcal{E}_{\mathrm{e}}}\frac{1}{|x_i-y_j|^{2a_{(ij)}}}\prod_{(ij)\in \mathcal{E}_{\mathrm{i}}}\frac{1}{|y_i-y_j|^{2b_{(ij)}}} \,,
\end{equation}
where $\ell$ is the number of loops, the $x_i$ are the external points, $\mathcal{E}_{\mathrm{e}},\mathcal{E}_{\mathrm{i}}$ refer to the sets of external and internal edges, respectively, and the $a_k,b_k$ are the propagator powers. 
As already noted in \cite{Duhr:2023bku}, these integrals generally evaluate to so-called Aomoto--Gelfand (AG) hypergeometric functions \cite{aomotoHGFs,aomotoKita}. This is similar to the situation in two dimensions, where Feynman integrals evaluate to single-valued bilinears in AG hypergeometric functions \cite{Duhr:2023bku}. This should be contrasted however, with the situation in higher dimensions, where the class of hypergeometric functions needed to capture general Feynman integrals needs to be enlarged to the class of $\mathcal{A}$-hypergeometric or GKZ-hypergeometric functions \cite{gkzToralManifolds,gkzEulerIntegrals,gkzTalk} (see also \cite{sstBook}), or rather reductions thereof, as shown in \cite{feynmanIntsAndGKZDModules,Vanhove:2018mto,delaCruz:2019skx,Klausen:2019hrg}.
In this section we will review AG hypergeometric functions and their connection to one-dimensional Feynman integrals. In particular we will see how the non-local symmetries that are the focus of this paper are included in the defining system of differential equations for AG functions. 


\subsection{Review of Aomoto--Gelfand Hypergeometric Functions}

Aomoto--Gelfand (AG) hypergeometric functions are defined by a point $Z$ in the Grassmannian 
\begin{equation}
	G(k+1,n+1)=\doublequotient{\mathrm{GL}(k+1,\mathbb{C})}{X}{(\mathbb{C}^{\times})^{n+1}} \,,
\end{equation}
for some integers $k\leqslant n$. Here we defined
\begin{equation}
	X=\left\{ M\in \mathrm{Mat}_{k+1,n+1}(\mathbb{C}) \, : \, \mathrm{rank}(M)=k+1 \right\} \,,
\end{equation}
and we are modding out the multiplication by $\mathrm{GL}(k+1,\mathbb{C})$ matrices from the left and the rescaling of all columns by non-zero complex numbers. The AG functions associated with 
\begin{equation}
	Z=\begin{pmatrix}
	z_{00} & z_{01} & \dots &z_{0n} \\
	z_{10} & z_{11} & \dots & z_{1n} \\
	\vdots & \vdots & & \vdots \\
	z_{k0} & z_{k1} & \dots & z_{kn}
	\end{pmatrix}\in G(k+1,n+1) \,,
\end{equation}
admit the integral representations
\begin{equation}
	F_{\gamma}(Z)=\int_{\gamma}\omega\prod_{j=0}^n(t_0z_{0j}+t_1z_{1j}+\dots + t_k z_{kj})^{\alpha_j} \,,
\end{equation}
where $\gamma$ is some suitable $k$-cycle in $\mathbb{CP}^k$ and $\omega$ is the usual (holomorphic) volume form on projective space
\begin{equation}
	\omega=\sum_{i=0}^k(-1)^i t_i \dd t_0 \wedge \dd t_1 \wedge \dots \wedge \dd t_{i-1} \wedge \dd t_{i+1} \wedge \dots \wedge \dd t_k \,.
\end{equation}
The $\alpha_i$ are parameters which should satisfy
\begin{equation}
	\sum_{j=0}^n \alpha_j=-(k+1) \,,
\end{equation}
for the integrand to be a valid differential form on $\mathbb{CP}^k$. Apart from this constraint we further assume the $\alpha_i$ to be generic. 

To make connection to Feynman integrals we choose the affine coordinates $y_i=t_i/t_0$, $i=1,\dots k$ and choose $Z$ of the form
\begin{equation}
	Z=\begin{pmatrix}
	1 & z_{01} & \dots &z_{0n} \\
	0 & z_{11} & \dots & z_{1n} \\
	\vdots & \vdots & & \vdots \\
	0 & z_{k1} & \dots & z_{kn}
	\end{pmatrix} \,,
\end{equation}
with the first column corresponding to the hyperplane at infinity. The integral then takes the form
\begin{equation}
	F_{\gamma}(Z)=\int_{\gamma}\dd^k y\prod_{j=1}^n(z_{0j}+y_1z_{1j}+\dots + y_k z_{kj})^{\alpha_j} \,.
\end{equation}

We will be interested in the case where the entries of $Z$ are all real. The columns of that matrix, or the linear factors in the integrand, then define hyperplanes in real space which dissect $\mathbb{R}^k$ into chambers bounded by them. Note that since we are now working in affine coordinates the hyperplane at infinity is distinguished and we refer to chambers bounded by infinity as \emph{unbounded}. The integration cycle $\gamma$ is then chosen as such a chamber (or more generally a linear combination of them). Note that the integrand is a multivalued function for generic $\alpha_j$, hence one should supplement the integral $F_{\gamma}(Z)$ by a choice of branch for all of the factors in the integrand. One hence typically promotes the integration cycle $\gamma$ to a \emph{loaded} or \emph{twisted cycle} which takes the form of a tuple made up of a cycle and a choice of branch for the integrand. In the following we will suppress the information on the branch. For a rigorous definition of twisted cycles, see e.g., \cite{aomotoKita,Mizera:2019gea}.

Since the Feynman integrals we are interested in are defined with absolute values in the integrand it is also natural to define a rescaled cycle $\hat{\gamma}$ as follows. We make some arbitrary but fixed choice of branch for the negative linear factors and then rescale the cycle by a compensating phase which allows us to flip all of the negative factors to be positive. This amounts to essentially taking absolute values of the factors as long as we restrict to the chamber. Note that something similar has been done in \cite{Britto:2021prf} in the context of string amplitudes.

If we now consider some general $\ell$-loop Feynman integral with $E$ propagators, we can immediately see that it has to evaluate to a sum of AG hypergeometric functions with $k=\ell, n=E$. The corresponding element of $Z$ encodes the external points $x_i$ and topology of the diagram, while the parameters $\alpha_j$ correspond to the propagator powers $a_j,b_j$. To see this we simply dissect all of $\mathbb{R}^\ell$ into the chambers defined by the propagators (note that these have real coefficients) and decompose the integral into a sum of all of these chambers. On each chamber the  integral then manifestly evaluates to an AG hypergeometric function integrated over the appropriately rescaled cycle $\hat{\gamma}$ as explained above
\begin{equation}
	I=\sum_{\hat{\gamma}}F_{\hat{\gamma}}(Z) \,.
\end{equation}

The $Z$ matrices we get from Feynman integrals will take a very particular form. For example for a track-like integral with internal vertices $y_1,\dots, y_{\ell}$ and external vertices $x_i^{(j)}$, $i=1,\dots, k_j$ connected to internal vertex $y_j$, we have
\begin{equation}
\label{eq:feynmanZMatrix}
	Z= \begin{pmatrix}
	1 & -x_1^{(1)} & \dots & -x_{k_1}^{(1)} & -x_{1}^{(2)} & \dots &  -x_{k_{\ell}}^{(\ell)} & 0 & 0 & \dots & 0   \\
	0 & 1 & \dots & 1 & 0 & \dots & 0 & 1  & 0 & \dots & 0  \\
	0 & 0 & \dots & 0 & 1 & \dots&  0 & -1  & 1 & \dots  & 0 \\
	0 & 0 & \dots & 0 & 0 & \dots & 0 & 0 & -1& \dots  & 0\\
	\vdots & \vdots & & \vdots & \vdots & & \vdots & \vdots & \vdots&& \vdots \\
	0 & 0 & \dots & 0 & 0 & \dots & 1 & 0 & 0 & \dots & -1
	\end{pmatrix} \,.
\end{equation}
Note that the rows (apart from the zeroth one) correspond to the internal vertices, while the columns (apart from the zeroth one) correspond to the external and internal propagators.

\subsection{Aomoto--Gelfand Differential Equations}
The equivalence relations or gauge symmetries of the Grassmannian $G(k+1,n+1)$ lead to differential equations satisfied by the integrals $F_{\gamma}(Z)$, when linearized. Alternatively the AG hypergeometric functions can be defined as the solutions to this system of differential equations. Explicitly this system reads \cite{aomotoKita}
\begin{align}
	&\sum_{j=0}^nz_{ij}\frac{\partial F}{\partial z_{pj}}=-\delta_{ip}F,\qquad 0\leqslant i,p \leqslant k \,, \label{eq:AGDE1} \\
	&\sum_{i=0}^k z_{ij}\frac{\partial F}{\partial z_{ij}}=\alpha_j F ,\qquad 0\leqslant j\leqslant n \,,  \label{eq:AGDE2}\\
	&\frac{\partial^2 F}{\partial z_{ip}\partial z_{jq}}=\frac{\partial^2F}{\partial z_{iq} \partial z_{jp}},\qquad 0 \leqslant i < j \leqslant k,\, 0\leqslant p< q \leqslant n \label{eq:AGDE3} \,.
\end{align} 
Note however that generally we will not have a generic $Z$ but some of its entries take fixed values. Even in the most general case one can use the gauge symmetries of the Grassmannian to fix many entries. To find the differential equations satisfied by the resulting function we hence need to find a way to `gauge-fix' the AG differential equation system. In simple cases one can achieve this in practice by solving the first order relations for the partial derivatives with respect to the $z_{ij}$ which have been fixed. Plugging these into the second order equations then yields second order differential equations solely in terms of the reduced set of variables. 
\paragraph{Example: Triangle Integral.}

Let us illustrate this gauge-fixing procedure on the example of the triangle integral
\begin{equation}
\label{eq:triangleInt}
	I_3=\int\frac{\dd x_0}{\sqrt{\pi}}\frac{1}{|x_{01}|^{2a_1}|x_{02}|^{2a_2}|x_{03}|^{2a_3}} \,.
\end{equation} 
From the integral representation it is clear that it will evaluate to a sum of AG functions defined by the $Z$ matrix
\begin{equation}
	Z_{I_3}=\begin{pmatrix} 1 & -x_1 & -x_2 & -x_3 \\ 0 & 1 & 1 & 1  \end{pmatrix} \in G(2,4) \,,
\end{equation}
with the parameters $\alpha_i=-2a_i$ for $i=1,2,3$ and $\alpha_0=-2+2a_1+2a_2+2a_3$. We can use the first order differential equations to solve for $\partial_{z_{i0}},\partial_{z_{1j}}$ with $i=0,1,\, j=0,\dots, 3$. We further find the constraints
\begin{align}
	&(\partial_{x_1}+\partial_{x_2}+\partial_{x_3})F=0 \,, \\
	&(x_1\partial_{x_1}+x_2\partial_{x_2}+x_3\partial_{x_3})F=(1-2a_1-2a_2-2a_3)F \,,
\end{align}
which we can identify as the translation and dilatation Ward identity, respectively. Their solution, as we could have of course anticipated, takes the form
\begin{equation}
	F(x_1,x_2,x_3)=|x_{12}|^{1-2(a_1+a_2+a_3)}f(\chi) \,,
\end{equation}
with 
\begin{equation}
	\chi=\frac{x_{12}}{x_{13}} \,.
\end{equation}
The second order differential equations then yield an ordinary differential equation for the function $f(\chi)$ given by
\begin{equation}
	\left[\chi(\chi-1)\partial_{\chi}^2+2(\chi(a_1+2a_2+a_3)-a_1-a_2)\partial_{\chi}+2a_2(2(a_1+a_2+a_3)-1) \right]f(\chi)=0 \,,
\end{equation}
which can immediately be identified with the Gauss differential equation. We could have of course gotten here more directly by first rewriting the original integral as a prefactor times an integral only depending in $\chi$, gauge-fixing the corresponding system of AG equations then immediately yields the Gauss differential equation, see for example \cite{Abe:2018bgl}.

\paragraph{Connection to $\levo{P}$-Symmetries.}
In the simple example we studied we saw that the AG differential equation system reduces to Gauss' differential equation for the simple triangle integral, which is the same result one gets from the $\levo{P}$-symmetries of the integral. While this is not surprising in this simple case one might wonder if more generally the system of AG equations and the $\levo{P}$ Ward identities are equivalent for one-dimensional Feynman integrals. In particular the equivalence would imply that the hypergeometric family to which the integral evaluates is fully fixed by its non-local symmetries. Note that this is essentially a simpler version of the question addressed in \cite{Levkovich-Maslyuk:2024zdy} investigating connections between GKZ systems and $\levo{P}$-symmetries in higher dimensions.

The main difficulty in establishing such a connection is the fact that we need to gauge fix the AG system before being able to compare to the $\levo{P}$-symmetries. We will hence not be able to prove the equivalence of the two systems but show how  the $\levo{P}$-symmetries emerge from the AG system corresponding to a Feynman integral. Before we study the $\levo{P}$-symmetries let us however first see how the dilatation and translation Ward identities emerge from the AG system, which imply that the result always takes the form of some prefactor carrying the dimension times a function of ratios of differences.

The dilatation Ward identity emerges quite easily by considering \eqref{eq:AGDE2} for column $0$, which simply reads $\partial_{z_{00}}F=\alpha_0 F$, and plugging it into \eqref{eq:AGDE1} for the pair of rows $(0,0)$ yielding the Ward identity
\begin{equation}
	\left(\sum_{j=0}^E z_{0j}\partial_{z_{0j}}+1\right) F=\left( \sum_i x_i\partial_{x_i}+\sum_i 2a_i-\ell \right)F = 0\,,
\end{equation}
as claimed. Here the sums over $i$ are over all external points $x_i$, with propagator powers $a_i$. To show the translation Ward identity we need to consider \eqref{eq:AGDE1} for the pair of rows $(i,0)$ and sum over all $i>0$. This yields
\begin{equation}
	\sum_{i=1}^{\ell}\sum_{j=0}^E z_{ij}\partial_{z_{0j}}F=-\sum_k\partial_{x_k}F+\sum_{\alpha}\partial_{z_{0\alpha}}F\sum_{i=1}^{\ell}{z_{i\alpha}}=0 \,.
\end{equation}
The sum in the first term is over all external points and the sum in the second term is over all columns $\alpha$ corresponding to internal propagators. For each such column we sum over $i=1,\dots ,\ell$, i.e.\ over all vertices. The $z_{i\alpha}$ is then only non-zero if the vertex $i$ is one of the ends of the propagator $\alpha$ and then it is $\pm 1$ depending on the orientation. Since however every propagator has precisely two end points this sum will always yield $(+1)+(-1)=0$ and hence we indeed recover the translation Ward identity.

After this warm-up let us now turn to the $\levo{P}$-symmetries, starting with the two-point symmetry. To see how this emerges, fix some vertex (and hence row) $a$ and two columns $i,j$ containing $x_i,x_j$, connected to vertex $a$. From the equations \eqref{eq:AGDE2} for these two columns we can solve for the two derivatives
\begin{equation}
	\partial_{z_{ai}}F=-(2a_i+x_i\partial_{x_i})F,\qquad \partial_{z_{aj}}F=-(2a_j+x_j\partial_{x_j})F \,.
\end{equation}
Plugging this into the second order equations  \eqref{eq:AGDE3} for rows $0,a$ and columns $i,j$ then yields
\begin{equation}
	\left[x_{ij}\partial_{x_i}\partial_{x_j}+2a_i\partial_{x_j}-2a_j\partial_{x_i} \right]F=0 \,,
\end{equation}
which precisely reproduces the two-point symmetry $\levo{P}_{ij}F=0$, c.f. \eqref{1D Pjk}.

As a less trivial example let us demonstrate how the bridge-vertex symmetry emerges from the system. This will also highlight some of the difficulties in proving full equivalence of the AG equations and the $\levo{P}$-symmetries. To this end consider three vertices $a,b,c$ where $a,c$ are end-vertices and there are propagators $\alpha,\beta$ from $a$ to $b$ and from $b$ to $c$, respectively. We will denote the external propagators (and hence the columns) connected to $a,b,c$ by $X_a,X_b,X_c$. Now consider the differential equations \eqref{eq:AGDE1} on rows $(a,a)$ and $(c,c)$
\begin{equation}
	\left[\sum_{j\in X_a}\partial_{z_{aj}}+\partial_{z_{a\alpha}}+1\right]F=0, \qquad \left[\sum_{j\in X_c}\partial_{z_{cj}}-\partial_{z_{c\beta}}+1\right]F=0 \,,
\end{equation}
and on the rows $(a,b)$ and $(c,b)$
\begin{equation}
	\left[\sum_{j\in X_a}\partial_{z_{bj}}+\partial_{z_{b\alpha}}\right]F=0, \qquad \left[\sum_{j\in X_c}\partial_{z_{bj}}-\partial_{z_{b\beta}}\right]F=0 \,.
\end{equation}
Further consider the equations \eqref{eq:AGDE2} for columns $\alpha,\beta$
\begin{equation}
	\left[\partial_{z_{a\alpha}}-\partial_{z_{b\alpha}}+2a_{\alpha}\right]F=0,\qquad \left[\partial_{z_{b\beta}}-\partial_{z_{c\beta}}+2a_{\beta}\right]F=0 \,.
\end{equation}
Using these relations we can deduce
\begin{align}
	&\sum_{i\in X_a}\sum_{j\in X_c}\left(\partial_{z_{0j}}\partial_{z_{ai}}-\partial_{z_{0i}}\partial_{z_{cj}} \right)F \nonumber \\
	&= \sum_{i\in X_a}\sum_{j\in X_c}\left(\partial_{z_{0j}}\partial_{z_{bi}}-\partial_{z_{0i}}\partial_{z_{bj}} \right)F
	-(2a_{\beta}-1)\sum_{j\in X_a}\partial_{z_{0j}}F +(2a_{\alpha}-1)\sum_{j\in X_c}\partial_{z_{0j}}F \,.
\end{align}
The first sum on the right-hand side vanishes due to \eqref{eq:AGDE3} while we can identify the others as momentum generators acting on the vertices $a$ and $c$. The left hand side yields the bilocal generator $\levo{P}_{ac}$ acting on vertices $a,c$ as before. We can hence conclude that
\begin{equation}
	\left[\levo{P}_{ac}-\frac{1}{2}(1-2a_{\beta})\gen{P}_a+\frac{1}{2}(1-2a_{\alpha})\gen{P}_c \right]F=0 \,,
\end{equation}
which is indeed the bridge-vertex symmetry of \cite{Loebbert:2024qbw}, specialized to $D=1$.

The  remaining symmetry relations of \Secref{sec:PhatForTrees}, which were proven in \cite{Loebbert:2024qbw} for integrals in any spacetime dimension $D$, can also be shown using similar arguments as above when specifying to $D=1$, see \Appref{sec:PHatFromAGDEs}. It may even be possible to argue that one cannot get more equations out of gauge-fixing the AG system, which would be a rigorous proof that the $\levo{P}$-symmetries fully determine the system of hypergeometric functions to which the Feynman integrals evaluate, at least in one dimension. We leave this for future work. The example shown here, however, already showcases some of the difficulties, as one generally will need some nontrivial linear combinations of the AG equations to connect to the non-local symmetries and one thus has to argue that one cannot find more combinations of the equations which reduce to differential equations in the $x_i$.


\section{From 1D to 2D}
\label{sec:1to2D}

We explain in this section how to use the result for one-dimensional integrals of the form
\begin{equation}
	I^\textrm{1D} = \int\prod_{j=1}^{\ell}\frac{\dd y_j}{\sqrt{\pi}} \prod_{(ij)}\frac{1}{|x_i-y_j|^{2a_{(ij)}}} \prod_{(ij)}\frac{1}{|y_i-y_j|^{2b_{(ij)}}}
\end{equation}
to straightforwardly compute the two-dimensional integral
\begin{equation}
	I^{\mathrm{2D}} = \int\prod_{j=1}^{\ell}\frac{\dd^2 w_j}{\pi} \prod_{(ij)}\frac{1}{|z_i-w_j|^{2a_{(ij)}}} \prod_{(ij)}\frac{1}{|w_i-w_j|^{2b_{(ij)}}}\, ,
\end{equation}
where the external points $x_k\in\mathbb{R}^2$ have been combined in
\begin{equation}
	z_k=x_k^1+\ii\, x_k^2,\qquad \bar{z}_k=x_k^1-\ii\, x_k^2\, ,
\end{equation}
and the integration measure is $\dd^2 w_j = \ii\, \dd w_j\wedge \dd \bar{w}_j/2$.


\paragraph{1D to 2D Algorithm.}

More precisely, we show that the final 2D result can be obtained from the final 1D result following the simple replacements
\begin{align}
	\mathcal{F}\left[\substack{2a+r,\dots \\ 2a'+r',\dots};\chi,\dots\right] &\longrightarrow \overline{\mathcal{F}}^{\mathrm{2D}}\left[\substack{a+r,\dots \\ a'+r',\dots};\bar\chi,\dots\right] \mathcal{F}^{\mathrm{2D}}\left[\substack{a+r,\dots \\ a'+r',\dots};\chi,\dots\right] \, ,\\
	\chi = x_{ij}/x_{kl} &\longrightarrow \chi = z_{ij}/z_{kl}\, ,\\
	|\chi|^{2a+r} &\longrightarrow |\chi|^{2a+2r}\, ,
\end{align}
where $a$, $a'$ are linear combination of propagator powers with coefficients in $\{-1,0,1\}$, the numbers $r,r'$ are integers, and the 1D and 2D hypergeometric series are the same up to normalisation: if
\begin{equation}
	\mathcal{F}\left[\substack{2a+r,\dots \\ 2a'+r',\dots};0,\dots,0\right] = \frac{A_{0}(a'+r'/2)\dots}{A_{0}(a+r/2)\dots}\, ,
\end{equation}
then
\begin{equation}
	\mathcal{F}^{\mathrm{2D}}\left[\substack{a+r,\dots \\ a'+r',\dots};0,\dots,0\right] = \frac{\Gamma(a+r)\dots}{\Gamma(a'+r')\dots} \quad \text{and} \quad \overline{\mathcal{F}}^{\mathrm{2D}}\left[\substack{a+r,\dots \\ a'+r',\dots};0,\dots,0\right] = \frac{\Gamma(1-a'-r')\dots}{\Gamma(1-a-r)\dots}\, .
\end{equation}
For the prefactor and the overall normalisation, we must also perform the replacements
\begin{equation}
	A_0(a+r/2) \longrightarrow A^{\mathrm{2D}}_{0}(a+r)\quad\text{and}\quad |x_{ij}|^{2a+r} \longrightarrow |z_{ij}|^{2a + 2r}\, ,
\end{equation}
where we have introduced the function (defined for integer $\ell$)
\begin{equation}
	A^{\mathrm{2D}}_\ell(a) = \frac{\Gamma\left(1+\frac{\ell}{2}-a\right)}{\Gamma\left(\frac{\ell}{2}+a\right)} = \frac{1}{A^{\mathrm{2D}}_\ell(1-a)} = (-1)^{\ell} A^{\mathrm{2D}}_{-\ell}(a)\, .
\end{equation}

\subsection{Two- and Three-Point Integrals}

The spectral transform in two dimensions involves spinning propagators that depend on two exponents $\ba$ and $\bba$ such that $\ba-\bba\in\mathbb{Z}$, and are given by
\begin{equation}
	[z]^{\ba} = |z|^{\ba+\bba} \e^{\ii(\ba-\bba)\theta}\quad\text{for}\quad z = |z|\e^{\ii\theta} \in\mathbb{C}\, .
\end{equation}
We stress that $\bba$ is not the complex conjugate of $\ba$. One can think of $[z]^{\ba}$ as $z^{\ba} \bar{z}^{\bba}$. We will often parametrise the exponents according to $(\ba,\bba) = (\ell/2+a,-\ell/2+a)$, where $(a,\ell)\in\mathbb{C}\times \mathbb{Z}$. We refer to propagators for which $\ba = \bba$, i.e.\ $\ell=0$, as scalar propagators. The spectral representation of a single propagator then reads
\begin{equation}\label{SoV prop 2d}
	\frac{1}{[x_{12}]^{\mathbf{a}}} = \sum_{\ell' = -\infty}^{+\infty}\! \int_{\mathbb{R} + \ii \eta} \! \frac{[x_{34}]^{\mathbf{a}}}{[x_{13} x_{24}]^{\mathbf{a}+\bu} [x_{14} x_{23}]^{-\bu}} \frac{A^{(2)}_\ell(a)}{A^{(2)}_{\ell'}(-\ii u)\, A^{(2)}_{\ell+\ell'}(a+\ii u)} \frac{\dd u}{2 \pi}\, .
\end{equation}
Note that even if we are interested in a 2D Feynman integral with only scalar propagators, computing it using the spectral transform requires to deal with spinning  propagators.

The star-triangle relation in 2 dimensions reads
\begin{equation}\label{star-triangle 2d}
	\includegraphicsbox[scale=.8]{FigThreePointStarArrows.pdf}
	\hspace{-4mm}
	=\!\int\! \frac{\pi^{-1} \dd^2 z_0}{\prod_{i=1}^3 [z_{i0}]^{\ba_i}} = \frac{\prod_{i=1}^3 A^{\mathrm{2D}}_{\ell_i}(a_i)}{[z_{12}]^{1-\ba_3} [z_{23}]^{1-\ba_1} [z_{31}]^{1-\ba_2}}
	\propto\hspace{-4mm}\includegraphicsbox[scale=.8]{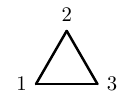},
\end{equation}
where the parameters $(a_i,\ell_i)\in\mathbb{C}\times \mathbb{Z}$ satisfy the constraints  $a_1 + a_2 + a_3 = 2$ and 
$\ell_1 + \ell_2 + \ell_3 = 0$. These constraints read equivalently $\ba_1+\ba_2+\ba_3 = \bba_1+\bba_2+\bba_3 = 2$.
Taking the limit $|z_3|\to\infty$ in the star-triangle relation yields the chain relation
\begin{equation}\label{chain 2d}
	\int \frac{\pi^{-1} \dd^2 z_0}{[z_{10}]^{\ba_1} [z_{02}]^{\ba_2}} = A^{\mathrm{2D}}_{\ell_1}(a_1) A^{\mathrm{2D}}_{\ell_2}(a_2) A^{\mathrm{2D}}_{-\ell_1-\ell_2}(2-a_1-a_2) \frac{1}{[z_{21}]^{\ba_1+\ba_2-1}}\, .
\end{equation}

In order to exemplify the procedure, we begin with the simplest non-trivial example, namely the non-conformal three-point integral
\begin{equation}
	I^{\mathrm{2D}}_3 = \hspace{-3mm} \includegraphicsbox[scale=.8]{FigThreePointStarArrows.pdf}
	\hspace{-2mm}
	=\!\int\! \frac{\pi^{-1} \dd^2 z_0}{\prod_{i=1}^3 [z_{i0}]^{\ba_i}}\, .
\end{equation}
We follow similar steps as in the one-dimensional case, but using the two-dimensional versions of the spectral representation of a propagator \eqref{SoV prop 2d} and of the chain relation \eqref{chain 2d}, to get
\begin{equation}\label{SoV star 2d}
	I^{\mathrm{2D}}_3 = \frac{A^{\mathrm{2D}}_{-\ell_2}(a_2) A^{\mathrm{2D}}_{\ell_3}(a_3)}{[z_{12}]^{\ba_1+\ba_2+\ba_3-1}} \sum_{\ell \in\mathbb{Z}}\! \int_{\mathbb{R} + \ii \eta}\!\!\!\! \frac{A^{\mathrm{2D}}_{\ell+\ell_1+\ell_3}(a_1+a_3+\ii u) A^{\mathrm{2D}}_{\ell+\sum_i\ell_i}(2-\textstyle{\sum_i} a_i-\ii u)}{A^{\mathrm{2D}}_{\ell}(-\ii u) A^{\mathrm{2D}}_{\ell_3+\ell}(a_3+\ii u)} \frac{[\chi]^{\bu} \dd u}{2\pi}\, ,
\end{equation}
where $(\bu,\bbu) = (\ell/2 + \ii u,-\ell/2+\ii u)$, $\chi=x_{13}/x_{12}$, and as usual each half-infinite series of poles lies entirely on one side of the integration contour. Closing the integration contour in the lower half-plane, we have to pick the residues of the simple poles in $-\ii(|\ell|/2+\mathbb{N})$ and in $-\ii(1-a_1-a_3+|\ell+\ell_1+\ell_3|/2+\mathbb{N})$. This gives\footnote{$\bar\chi$ is the complex conjugate of $\chi$.}
\begin{multline}
	I^{\mathrm{2D}}_3 = \frac{A^{\mathrm{2D}}_{-\ell_2}(a_2) A^{\mathrm{2D}}_{\ell_3}(a_3)}{[z_{12}]^{\ba_1+\ba_2+\ba_3-1}} \bigg[ \cGGtb{2}{1}{\bba_3,\sum_i \bba_i - 1}{\bba_1+\bba_3}{\bar\chi} \cGGt{2}{1}{\ba_3,\sum_i \ba_i - 1}{\ba_1+\ba_3}{\chi}\\
	+ [\chi]^{1-\ba_1-\ba_3} \cGGtb{2}{1}{\bba_2,1-\bba_1}{2-\bba_1-\bba_3}{\bar \chi} \cGGt{2}{1}{\ba_2,1-\ba_1}{2-\ba_1-\ba_3}{\chi}\bigg]\, ,
\end{multline}
where
\begin{align}
	\cGGt{2}{1}{a,b}{c}{x} &= \frac{\Gamma(a) \Gamma(b)}{\Gamma(c)} \GG{2}{1}{a,b}{c}{x}\, ,\\
	\cGGtb{2}{1}{a,b}{c}{x} &= \frac{\Gamma(1-c)}{\Gamma(1-a) \Gamma(1-b)} \GG{2}{1}{a,b}{c}{x}\, .
\end{align}
Note that when $\ell_1=\ell_2=\ell_3 = 0$, this result can be obtained from the one-dimensional result using the procedure described at the beginning of this section.


\subsection{General Integrals}
\label{sec:genints2D}

In two dimensions, we study integrals of the form
\begin{equation}
	I^{\mathrm{2D}}=\int\prod_{j=1}^{\ell}\frac{\dd^2 w_j}{\pi} \prod_{(ij)}\frac{1}{[z_i-w_j]^{2\ba_{ij}}} \prod_{(ij)}\frac{1}{[w_i-w_j]^{2\mathbf{b}_{ij}}} \,.
\end{equation}
We see that the integrand factorises into a holomorphic and an anti-holomorphic copy which each essentially look like the integrands of one-dimensional Feynman integrals with the propagator powers rescaled by $1/2$. We will refer to this as the \emph{double copy} \cite{Brown:2018omk} of the one-dimensional integral. \\


\paragraph{$\levo{P}$-Symmetries and Double Copy.}
It turns out that there is also factorisation in the result of the integration, as can be deduced from the differential equations (see also \cite{Duhr:2023bku} for a different argument). 
To see this consider the operators 
\begin{equation}
	\widehat{\gen{P}}_{jk}^\mathrm{2D}=\frac{i}{2}\brk[s]*{(z_j-z_k)\partial_{z_j}\partial_{z_k}-\ba_k\partial_{z_j}+\ba_j\partial_{z_k}} \,, \qquad
	\widehat{\overline{\gen{P}}}_{jk}^\mathrm{2D}=\frac{i}{2}\brk[s]*{(\bar{z}_j-\bar{z}_k)\partial_{\bar{z}_j}\partial_{\bar{z}_k}-\bba_k\partial_{\bar{z}_j}+\bba_j\partial_{\bar{z}_k}} \,.
\end{equation}
Note that these take precisely the same form as the one-dimensional $\levo{P}_{jk}$ with the propagator powers $2a$ replaced by the holomorphic or anti-holomorphic propagator power $\ba$ or $\bba$, respectively. It can be seen analogously to the one-dimensional case, that these annihilate a product of holomorphic or anti-holomorphic propagators, respectively. 
Similarly the proofs of \cite{Loebbert:2024qbw} for the partial $\levo{P}$ symmetries (cf.\ \secref{sec:Phat}) go through  for this generalization to differing holomorphic and anti-holomorphic propagator powers; these only rely on integration by parts which can be performed independently in the holomorphic and anti-holomorphic coordinate due to the factorized structure of the integrand. Hence we see that the symmetries of a one-dimensional integral immediately give rise to two copies of this symmetry, a holomorphic and an anti-holomorphic one, in two dimensions. 
As a consequence, given a basis for the $m$-dimensional solution space of the differential equations satisfied by some one-dimensional $n$-point Feynman integral, we can immediately infer the basis for the corresponding two-dimensional problem.

 To be explicit, consider some one-dimensional, $n$-point Feynman integral $I^{\mathrm{1D}}$ that we decompose as
\begin{equation}
	I^{\mathrm{1D}} = V^{\mathrm{1D}}(x_1,\dots ,x_n)\, \phi^{\mathrm{1D}}(\chi_1,\dots ,\chi_{n-2})
\end{equation}
for some suitable prefactor $V^{\mathrm{1D}}$ and function $\phi^{\mathrm{1D}}$ of the ratios $\chi_j$. Following our bootstrap algorithm we can compute a basis for the solution space of $\phi^{\mathrm{1D}}$ taking the form  
\begin{equation}
	\{\chi^{r^{(1)}} f_1(\chi),\dots ,\chi^{r^{(m)}} f_m(\chi) \} \,, 
\end{equation}
with the abbreviation $\chi^{r^{(j)}}=\chi_1^{r_1^{(j)}}\dots \chi_{n-2}^{r_{n-2}^{(j)}}$, for some power series $f_j$ and indicials $r^{(j)}$. We now consider the corresponding two-dimensional Feynman integral, which we decompose as
\begin{equation}
	I^{\mathrm{2D}} = V^{\mathrm{2D}}(z_1,\bar{z}_1,\dots ,z_n,\bar{z}_n) \phi^{\mathrm{2D}}(\chi_1,\dots ,\chi_{n-2},\bar{\chi}_1,\dots \bar{\chi}_{n-2}) \,.
\end{equation}
Here $V^{\mathrm{2D}}$ is obtained from $V^{\mathrm{1D}}$ as follows:
\begin{equation}
	V^{\mathrm{1D}}(x_1,\dots ,x_n) = \prod_{(ij)} |x_i-x_j|^{2a_{ij}+r} \rightarrow V^{\mathrm{2D}}(z_1,\bar{z}_1,\dots ,z_n,\bar{z}_n) = \prod_{(ij)} [z_i-z_j]^{\ba_{ij}+r}\, ,
\end{equation}
where $r$ is an integer, $a_{ij}$ is a linear combination of the 1D propagator exponents $a_k$ and $b_k$, and $\ba_{ij},\bba_{ij}$ are the same linear combinations of the 2D exponents $\ba_k$ and $\mathbf{b}_k$ or $\bba_k$ and $\mathbf{\bar{b}}_k$. The two-dimensional variables $\chi_i,\bar{\chi}_i$ are the same as the one-dimensional variables with $x_j$ replaced by $z_j$ or $\bar{z}_j$, respectively. Then our above argument implies that the function $\phi^{\mathrm{2D}}$ satisfies differential equations which split into holomorphic and anti-holomorphic copies of the differential equations satisfied by $\phi^{\mathrm{1D}}$ (with halved propagator powers). This means that $\phi^{\mathrm{2D}}$ belongs to the $m^2$-dimensional solution space of these differential equations and can be expanded as
\begin{equation}\label{2Dphi basis}
	\phi^{\mathrm{2D}} = \sum_{i,j=1}^m c_{i,j} \left[\chi^{r^{(i)}} f_i(\chi)\right]\!\big|_{\substack{a \rightarrow \ba/2 \\ b\rightarrow \mathbf{b}/2}} \left[\bar{\chi}^{r^{(j)}}f_j(\bar{\chi})\right]\!\big|_{\substack{a \rightarrow \bba/2 \\ b\rightarrow \mathbf{\bar{b}}/2}} \,.
\end{equation}
for some constant coefficients $c_{i,j}$ that we must still determine. The fact that the basis of the 2D solution space is given by precisely these $m^2$ functions can be seen as follows. These basis functions certainly solve the 2D differential equations. Furthermore they span the full solution space since the $\chi,\bar{\chi}$ dependence is fully factorized. Indeed, consider first the holomorphic system and write down the known solution
\begin{equation}
	\phi^{\mathrm{2D}}=\sum_{i=1}^m c_i(\bar{\chi})\chi^{r^{(i)}}f_i(\chi)\big|_{\substack{a \rightarrow a/2 \\ b\rightarrow b/2}} \,.
\end{equation}
The coefficients $c_i$ in an expansion in these 1D basis functions now depend on the anti-holomorphic variables $\bar \chi$, which need to solve the 1D differential equations themselves. Hence, by again using the known solution of these 1D differential equations we find the above set of basis functions. Finally note that there can be no non-trivial linear relations between these functions as they would lead to non-trivial linear relations of the 1D basis functions upon specialization of, say, only the anti-holomorphic variables, which cannot exist.\footnote{In this argument, we are treating the holomorphic and anti-holomorphic variables as independent which is valid since we are only considering the differential equations which we can formally view as differential equations in both the holomorphic and anti-holomorphic variables since the relation via complex conjugation does not play a role here.}

We will now use arguments based on the spectral transform to show that $c_{i,j}=0$ whenever $i\neq j$, and to determine the diagonal coefficients $c_{i,i}$ for scalar propagators.
Given the similar forms of the spectral transform and the star-triangle relation in one and two dimensions, see \eqref{SoV prop}, \eqref{star-triangle} and \eqref{SoV prop 2d}, \eqref{star-triangle 2d}, it is clear that the spectral representation for 2D integrals is the same as in 1D up to the replacements
\begin{equation}\label{1Dto2D}
	A_{[\ep_u+\ep_a]}(\ii u+a+r/2) \longrightarrow A^{\mathrm{2D}}_{\ell_u+\ell_a}(\ii u+a+r) \, ,\qquad \sgn^{\ep_j}(\chi_j)|\chi_j|^{2\ii u_j} \longrightarrow [\chi_j]^{\bu_j}\, ,
\end{equation}
where $u$ stands for any linear combination of the spectral variables $u_j$, $a$ stands for any linear combination of the external and internal propagator exponents $a_i$ and $b_i$, and $r\in\mathbb{Z}$. This procedure only generates the correct 2D integrand up to a sign. Part of this sign depends only on the 2D propagator exponents (namely $\ba_i - \bba_i$ and $\mathbf{b}_i - \mathbf{\bar{b}}_i$), and can be absorbed in the overall prefactor, which depends on the precise definition of the integral anyway. However, this sign could also contain the spin associated to the spectral variables $\bu_i$. But it does not matter because we also know that the result must be of the double-copy form \eqref{2Dphi basis}. 

Let us examine a little more closely the possible form of the residues. First, we consider those coming from some $A^{\mathrm{2D}}_{\ell}(1+ \ii u)$, i.e. located at $u \in -\ii(|\ell|/2+\mathbb{N})$. For $u \to -\ii(|\ell|/2+m)$, we have
\begin{equation}
	-\ii A^{\mathrm{2D}}_{\ell}(1+ \ii u) \sim \frac{1}{u +\ii(|\ell|/2+m)}\times \frac{(-1)^q}{p!\, q!} \, ,
\end{equation}
where we introduced the non-negative integers $(p,q) = ((|\ell|+\ell)/2+m, (|\ell|-\ell)/2+m)$. Under this change of summation indices, the sums become
\begin{equation}
	\sum_{\ell\in\mathbb{Z}}\sum_{m\geqslant 0} = \sum_{p,q\geqslant 0}\, .
\end{equation}
The other parts of the integrand are regular when $u \to -\ii(|\ell|/2+m)$ and are of the form
\begin{equation}
	A^{\mathrm{2D}}_{\ell+\ell_a}(\ii u + a) \to A^{\mathrm{2D}}_{\ell_a}(a) \frac{(-1)^q}{(\ba)_p (\bba)_q}\, ,\quad \text{or}\quad A^{\mathrm{2D}}_{-\ell+\ell_a}(-\ii u + a) \to A^{\mathrm{2D}}_{\ell_a}(a)  (-1)^p (1-\ba)_p (1-\bba)_q\, ,
\end{equation}
as well as
\begin{equation}
	[\chi]^{\bu} \to \chi^p \bar\chi^q\, .
\end{equation}
More generally, the residues could come from some $A^{\mathrm{2D}}_{\ell_u+\ell_a}(\ii u+a+2r)$, and they would then be located at $u \in -\ii(|\ell+\ell_a|/2+1-2r-a+\mathbb{N})$. For such a series of residues, we would first make the change of variable $\ell\to\ell-\ell_a$, and that would bring us back to the previous situation: the residue would still be factorised. It is thus clear that the sums over $p$ and $q$ factorise, and the signs must combine such that these sums only differ by the exchange $\ba\leftrightarrow \bba$ and $\chi\leftrightarrow\bar\chi$. This is because in this procedure, each rapidity $u$ must appear in an even number of functions $A^{\mathrm{2D}}$, and given the form of the residues shown above the associated sign can then only be $+1$ or $(-1)^{p+q} = (-1)^\ell$. Comparing with the expansion \eqref{2Dphi basis}, this shows that all the off-diagonal coefficients vanish. Moreover, the diagonal coefficients can also be read off from the residues given above.

For two-dimensional integrals with only scalar propagators, we have thus proved the procedure described at the beginning of this section. However, when we consider spinning propagators, these simple replacement rules are not enough since they come with a sign ambiguity: it does not make sense to try to apply $A_0(a+r/2) \longrightarrow A^{\mathrm{2D}}_{\ell_a}(a+r)$ since we have $A_0(a+r/2) = A_0^{-1}((1-r)/2-a)$ for the 1D function but $A^{\mathrm{2D}}_{\ell_a}(a+r) = (-1)^{\ell_a} A^{\mathrm{2D},-1}_{-\ell_a}(1-r-a)$.


\section{Conformal Double-Box Integral}
\label{sec:confdoubbox}

In this last section, we compute a particularly interesting example, namely the conformal double-box integral in one and two dimensions. We will first compute this integral in 1D using the spectral transform method of \Secref{sec:spectransf}. We will then compute it in 2D for the generalized case of spinning propagators using the same method, and we verify that for scalar propagators, the result reduces to the expression obtained from the recipe given in the previous \Secref{sec:1to2D}.

\subsection{Conformal Double Box in 1D}

Let us consider the conformal version of the double-box integral computed in \secref{sec:DoubleBoxNonConf}, namely
\begin{equation}
	I_{3,3}^\text{conf}
	=\includegraphicsbox{FigDoubleBoxConformal.pdf}
	= \int \frac{\pi^{-1} \dd x_0\, \dd x_{0'}}{|x_{10}|^{2a_1} |x_{60}|^{2a_6} |x_{50}|^{2a_5} |x_{00'}|^{2b} |x_{40'}|^{2a_4} |x_{30'}|^{2a_3} |x_{20'}|^{2a_2}}\, ,
\end{equation}
where the two conformal integration vertices are indicated by circles. Conformal symmetry requires the previously independent parameters to satisfy the constraints
\begin{equation}
	a_1 + a_5 + a_6 = a_2 + a_3 + a_4 = 1-b\, .
\end{equation}
Starting from the explicit result \eqref{eq:resultDoubleBox} for the generic double box integral, one notices that the two terms involving $\mathcal{B}_3$ and the two terms involving $\mathcal{B}_5$ drop out, as their coefficients vanish in this conformal configuration. However, this gives the result for this 5-parameter integral in terms of 7-parameter hypergeometric functions. 

We can actually do better and write it in terms of 5-parameter hypergeometric functions. This requires to apply the spectral representation in a way that takes into account from the start the conformal nature of the integral. We begin by applying it to $|x_{10}|^{-2a_1}$: we use~\eqref{SoV prop} with the replacements $(x_1,x_2,x_3,x_4,a,u,\ep)\rightarrow (x_0,x_1,x_5,x_6,a_1,u_3,\ep_3)$. After that, we perform the integral over $x_0$ using the star-triangle relation \eqref{star-triangle}. The remaining integral over $x_{0'}$ is a conformal pentagon integral involving two spinning propagators, those connected to $x_6$ and $x_5$. We thus use the general spectral representation \eqref{SoV prop sign} with the replacements $(x_1,x_2,x_3,x_4,a,\ep,u,\ep')\rightarrow (x_{0'},x_6,x_3,x_4,1/2-a_1-a_5-\ii u_3,\ep_3,u_2,\ep_2)$, and the simpler representation \eqref{SoV prop} with the replacements $(x_1,x_2,x_3,x_4,a,u,\ep)\rightarrow (x_{0'},x_2,x_4,x_3,a_2,u_1,\ep_1)$. We can then finally perform the integral over $x_{0'}$ using the star-triangle relation. Doing so we arrive at
\begin{align}
	I_{3,3}^\text{conf} &= A_0(b) A_0(a_1) A_0(a_2)\frac{|x_{34}|^{2(a_2+b)-1} |x_{35}|^{2(a_2+a_4)-1} |x_{46}|^{2(a_1+a_5)-1} |x_{56}|^{2(a_1+b)-1}}{|x_{16}|^{2a_1} |x_{23}|^{2a_2} |x_{45}|^{2(a_1+a_2+a_4+a_5 + b-1)}} \nonumber\\
	&\times \prod_{j=1}^3 \sum_{\ep_j = 0}^1 \sgn^{\ep_j}(\chi_j)\int_{\mathbb{R}+\ii \eta} \frac{\dd u_j}{2\sqrt{\pi}} |\chi_j|^{2\ii u_j} (-1)^{\ep_1\ep_3} \nonumber\\
	&\times\frac{A_{\ep_1}(\tilde{a}_2-\ii u_1) A_{\ep_3}(\tilde{a}_1-\ii u_3) A_{[\ep_1+\ep_2]}(a_2+a_4+\ii u_{12}) A_{[\ep_2+\ep_3]}(a_1+a_5+\ii u_{32})}{A_{\ep_1}(-\ii u_1) A_{\ep_2}(-\ii u_2) A_{\ep_3}(-\ii u_3) A_{[\sum_j \ep_j]}(b+\sum_{i\neq 3,6} a_i - 1 + \ii(u_1+u_3-u_2))}\, .
\end{align}
where the cross ratios are
\begin{equation}\label{cross-ratios double box}
	\chi_1 = \frac{x_{24} x_{35}}{x_{23} x_{45}}\, ,\qquad \chi_2 = \frac{x_{36} x_{45}}{x_{35} x_{46}}\, ,\qquad\chi_3 = \frac{x_{46} x_{15}}{x_{45} x_{16}}\, .
\end{equation}
Assuming that all the cross ratios are small enough, we compute the integrals as sums over residues and get
\begin{align}
	&I_{3,3}^\text{conf} = A_0(b) A_0(a_1) A_0(a_2)\frac{|x_{34}|^{2(a_2+b)-1} |x_{35}|^{2(a_2+a_4)-1} |x_{46}|^{2(a_1+a_5)-1} |x_{56}|^{2(a_1+b)-1}}{|x_{16}|^{2a_1} |x_{23}|^{2a_2} |x_{45}|^{2(\sum_{i\neq 3,6} a_i + b-1)}} \nonumber\\
	&\times\bigg[\CC{1}{2a_2,2a_1;2(b+\sum_{i\neq 3,6} a_i - 1)}{2(a_2+a_4),2(a_1+a_5)}{\chi_1,-\chi_2,\chi_3} \nonumber\\
	&+ |\chi_1|^{1-2(a_2+a_4)} |\chi_3|^{1-2(a_1+a_5)} \CC{2}{2\tilde{a}_4,2\tilde{a}_5;2b}{2(\tilde{a}_2+\tilde{a}_4),2(\tilde{a}_1+\tilde{a}_5)}{\chi_1,-\chi_1\chi_2\chi_3,\chi_3} \nonumber\\
	&+ |\chi_2|^{2(b+\sum_{i\neq 3,6} a_i-1)} \CC{3}{2a_2,2a_1;2\tilde{a}_3,2\tilde{a}_6}{2(b+\sum_{i\neq 3,6} a_i)-1}{\chi_1\chi_2,\chi_2,\chi_2\chi_3} \nonumber\\
	&+  |\chi_3|^{1-2(a_1+a_5)} \CC{4}{2a_2;2\tilde{a}_3,2\tilde{a}_5}{2(a_2+a_4),2(\tilde{a}_1+\tilde{a}_5)}{\chi_1,-\chi_2\chi_3,\chi_3} \nonumber\\
	&+  |\chi_1|^{1-2(a_2+a_4)} \CC{4}{2a_1;2\tilde{a}_6,2\tilde{a}_4}{2(a_1+a_5),2(\tilde{a}_2+\tilde{a}_4)}{\chi_3,-\chi_1\chi_2,\chi_1}\bigg]\,,
\end{align}
where, as usual, the definition of the series is relegated to \Appref{app:defs}.


\subsection{Conformal Double Box in 2D}

In two dimensions, we define the conformal double box with spinning propagators (arrows) according to
\begin{equation}
	I_{3,3}^{2\text{D,conf}}
	=\includegraphicsbox{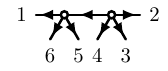}
		= \int \frac{\pi^{-2} \dd^2 x_0\, \dd^2 x_{0'}}{[x_{10}]^{\ba_1} [x_{60}]^{\ba_6} [x_{50}]^{\ba_5} [x_{00'}]^{\mathbf{b}} [x_{40'}]^{\ba_4} [x_{30'}]^{\ba_3} [x_{20'}]^{\ba_2}}\, ,
\end{equation}
with the conformal constraints
\begin{align}
	\ba_1 + \ba_5 + \ba_6 &= \ba_2 + \ba_3 + \ba_4 = 2-\mathbf{b}\, ,\nonumber\\
	\bba_1 + \bba_5 + \bba_6 &= \bba_2 + \bba_3 + \bba_4 = 2-\mathbf{\bar{b}}\, .
\end{align}
Following the exact same steps as for the one-dimensional case, we arrive at the following spectral representation:
\begin{align}
	I_{3,3}^{2\text{D,conf}} &= A^{\mathrm{2D}}_{\ell_b}(b) A^{\mathrm{2D}}_{\ell_1}(a_1) A^{\mathrm{2D}}_{\ell_2}(a_2) \frac{[x_{34}]^{\ba_2+\mathbf{b}-1} [x_{53}]^{\ba_2+\ba_4-1} [x_{46}]^{\ba_1+\ba_5-1} [x_{56}]^{\ba_1+\mathbf{b}-1}}{[x_{16}]^{\ba_1} [x_{23}]^{\ba_2} [x_{54}]^{\ba_1+\ba_2+\ba_4+\ba_5 + \mathbf{b}-2}} (-1)^{\ell_1+\ell_5} \nonumber\\
	&\times \prod_{j=1}^3 \sum_{\ell'_j \in \mathbb{Z}} \int_{\mathbb{R}+\ii \eta} \frac{\dd u_j}{2 \pi} [\chi_j]^{\bu_j} (-1)^{\ell'_2} \frac{A^{\mathrm{2D}}_{\ell_2+\ell'_1}(1-a_2-\ii u_1) A^{\mathrm{2D}}_{\ell_1+\ell'_3}(1-a_1-\ii u_3)}{A^{\mathrm{2D}}_{\ell'_1}(-\ii u_1) A^{\mathrm{2D}}_{\ell'_2}(-\ii u_2) A^{\mathrm{2D}}_{\ell'_3}(-\ii u_3)}\nonumber\\
	&\times \frac{A^{\mathrm{2D}}_{\ell_2+\ell_4+\ell'_1-\ell'_2}(a_2+a_4+\ii u_{12}) A^{\mathrm{2D}}_{\ell_1+\ell_5+\ell'_3-\ell'_2}(a_1+a_5+\ii u_{32})}{A^{\mathrm{2D}}_{\ell_b+\sum_{i\neq 3,6}\ell_i +\ell'_1+\ell'_3-\ell'_2}(b+\sum_{i\neq 3,6} a_i - 2 + \ii(u_1+u_3-u_2))}\, ,
\end{align}
where the cross ratios are still given by \eqref{cross-ratios double box}. As expected following the discussion in the previous \Secref{sec:1to2D}, the two-dimensional integrand is obtained from the one-dimensional one through the simple replacements \eqref{1Dto2D}, up to a sign. Then, computing the integrals as sums over residues gives
\begin{align}
	&I_{3,3}^{2\text{D,conf}} = A^{\mathrm{2D}}_{\ell_b}(b) A^{\mathrm{2D}}_{\ell_1}(a_1) A^{\mathrm{2D}}_{\ell_2}(a_2) \frac{[x_{34}]^{\ba_2+\mathbf{b}-1} [x_{53}]^{\ba_2+\ba_4-1} [x_{46}]^{\ba_1+\ba_5-1} [x_{56}]^{\ba_1+\mathbf{b}-1}}{[x_{16}]^{\ba_1} [x_{23}]^{\ba_2} [x_{54}]^{\ba_1+\ba_2+\ba_4+\ba_5 + \mathbf{b}-2}} (-1)^{\ell_1+\ell_5} \nonumber\\
	&\times\bigg[\CCtb{1}{\bba_2,\bba_1;\mathbf{\bar{b}}+\sum_{i\neq 3,6}\bba_i - 2}{\bba_2+\bba_4,\bba_1+\bba_5}{\bar\chi_1,-\bar\chi_2,\bar\chi_3} \CCt{1}{\ba_2,\ba_1;\mathbf{b}+\sum_{i\neq 3,6}\ba_i - 2}{\ba_2+\ba_4,\ba_1+\ba_5}{\chi_1,-\chi_2,\chi_3} \nonumber\\
	&+ [\chi_1]^{1-\ba_2-\ba_4} [\chi_3]^{1-\ba_1-\ba_5} \CCtb{2}{1-\bba_4,1-\bba_5;\mathbf{\bar{b}}}{2-\bba_2-\bba_4,2-\bba_1-\bba_5}{\bar\chi_1,-\bar\chi_1\bar\chi_2\bar\chi_3,\bar\chi_3}\nonumber\\
	&\hspace*{8cm}\times\CCt{2}{1-\ba_4,1-\ba_5;\mathbf{b}}{2-\ba_2-\ba_4,2-\ba_1-\ba_5}{\chi_1,-\chi_1\chi_2\chi_3,\chi_3} \nonumber\\
	&+ (-1)^{\ell_b}[\chi_2]^{\mathbf{b}+\sum_{i\neq 3,6}\ba_i-1} \CCtb{3}{\bba_2,\bba_1;1-\bba_3,1-\bba_6}{\mathbf{\bar{b}}+\sum_{i\neq 3,6}\bba_i-1}{\bar\chi_1\bar\chi_2,\bar\chi_2,\bar\chi_2\bar\chi_3}\nonumber\\
	&\hspace*{8cm}\times\CCt{3}{\ba_2,\ba_1;1-\ba_3,1-\ba_6}{\mathbf{b}+\sum_{i\neq 3,6}\ba_i-1}{\chi_1\chi_2,\chi_2,\chi_2\chi_3} \nonumber\\
	&+  [\chi_3]^{1-\ba_1-\ba_5} \CCtb{4}{\bba_2;1-\bba_3,1-\bba_5}{\bba_2+\bba_4,2-\bba_1-\bba_5}{\bar\chi_1,-\bar\chi_2\bar\chi_3,\bar\chi_3} \CCt{4}{\ba_2;1-\ba_3,1-\ba_5}{\ba_2+\ba_4,2-\ba_1-\ba_5}{\chi_1,-\chi_2\chi_3,\chi_3} \nonumber\\
	&+  [\chi_1]^{1-\ba_2-\ba_4} \CCtb{4}{\bba_1;1-\bba_6,1-\bba_4}{\bba_1+\bba_5,2-\bba_2-\bba_4}{\bar\chi_3,-\bar\chi_1\bar\chi_2,\bar\chi_1} \CCt{4}{\ba_1;1-\ba_6,1-\ba_4}{\ba_1+\ba_5,2-\ba_2-\ba_4}{\chi_3,-\chi_1\chi_2,\chi_1}\bigg]\,.
\end{align}
It is clear that when all the propagators are scalar, i.e.\ $\ell_i = \ell_b = 0$ or $\ba_i=\bba_i$, $\mathbf{b}=\mathbf{\bar b}$, this result could have been obtained directly from the one-dimensional result performing the simple replacements detailed in \secref{sec:1to2D}. For the spinning case, we note in particular the $(-1)^{\ell_b}$ in front of the third term, which we could not predict with our general arguments given in \Secref{sec:genints2D}.


\section{Outlook}
\label{sec:outlook}

In this paper we have demonstrated that track Feynman integrals in one and two dimensions are fully fixed by sets of differential operators associated with the above $\levo{P}$-symmetries. Similar to Gelfand--Kapranov--Zelevinsky (GKZ) or Aomoto--Gelfand (AG) hypergeometric functions, we can thus understand these spacetime differential operators as defining the respective systems of Feynman integrals. This proof of concept suggests to further investigate the constraining power of the $\levo{P}$-symmetries for practical purposes. 

While the present paper focuses on Feynman integrals in one and two dimensions, the $\levo{P}$-symmetries are present for diagrams in any spacetime dimension, see \cite{Loebbert:2024qbw}. It is thus natural to employ the refined insights on the above bootstrap in order to tackle further examples of higher dimensional integrals, see \cite{Loebbert:2019vcj,Loebbert:2020glj,Corcoran:2020epz,Loebbert:2020aos} for similar investigations. Along these lines it would be interesting to better understand the role of the dimensional recursions for track integrals studied in \cite{Loebbert:2024fsj}, which in particular connect two-dimensional results to four dimensions.

While most planar graph topologies relevant for real-world physics form part of the track diagrams considered in this paper, conceptually it would be interesting to further explore the impact of the $\levo{P}$-symmetries for more general (position-space) trees as well as for Feynman diagrams including loops (i.e.\ loops of loops in the dual momentum space). Given certain constraints on the propagator powers, at least the full $\levo{P}$-symmetry is present for such loop graphs \cite{Loebbert:2025abz}; the question for the existence of partial $\levo{P}$-symmetries \`a la \cite{Loebbert:2024qbw} remains another interesting open problem.

Notably, the above nonlocal $\levo{P}$-symmetries only annihilate Feynman integrals with non-coinciding external points, thus excluding for instance the cases of conformal ladder or Basso--Dixon graphs that lead to inhomogeneous versions of $\levo{P}$ Ward identities \cite{Corcoran:2021gda}. It would be interesting to identify a spacetime formulation of (potentially higher order) annihilating differential operators that can be used to bootstrap examples of the latter graphs efficiently. Here the connections to partition functions \cite{Karydas:2023ufs,Karydas:2025tfs} or Toda-like models \cite{Loebbert:2024fsj} are very inspirational. We stress that the method of separation of variables, which extends the spectral transform used in the present paper, has been successfully applied to graphs with coinciding points \cite{Basso:2017jwq,Derkachov:2018rot,Derkachov:2019tzo,Derkachov:2021ufp,Alfimov:2023vev} (cf.\ the example of comb-channel conformal partial waves in \Secref{sec:comb}). While the spectral transform turned out to be particularly efficient in one and two dimensions, it would be interesting to understand how far higher dimensional track integrals can be computed in a similar fashion (cf.\ the example of the three-point integral in \Appref{app:higherd}).

Notably, the transition from 1D to 2D Feynman integrals outlined in \secref{sec:1to2D} shows that the intersection matrix, which essentially combines two copies of $\levo{P}$ invariants in one dimension into the two-dimensional integral, is always diagonal in our hypergeometric solution basis. Moreover, we have normalized the solution basis of the $\levo{P}$ equations in such a way that all relative coefficients in the linear combinations representing the 1D integrals are trivial. Capturing these features from the perspective of symmetries and intersection theory is interesting.

The $\levo{P}$-bootstrap employed here resembles the construction of conformal blocks via conformal Casimir operators. It would be interesting to further explore the connections between both approaches starting e.g.\ with the case of conformal blocks in one and two dimensions as computed in \cite{Rosenhaus:2018zqn,Fortin:2020zxw,Fortin:2023xqq}. In particular, the role of $\levo{P}$ for Yangian symmetry and integrability should be further compared to the approach via integrable Calogero--Sutherland and Gaudin models investigated in \cite{Isachenkov:2016gim,Buric:2020dyz,Eberhardt:2020ewh,Buric:2021ywo,Buric:2021ttm,Buric:2021kgy}.

As stressed above, the choice of variables defining the hypergeometric systems in the paper at hand is crucial for identifying an optimal form of the result. It would be interesting to better understand this choice from a mathematical perspective, which might imply a hypergeometric version of the points of maximal unipotent monodromy (MUM) on Calabi--Yau (CY) geometries.  Identification of these points is crucial for controlling Feynman integrals with CY structure.
Here relations to the recently investigated hypergeometric structures of Feynman integrals, see e.g.\ \cite{Pal:2021llg}, \cite{S:2024zqp} and \cite{Alkalaev:2025fgn,Alkalaev:2025zhg} might be instructive.

Notably, the $\levo{P}$-symmetries also extend to Feynman graphs with massive propagators \cite{Loebbert:2020hxk,Loebbert:2020glj,Loebbert:2024qbw,Loebbert:2025abz}. Bootstrapping such graphs in one and two dimensions is an obvious follow-up question. This might also lead to the identification of a massive version of the separation of variables method which so far has not been worked out.

Finally, Witten diagrams in curved space share many similarities with Feynman integrals. In particular, it was argued that contact Witten diagrams have the same Yangian $\levo{P}$-symmetries as one-loop Feynman integrals \cite{Rigatos:2022eos}. It would be interesting to understand how these symmetries extend to more general classes of Witten diagrams and can be used to bootstrap these in terms of hypergeometric functions along the lines of the present paper, cf.\ e.g.\ the recent work \cite{Herderschee:2025znl}.

\subsection*{Acknowledgments}

We are grateful to Janis D\"ucker, Albrecht Klemm and Julian Piribauer for helpful discussions. We are furthermore grateful to Souvik Bera and Tanay Pathak for pointing us to \cite{srivastavaBook} and spotting typos in \eqref{eq:G2F1Identity} in an earlier version of the preprint. Also we would like to thank Claude Duhr for pointing us to \cite{brownUVars}. The work of SFS is funded by the European Union (ERC Consolidator Grant LoCoMotive 101043686).
Views and opinions expressed are however those of the author(s) only and do not necessarily reflect
those of the European Union or the European Research Council. Neither the European Union nor the
granting authority can be held responsible for them. The work of GF and FL is funded by the Deutsche Forschungsgemeinschaft (DFG, German Research Foundation) -- Projektnummer  508889767.

\begin{appendix}

\section{Spectral Transform in Higher Dimension}
\label{app:higherd}

We first give the spectral transform in one dimension for the ``spinning'' propagator
\begin{equation}\label{SoV prop sign}
	\frac{\sgn^{\ep}(x_{12})}{|x_{12}|^{2 a}} = \sum_{\ep' = 0}^1\! \int_{\mathbb{R} + \ii \eta} \! \frac{\sgn^{\ep}(x_{34}) \sgn^{\ep+\ep'}(x_{13} x_{24}) \sgn^{\ep'}(x_{14} x_{23})}{(|x_{13}| |x_{24}|)^{2(a+\ii u)} (|x_{14}| |x_{23}|)^{-2\ii u}} \frac{|x_{34}|^{2a} (-1)^{\ep \ep'} A_{\ep}(a)}{A_{\ep'}(-\ii u)\, A_{[\ep+\ep']}(a+\ii u)} \frac{\dd u}{2\sqrt{\pi}}\, ,
\end{equation}
where we use the notation $[\ep+\ep']$ defined in \eqref{kappasum}.

The two-dimensional analogue was given in \Secref{sec:1to2D}, and in higher dimension we have
\begin{multline}\label{SoV prop d}
	\frac{1}{|x_{12}|^{2a}} = \Gamma\!\left(\frac{D-2}{2}\right) \sum_{\ell = 0}^{+\infty} \left(\frac{D-2}{2}+ \ell\right) C_\ell^{\left(\frac{D-2}{2}\right)}(\cos\theta)\\
	\times \int_{\mathbb{R} + \ii \eta} \! \frac{|x_{34}|^{2a}}{(|x_{13}| |x_{24}|)^{2(a+\ii u)} (|x_{14}| |x_{23}|)^{-2\ii u}} \frac{A^{(D)}_0(a)}{A^{(D)}_\ell(-\ii u)\, A^{(D)}_\ell(a+\ii u)} \frac{\dd u}{2 \pi}\, ,
\end{multline}
where we use the function
\begin{equation}
	A^{(D)}_\ell (u) = \frac{\Gamma\!\left(\frac{D+\ell}{2}-u\right)}{\Gamma\!\left(\frac{\ell}{2}+u\right)}\, ,
\end{equation}
the $C_\ell^{(\lambda)}$ are Gegenbauer polynomials, and
\begin{equation}
	\cos\theta = x^2_{34} \frac{|x_{13}| |x_{23}|}{|x_{14}| |x_{24}|} \left(\frac{x_{13}}{x^2_{13}} - \frac{x_{43}}{x^2_{43}}\right)\cdot\left(\frac{x_{23}}{x^2_{23}} - \frac{x_{43}}{x^2_{43}}\right)\, .
\end{equation}
Notice that this is such that $\pm\theta$ is the phase of the complex number $\chi$ defined through
\begin{equation}
	\chi \bar\chi = \frac{x^2_{14} x^2_{23}}{x^2_{13} x^2_{24}}\quad \text{and}\quad (1-\chi)(1-\bar\chi) = \frac{x^2_{12} x^2_{34}}{x^2_{13} x^2_{24}}\, .
\end{equation}

The proof of formula \eqref{SoV prop d} reads as follows. We first move some propagators from the right- to the left-hand side, so that the equation becomes
\begin{equation}
	\frac{1}{|1-\chi|^{2 a}} = \sum_{\ell = 0}^{+\infty} \left(\frac{D-2}{2}+ \ell\right) C_\ell^{\left(\frac{D-2}{2}\right)}(\cos\theta) \int_{\mathbb{R} + \ii \eta} \! |\chi|^{2\ii u} \frac{\Gamma\!\left(\frac{D-2}{2}\right)  A^{(D)}_0(a)}{A^{(D)}_\ell(-\ii u)\, A^{(D)}_\ell(a+\ii u)} \frac{\dd u}{2 \pi}\, .
\end{equation}
We then compute the integral as a sum over residues. Assuming, for instance, that $|\chi|<1$ we have to pick the residues of the simple poles at $u\in -\ii\mathbb{N}$ and we get
\begin{equation}
	\frac{1}{|1-\chi|^{2 a}} = \sum_{\ell,m = 0}^{+\infty} |\chi|^{\ell+2m} \left(\frac{D-2}{2}+ \ell\right) \frac{(a)_{\ell+m} (a+\frac{2-D}{2})_{m}}{m! \left(\frac{D-2}{2}\right)_{\ell+m+1}} C_\ell^{\left(\frac{D-2}{2}\right)}(\cos\theta)\, .
\end{equation}
We then perform the change of summation indices from $(\ell,m)$ to $(n,p)=(\ell+2m,m)$ and compute the sum over $p$ using a summation formula for the Gegenbauer polynomials to obtain
\begin{equation}
	\frac{1}{|1-\chi|^{2 a}} = \sum_{n = 0}^{+\infty} |\chi|^{n} C_n^{(a)}(\cos\theta)\,.
\end{equation}
Since $|1-\chi|^2 = 1 - 2|\chi|\cos\theta  + |\chi|^2$, this last equation is nothing else than the generating function for the Gegenbauer polynomials, and \eqref{SoV prop d} is thus verified.

\paragraph{Three-Point Integral in Higher Dimensions.}

The spectral representation of the three-point star integral reads
\begin{multline}
	I_3^{(D)} = \frac{A^{(D)}_0(a_2)\, A^{(D)}_0(a_3)}{|x_{12}|^{2(a_1+a_2+a_3)-d}} \Gamma\!\left(\frac{D-2}{2}\right) \sum_{\ell = 0}^{+\infty} \left(\frac{D-2}{2}+ \ell\right) C_\ell^{\left(\frac{D-2}{2}\right)}(\cos\theta)\\
	\times \int_{\mathbb{R} + \ii \eta} \! r^{2\ii u} \frac{A^{(D)}_\ell(a_1+a_3+\ii u)\, A^{(D)}_\ell(D-\sum_i a_i-\ii u)}{A^{(D)}_\ell(-\ii u)\, A^{(D)}_\ell(a_3+\ii u)} \frac{\dd u}{2 \pi}\, ,
\end{multline}
for $r = |x_{13}|/|x_{12}|$ and $\cos\theta = x_{12}\cdot x_{13}/|x_{13}| |x_{12}|$. Assuming that $r<1$ and computing the integral gives
\begin{equation}
	I_3^{(D)} = \frac{A^{(D)}_0(a_2)\, A^{(D)}_0(a_3)}{|x_{12}|^{2(a_1+a_2+a_3)-d}} \bigg[ \mathcal{F}^{(D)}\!\left[\substack{a_3,\sum_i a_i - D/2 \\ a_1+a_3};r,\cos\theta\right] + r^{D-2(a_1+a_3)} \mathcal{F}^{(D)}\!\left[\substack{a_2,D/2-a_1 \\ D-a_1-a_3};r,\cos\theta\right]\!\bigg]\, ,
\end{equation}
where
\begin{multline}
	\mathcal{F}^{(D)}\!\left[\substack{a,b \\ c};r,t\right] = \frac{A^{(D)}_0(c)}{A^{(D)}_0(a)\, A^{(D)}_0(b)} \sum_{\ell,m=0}^{+\infty} \frac{(a)_{\ell+m}\, (a+(2-D)/2)_{m}\, (b)_{\ell+m}\, (b+(2-D)/2)_{m}}{(D/2)_{\ell+m}\, m!\, (c)_{\ell+m}\, (c+(2-D)/2)_{m}} \\
	\times \frac{2\ell+D-2}{D-2} r^{\ell+2m} C_\ell^{\left(\!\frac{D-2}{2}\!\right)}\!(t)\, .
\end{multline}

\section{Useful Series and Relations}
\label{app:defs}

In this appendix we will define the various hypergeometric series used in the main text. Note that we will be using non-caligraphic letters to denote the series with constant term being $1$, while we use caligraphic letters for conveniently rescaled versions. For the non-standard series that we define here we will only be using the rescaled/caligraphic versions. 

\paragraph{Standard Series.}

The Appell function $F_1$ is defined for $|x|<1$ and $|y|<1$ by
\begin{equation}
\label{eq:defF1}
	\FF{1}{a;b_1,b_2}{c}{x,y}  = \sum_{m,n=0}^{+\infty} \frac{(a)_{m+n} (b_1)_{m} (b_2)_{n}}{(c)_{m+n}} \frac{x^m y^n}{m! n!}\, ,
\end{equation}
and the Appell function $F_2$ is defined for $|x|+|y|<1$ by
\begin{equation}
	\FF{2}{a;b_1,b_2}{c_1,c_2}{x,y} = \sum_{m,n=0}^{+\infty} \frac{(a)_{m+n} (b_1)_{m} (b_2)_{n}}{(c_1)_m (c_2)_n} \frac{x^m y^n}{m! n!}\, .
\end{equation}
The Horn function $G_2$ is defined for $|x|<1$ and $|y|<1$ by
\begin{equation}
	\GH{2}{a_1,a_2}{b_1, b_2}{x,y} = \sum_{m,n=0}^{+\infty} (a_1)_m (a_2)_n (b_1)_{n-m} (b_2)_{m-n} \frac{x^m y^n}{m! n!}\, .
\end{equation}
The family of Lauricella $F_D$ functions is defined for $|x_1|,\dots, |x_n|<1$ by
\begin{equation}
	\FL{D}{n}{a,b_1,\dots b_n}{c}{x_1,\dots ,x_n}=\sum_{m_1,\dots m_n=0}^{\infty}\frac{(a)_{m_1+\dots m_n}(b_1)_{m_1}\dots (b_n)_{m_n}}{(c)_{m_1+\dots m_n}}\prod_{j=1}^n\frac{x_j^{m_j}}{m_j!} \,.
\end{equation}
For $|x|>|y|>1$ and $x,y$ not in $\mathbb{R}^+$, the (analytic continuation of the) Appell $F_1$ function satisfies \cite{olsson1964integration}
\begin{align}\label{G2 to F1}
	\FF{1}{a,b_1,b_2}{c}{x,y} &= (-x)^{-b_1}(-y)^{-b_2} \frac{\Gamma(c) \Gamma(a-b_1-b_2)}{\Gamma(a) \Gamma(c-b_1-b_2)} \FF{1}{b_1+b_2-c+1,b_1,b_2}{-a+b_1+b_2+1}{\frac{1}{x},\frac{1}{y}}\nonumber\\
	&+ (-x)^{-b_1} (-y)^{b_1-a} \frac{\Gamma(c) \Gamma(a-b_1) \Gamma(b_1+b_2-a)}{\Gamma(a) \Gamma(b_2) \Gamma(c-a)} \GH{2}{b_1,a-c+1}{a-b_1, b_1+b_2-a}{-\frac{y}{x},-\frac{1}{y}} \nonumber\\
	&+ (-x)^{-a} \frac{\Gamma(c) \Gamma(b_1-a)}{\Gamma(b_1) \Gamma(c-a)} \FF{1}{a,a-c+1,b_2}{a-b_1+1}{\frac{1}{x},\frac{y}{x}}\, .
\end{align}

\paragraph{Triangle-Box Integral.}
\begin{align}
&\HH{1}{a_1,a_2;b}{c_1,c_2;d}{x_1,x_2,x_3} \\
&\quad= \frac{A_0(\sfrac{c_2}{2})A_0(\sfrac{d}{2})}{A_0(\sfrac{a_1}{2})A_0(\sfrac{a_2}{2})A_0(\sfrac{b}{2})A_0(\sfrac{c_1}{2})}\sum_{m_1,m_2,m_3=0}^{\infty} (a_1)_{m_1} (a_2)_{m_3}\frac{(b)_{m_2+m_3} (c_1)_{m_1-m_2}}{(c_2)_{m_1-m_2} (d)_{m_3}} \prod_{i=1}^3 \frac{x_i^{m_i}}{m_i!}\, ,  \nonumber\\
	&\HH{2}{a_1,a_2;b;c}{d;e}{x_1,x_2,x_3}  \\
	&\quad =\frac{A_0(\sfrac{d}{2})A_0(\sfrac{e}{2})}{A_0(\sfrac{a_1}{2})A_0(\sfrac{a_2}{2})A_0(\sfrac{b}{2})A_0(\sfrac{c}{2})}  \sum_{m_1,m_2,m_3=0}^{\infty}  (a_1)_{m_1} (a_2)_{m_3} \frac{(b)_{m_1+m_2} (c)_{m_2+m_3}}{(d)_{m_1+m_2} (e)_{m_3}} \prod_{i=1}^3 \frac{x_i^{m_i}}{m_i!}\, , \nonumber\\
&\HH{3}{a_1,a_2,a_3;b}{c; d}{x_1,x_2,x_3}  \\
&\quad =\frac{A_0(\sfrac{c}{2})A_0(\sfrac{d}{2})}{A_0(\sfrac{a_1}{2})A_0(\sfrac{a_2}{2})A_0(\sfrac{a_3}{2})A_0(\sfrac{b}{2})} \sum_{m_1,m_2,m_3=0}^{\infty}  \frac{\prod_{i=1}^3(a_i)_{m_i} (b)_{m_1+m_2+m_3}}{(c)_{m_1+m_2} (d)_{m_3}} \prod_{i=1}^3 \frac{x_i^{m_i}}{m_i!}\, , \nonumber
\end{align}

\paragraph{Double-Box Integral.}
\begin{align}
	&\BB{1}{a_1,a_2;b}{c_1,c_2,c_3,c_4}{x_1,x_2,x_3,x_4}  = \frac{A_0(\sfrac{c_2}{2})A_0(\sfrac{c_4}{2})}{A_0(\sfrac{a_1}{2})A_0(\sfrac{a_2}{2})A_0(\sfrac{b}{2})A_0(\sfrac{c_1}{2})A_0(\sfrac{c_3}{2})} \\
	&\quad \times\sum_{m_1,m_2,m_3,m_4=0}^{\infty} (a_1)_{m_1} (a_2)_{m_3}\frac{(b)_{m_2+m_4} (c_1)_{m_1-m_2} (c_3)_{m_3-m_4}}{(c_2)_{m_1-m_2} (c_4)_{m_3-m_4}} \prod_{i=1}^4 \frac{x_i^{m_i}}{m_i!}\, , \nonumber \\
	&\BB{2}{a_1,a_2;b;c}{d_1,d_2,d_3}{x_1,x_2,x_3,x_4} = \frac{A_0(\sfrac{d_1}{2})A_0(\sfrac{d_3}{2})}{A_0(\sfrac{a_1}{2})A_0(\sfrac{a_2}{2})A_0(\sfrac{b}{2})A_0(\sfrac{c}	{2})A_0(\sfrac{d_2}{2})} \\
	&\quad\times \sum_{m_1,m_2,m_3,m_4=0}^{\infty} (a_1)_{m_1} (a_2)_{m_3}\frac{(b)_{m_2+m_4} (c)_{m_1+m_2} (d_2)_{m_3-m_4}}{(d_1)_{m_1+m_2} (d_3)_{m_3-m_4}} \prod_{i=1}^4 \frac{x_i^{m_i}}{m_i!}\, , \nonumber \\
	&\BB{3}{a_1,a_2,a_3;b}{c_1,c_2;d}{x_1,x_2,x_3,x_4}  = \frac{A_0(\sfrac{c_2}{2})A_0(\sfrac{d}{2})}{A_0(\sfrac{a_1}{2})A_0(\sfrac{a_2}{2})A_0(\sfrac{a_3}{2})A_0(\sfrac{b}{2})A_0(\sfrac{c_1}{2})} \\
	&\quad\times \sum_{m_1,m_2,m_3,m_4=0}^{\infty} \prod_{i=1}^3(a_i)_{m_i}\frac{(b)_{m_1+m_2+m_4} (c_1)_{m_3-m_4}}{(c_2)_{m_3-m_4} (d)_{m_1+m_2}} \prod_{i=1}^4 \frac{x_i^{m_i}}{m_i!}\, , \nonumber  \\
	&\BB{4}{a_1,a_2;b;c_1,c_2}{d_1,d_2}{x_1,x_2,x_3,x_4}  = \frac{A_0(\sfrac{d_1}{2})A_0(\sfrac{d_2}{2})}{A_0(\sfrac{a_1}{2})A_0(\sfrac{a_2}{2})A_0(\sfrac{b}{2})A_0(\sfrac{c_1}{2})A_0(\sfrac{c_2}{2})} \\
	&\quad\times \sum_{m_1,m_2,m_3,m_4=0}^{\infty} (a_1)_{m_1} (a_2)_{m_3}\frac{(b)_{m_2+m_4} (c_1)_{m_1+m_2} (c_2)_{m_3+m_4}}{(d_1)_{m_1+m_2} (d_2)_{m_3+m_4}} \prod_{i=1}^4 \frac{x_i^{m_i}}{m_i!}\, , \nonumber \\
	&\BB{5}{a_1,a_2,a_3;b;c}{d_1,d_2}{x_1,x_2,x_3,x_4}  =\frac{A_0(\sfrac{d_1}{2})A_0(\sfrac{d_2}{2})}{A_0(\sfrac{a_1}{2})A_0(\sfrac{a_2}{2})A_0(\sfrac{a_3}{2})A_0(\sfrac{b}{2})A_0(\sfrac{c}{2})} \\
	&\quad\times\sum_{m_1,m_2,m_3,m_4=0}^{\infty} \prod_{i=1}^3(a_i)_{m_i} \frac{(b)_{m_1+m_2+m_4} (c)_{m_3+m_4}}{(d_1)_{m_1+m_2} (d_2)_{m_3+m_4}} \prod_{i=1}^4 \frac{x_i^{m_i}}{m_i!}\, , \nonumber \\
	&\BB{6}{a_1,a_2,a_3,a_4;b}{c_1, c_2}{x_1,x_2,x_3,x_4} =\frac{A_0(\sfrac{c_1}{2})A_0(\sfrac{c_2}{2})}{A_0(\sfrac{a_1}{2})A_0(\sfrac{a_2}{2})A_0(\sfrac{a_3}{2})A_0(\sfrac{a_4}	{2})A_0(\sfrac{b}{2})} \\
	&\quad \times \sum_{m_1,m_2,m_3,m_4=0}^{\infty} \frac{\prod_{i=1}^4(a_i)_{m_i} (b)_{m_1+m_2+m_3+m_4}}{(c_1)_{m_1+m_2} (c_2)_{m_3+m_4}} 		\prod_{i=1}^4 \frac{x_i^{m_i}}{m_i!}\, . \nonumber  
\end{align}

\paragraph{Triangle-Pentagon Integral.}
\begin{align}
	&\HH{4}{a_1,a_2,a_3;b}{c_1,c_2;d}{x_1,x_2,x_3,x_4}=
	\frac{A_0(\sfrac{c_2}{2})A_0(\sfrac{d}{2})}{\prod_{i=1}^3A_0(\sfrac{a_i}{2})A_0(\sfrac{b}{2})A_0(\sfrac{c_1}{2})} \\
	&\quad\times\sum_{m_1,m_2,m_3,m_4=0}^{\infty} \frac{(a_1)_{m_1}(a_2)_{m_2}(a_3)_{m_4}(b)_{m_3+m_4}(c_1)_{m_1+m_2-m_3}}{(c_2)_{m_1+m_2-m_3}(d)_{m_4}}\prod_{i=1}^4\frac{x_i^{m_i}}{m_i!} \,,\notag \\
	&\HH{5}{a_1,a_2,a_3;b}{c_1,c_2;d}{x_1,x_2,x_3,x_4}=
	\frac{A_0(\sfrac{c_2}{2})A_0(\sfrac{d}{2})}{\prod_{i=1}^3A_0(\sfrac{a_i}{2})A_0(\sfrac{b}{2})A_0(\sfrac{c_1}{2})} \\
	&\quad\times\sum_{m_1,m_2,m_3,m_4=0}^{\infty}\frac{(a_1)_{m_1}(a_2)_{m_2}(a_3)_{m_4}(b)_{m_3+m_4}(c_1)_{m_1-m_2-m_3}}{(c_2)_{m_1-m_2-m_3}(d)_{m_4}}\prod_{i=1}^4\frac{x_i^{m_i}}{m_i!} \,,\notag \\
	&\HH{6}{a_1,a_2,a_3,a_4;b}{c;d}{x_1,x_2,x_3,x_4}=
	\frac{A_0(\sfrac{c}{2})A_0(\sfrac{d}{2})}{\prod_{i=1}^4A_0(\sfrac{a_i}{2})A_0(\sfrac{b}{2})} \\
	&\quad\times \sum_{m_1,m_2,m_3,m_4=0}^{\infty}\frac{\prod_{i=1}^4(a_i)_{m_i}(b)_{m_1+m_2+m_3+m_4}}{(c)_{m_1+m_2+m_3}(d)_{m_4}}\prod_{i=1}^4\frac{x_i^{m_i}}{m_i!} \,, \notag\\
	&\HH{7}{a_1,a_2,a_3;b}{c_1,c_2;d}{x_1,x_2,x_3,x_4}=
	\frac{A_0(\sfrac{c_2}{2})A_0(\sfrac{d}{2})}{\prod_{i=1}^3A_0(\sfrac{a_i}{2})A_0(\sfrac{b}{2})A_0(\sfrac{c_1}{2})} \\
	&\quad\times\sum_{m_1,m_2,m_3,m_4=0}^{\infty}\frac{(a_1)_{m_1}(a_2)_{m_2}(a_3)_{m_4}(b)_{m_3+m_4}(c_1)_{m_1+m_2+m_3}}{(c_2)_{m_1+m_2+m_3}(d)_{m_4}}\prod_{i=1}^4\frac{x_i^{m_i}}{m_i!} \,, \notag
\end{align}

\paragraph{Triangle-Triangle-Box Integral.}
\begin{align}
	&\HH{8}{a_1,a_2,a_3,a_4}{b_1,b_2;c;d}{x_1,x_2,x_3,x_4}=
	\frac{A_0(\sfrac{b_2}{2})A_0(\sfrac{c}{2})A_0(\sfrac{d}{2})}{A_0(\sfrac{a_1}{2})A_0(\sfrac{a_2}{2})A_0(\sfrac{a_3}{2})A_0(\sfrac{a_4}{2})A_0(\sfrac{b_1}{2})} \\
	&\quad\times\sum_{m_1,m_2,m_3,m_4=0}^{\infty} \frac{(a_1)_{m_1}(a_2)_{m_1}(a_3)_{m_2}(a_4)_{m_4}(b_1)_{m_3-m_4}}{(b_2)_{m_3-m_4}(c)_{m_1-m_2}(d)_{m_2-m_3}} \prod_{i=1}^4 \frac{x_i^{m_i}}{m_i!} \,, \nonumber  \\
	&\HH{9}{a_1,a_2;b_1,b_2,b_3}{c_1,c_2;d}{x_1,x_2,x_3,x_4}=
	\frac{A_0(\sfrac{b_3}{2})A_0(\sfrac{c_2}{2})A_0(\sfrac{d}{2})}{A_0(\sfrac{a_1}{2})A_0(\sfrac{a_2}{2})A_0(\sfrac{b_1}{2})A_0(\sfrac{b_2}{2})A_0(\sfrac{c_1}{2})} \\
	&\quad\times\sum_{m_1,m_2,m_3,m_4=0}^{\infty}\frac{(a_1)_{m_2}(a_2)_{m_4}(b_1)_{m_1+m_2}(b_2)_{m_1+m_2}(c_1)_{m_3-m_4}}{(b_3)_{m_1+m_2}(c_2)_{m_3-m_4}(d)_{m_2-m_3}}\prod_{i=1}^4 \frac{x_i^{m_i}}{m_i!} \,, \nonumber  \\
	&\HH{10}{a_1,a_2,a_3;b_1,b_2}{c_1,c_2;d}{x_1,x_2,x_3,x_4}=
	\frac{A_0(\sfrac{b_2}{2})A_0(\sfrac{c_2}{2})A_0(\sfrac{d}{2})}{A_0(\sfrac{a_1}{2})A_0(\sfrac{a_2}{2})A_0(\sfrac{a_3}{2})A_0(\sfrac{b_1}{2})A_0(\sfrac{c_1}{2})} \\
	&\quad\times\sum_{m_1,m_2,m_3,m_4=0}^{\infty}\frac{(a_1)_{m_1}(a_2)_{m_1}(a_3)_{m_4}(b_1)_{m_2+m_3}(c_1)_{m_3-m_4}}{(b_2)_{m_2+m_3}(c_2)_{m_3-m_4}(d)_{m_1-m_2-m_3}} \prod_{i=1}^4 \frac{x_i^{m_i}}{m_i!} \,, \nonumber  \\
	&\HH{11}{a_1,a_2,a_3,a_4,a_5}{b;c;d}{x_1,x_2,x_3,x_4}=
	\frac{A_0(\sfrac{b}{2})A_0(\sfrac{c}{2})A_0(\sfrac{d}{2})}{A_0(\sfrac{a_1}{2})A_0(\sfrac{a_2}{2})A_0(\sfrac{a_3}{2})A_0(\sfrac{a_4}{2})A_0(\sfrac{a_5}{2})} \\
	&\quad\times\sum_{m_1,m_2,m_3,m_4=0}^{\infty} \frac{(a_1)_{m_1}(a_2)_{m_1}(a_3)_{m_2}(a_4)_{m_3}(a_5)_{m_4}}{(b)_{m_1-m_2}(c)_{m_3+m_4}(d)_{m_2-m_3-m_4}} \prod_{i=1}^4 \frac{x_i^{m_i}}{m_i!} \,, \nonumber  \\
	&\HH{12}{a_1,a_2,a_3,a_4}{b_1,b_2;c;d}{x_1,x_2,x_3,x_4}=
	\frac{A_0(\sfrac{b_2}{2})A_0(\sfrac{c}{2})A_0(\sfrac{d}{2})}{A_0(\sfrac{a_1}{2})A_0(\sfrac{a_2}{2})A_0(\sfrac{a_3}{2})A_0(\sfrac{a_4}{2})A_0(\sfrac{b_1}{2})} \\
	&\quad\times\sum_{m_1,m_2,m_3,m_4=0}^{\infty} \frac{(a_1)_{m_1}(a_2)_{m_1}(a_3)_{m_2}(a_4)_{m_4}(b_1)_{m_3+m_4}}{(b_2)_{m_3+m_4}(c)_{m_1-m_2}(d)_{m_2-m_3}} \prod_{i=1}^4 \frac{x_i^{m_i}}{m_i!} \,, \nonumber  \\
	&\HH{13}{a_1,a_2,a_3;b_1,b_2}{c_1,c_2;d}{x_1,x_2,x_3,x_4}=
	\frac{A_0(\sfrac{b_2}{2})A_0(\sfrac{c_2}{2})A_0(\sfrac{d}{2})}{A_0(\sfrac{a_1}{2})A_0(\sfrac{a_2}{2})A_0(\sfrac{a_3}{2})A_0(\sfrac{b_1}{2})A_0(\sfrac{c_1}{2})} \\
	&\quad\times\sum_{m_1,m_2,m_3,m_4=0}^{\infty}\frac{(a_1)_{m_1}(a_2)_{m_1}(a_3)_{m_4}(b_1)_{m_2+m_3}(c_1)_{m_3+m_4}}{(b_2)_{m_2+m_3}(c_2)_{m_3+m_4}(d)_{m_1-m_2-m_3}} \prod_{i=1}^4 \frac{x_i^{m_i}}{m_i!} \,, \nonumber  \\
	&\HH{14}{a_1,a_2,a_3,a_4}{b_1,b_2;c;d}{x_1,x_2,x_3,x_4}=
	\frac{A_0(\sfrac{b_2}{2})A_0(\sfrac{c}{2})A_0(\sfrac{d}{2})}{A_0(\sfrac{a_1}{2})A_0(\sfrac{a_2}{2})A_0(\sfrac{a_3}{2})A_0(\sfrac{a_4}{2})A_0(\sfrac{b_1}{2})} \\
	&\quad\times\sum_{m_1,m_2,m_3,m_4=0}^{\infty} \frac{(a_1)_{m_1}(a_2)_{m_1}(a_3)_{m_3}(a_4)_{m_4}(b_1)_{m_2+m_3+m_4}}{(b_2)_{m_2+m_3+m_4}(c)_{m_3+m_4}(d)_{m_1-m_2-m_3-m_4}} \prod_{i=1}^4 \frac{x_i^{m_i}}{m_i!} \,, \nonumber  \\
	&\HH{15}{a_1,a_2;b_1,b_2,b_3}{c_1,c_2;d}{x_1,x_2,x_3,x_4}=
	\frac{A_0(\sfrac{b_3}{2})A_0(\sfrac{c_2}{2})A_0(\sfrac{d}{2})}{A_0(\sfrac{a_1}{2})A_0(\sfrac{a_2}{2})A_0(\sfrac{b_1}{2})A_0(\sfrac{b_2}{2})A_0(\sfrac{c_1}{2})} \\
	&\quad\times\sum_{m_1,m_2,m_3,m_4=0}^{\infty} \frac{(a_1)_{m_2}(a_2)_{m_4}(b_1)_{m_1+m_2}(b_2)_{m_1+m_2}(c_1)_{m_3+m_4}}{(b_3)_{m_1+m_2}(c_2)_{m_3+m_4}(d)_{m_2-m_3}}  \prod_{i=1}^4 \frac{x_i^{m_i}}{m_i!} \,, \nonumber  
\end{align}
\begin{align}
	&\HH{16}{a_1,a_2,a_3;b_1,b_2,b_3}{c;d}{x_1,x_2,x_3,x_4}=
	\frac{A_0(\sfrac{b_3}{2})A_0(\sfrac{c}{2})A_0(\sfrac{d}{2})}{A_0(\sfrac{a_1}{2})A_0(\sfrac{a_2}{2})A_0(\sfrac{a_3}{2})A_0(\sfrac{b_1}{2})A_0(\sfrac{b_2}{2})} \\
	&\quad\times\sum_{m_1,m_2,m_3,m_4=0}^{\infty} \frac{(a_1)_{m_2}(a_2)_{m_3}(a_3)_{m_4}(b_1)_{m_1+m_2}(b_2)_{m_1+m_2}}{(b_3)_{m_1+m_2}(c)_{m_3+m_4}(d)_{m_2-m_3-m_4}}  \prod_{i=1}^4 \frac{x_i^{m_i}}{m_i!} \,, \nonumber  \\
	&\HH{17}{a;b_1,b_2;c_1,c_2}{d_1,d_2,d_3}{x_1,x_2,x_3,x_4}=
	\frac{A_0(\sfrac{b_2}{2})A_0(\sfrac{c_2}{2})A_0(\sfrac{d_3}{2})}{A_0(\sfrac{a}{2})A_0(\sfrac{b_1}{2})A_0(\sfrac{c_1}{2})A_0(\sfrac{d_1}{2})A_0(\sfrac{d_2}{2})} \\
	&\quad\times\sum_{m_1,m_2,m_3,m_4=0}^{\infty} \frac{(a)_{m_4}(b_1)_{m_2+m_3}(c_1)_{m_3-m_4}(d_1)_{m_1+m_2+m_3}(d_2)_{m_1+m_2+m_3}}{(b_2)_{m_2+m_3}(c_2)_{m_3-m_4}(d_3)_{m_1+m_2+m_3}}  \prod_{i=1}^4 \frac{x_i^{m_i}}{m_i!} \,, \nonumber  \\
	&\HH{18}{a;b_1,b_2;c_1,c_2}{d_1,d_2,d_3}{x_1,x_2,x_3,x_4}=
	\frac{A_0(\sfrac{b_2}{2})A_0(\sfrac{c_2}{2})A_0(\sfrac{d_3}{2})}{A_0(\sfrac{a}{2})A_0(\sfrac{b_1}{2})A_0(\sfrac{c_1}{2})A_0(\sfrac{d_1}{2})A_0(\sfrac{d_2}{2})} \\
	&\quad\times\sum_{m_1,m_2,m_3,m_4=0}^{\infty} \frac{(a)_{m_4}(b_1)_{m_2+m_3}(c_1)_{m_3+m_4}(d_1)_{m_1+m_2+m_3}(d_2)_{m_1+m_2+m_3}}{(b_2)_{m_2+m_3}(c_2)_{m_3+m_4}(d_3)_{m_1+m_2+m_3}} \prod_{i=1}^4 \frac{x_i^{m_i}}{m_i!} \,, \nonumber  \\
	&\HH{19}{a_1,a_2;b_1,b_2}{c_1,c_2,c_3;d}{x_1,x_2,x_3,x_4}=
	\frac{A_0(\sfrac{b_2}{2})A_0(\sfrac{c_3}{2})A_0(\sfrac{d}{2})}{A_0(\sfrac{a_1}{2})A_0(\sfrac{a_2}{2})A_0(\sfrac{b_1}{2})A_0(\sfrac{c_1}{2})A_0(\sfrac{c_2}{2})} \\
	&\quad\times\sum_{m_1,m_2,m_3,m_4=0}^{\infty} \frac{(a_1)_{m_3}(a_2)_{m_4}(b_1)_{m_2+m_3+m_4}(c_1)_{m_1+m_2+m_3+m_4}(c_2)_{m_1+m_2+m_3+m_4}}{(b_2)_{m_2+m_3+m_4}(c_3)_{m_1+m_2+m_3+m_4}(d)_{m_3+m_4}}  \prod_{i=1}^4 \frac{x_i^{m_i}}{m_i!} \,. \nonumber 
\end{align}

\paragraph{Triangle-Box-Triangle Integral.}
\begin{align}
	&\HH{20}{a_1,a_2;b_1,b_2}{c_1,c_2;d_1,d_2}{x_1,x_2,x_3,x_4}=
	\frac{A_0(\sfrac{c_2}{2})A_0(\sfrac{d_1}{2})A_0(\sfrac{d_2}{2})}{A_0(\sfrac{a_1}{2})A_0(\sfrac{a_2}{2})A_0(\sfrac{b_1}{2})A_0(\sfrac{b_2}{2})A_0(\sfrac{c_1}{2})}  \\
	&\quad\times\sum_{m_1,m_2,m_3,m_4=0}^{\infty}\frac{(a_1)_{m_1}(a_2)_{m_4}(b_1)_{m_1+m_2}(b_2)_{m_3+m_4}(c_1)_{m_2-m_3-m_4}}{(d_1)_{m_1}(d_2)_{m_4}(c_2)_{m_2-m_3-m_4}} \prod_{i=1}^4 \frac{x_i^{m_i}}{m_i!} \,, \nonumber  \\
	&\HH{21}{a_1,a_2,a_3;b}{c;d_1,d_2;e}{x_1,x_2,x_3,x_4}=
	\frac{A_0(\sfrac{d_1}{2})A_0(\sfrac{d_2}{2})A_0(\sfrac{e}{2})}{A_0(\sfrac{a_1}{2})A_0(\sfrac{a_2}{2})A_0(\sfrac{a_3}{2})A_0(\sfrac{b}{2})A_0(\sfrac{c}{2})} \\
	&\quad\times\sum_{m_1,m_2,m_3,m_4=0}^{\infty} \frac{(a_1)_{m_1}(a_2)_{m_2}(a_3)_{m_4}(b)_{m_3+m_4}(c)_{m_1+m_2+m_3+m_4}}{(d_1)_{m_1}(d_2)_{m_4}(e)_{m_2+m_3+m_4}}\prod_{i=1}^4 \frac{x_i^{m_i}}{m_i!} \,, \nonumber  \\
	&\HH{22}{a_1,a_2,a_3;b_1,b_2}{c_1,c_2;d}{x_1,x_2,x_3,x_4} = 
	\frac{A_0(\sfrac{c_1}{2})A_0(\sfrac{c_2}{2})A_0(\sfrac{d}{2})}{A_0(\sfrac{a_1}{2})A_0(\sfrac{a_2}{2})A_0(\sfrac{a_3}{2})A_0(\sfrac{b_1}{2})A_0(\sfrac{b_2}{2})}  \\
	&\quad\times\sum_{m_1,m_2,m_3,m_4=0}^{\infty} \frac{(a_1)_{m_1}(a_2)_{m_3}(a_3)_{m_4}(b_1)_{m_1+m_2}(b_2)_{m_2+m_3}}{(c_1)_{m_1}(c_2)_{m_4}(d)_{m_2+m_3-m_4}} \prod_{i=1}^4 \frac{x_i^{m_i}}{m_i!} \,. \nonumber  
\end{align}

\begin{samepage}
\paragraph{Polygon Integrals.}
\begin{align}
	&\PP{n}{k}{a_1,\dots ,a_{n-2}}{b_1,b_2}{x_1,\dots ,x_{n-2}}\\
	&\quad =\frac{A_0(\sfrac{b_2}{2})}{A_0(\sfrac{b_1}{2})\textstyle\prod_{i=1}^{n-2} A_0(\sfrac{a_i}{2})} 
	  \sum_{m_1,\dots ,m_{n-2}=0}^{\infty} \frac{\textstyle\prod_{i=1}^{n-2}(a_i)_{m_i}(b_1)_{\sum_{i=1}^{k-1}m_i-\sum_{i=k}^{n-2}m_i}}{(b_2)_{\sum_{i=1}^{k-1}m_i-\sum_{i=k}^{n-2}m_i}}\prod_{i=1}^{n-2}\frac{x_i^{m_i}}{m_i!}\,, \notag 
\end{align}
for $k=2,\dots ,n-2$.
\end{samepage}

\paragraph{Triangle-Track Integrals.}
\begin{multline}
	\TT{w}{a;\vec{b}}{\vec{c};\vec{d};e}{\vec{x}} = \frac{\prod_{i: w_i=0}A_0(\sfrac{c_i}{2})\prod_{i: w_i=1}A_0(\sfrac{d_i}{2}) A_0(\sfrac{e}{2})}{A_0(\sfrac{a}{2}) \prod_{i=1}^\ell A_0(\sfrac{b_i}{2})} \\
	\times\!\!\! \sum_{m_1,\dots ,m_{\ell}=0}^{\infty} \frac{(a)_{m_\ell+\sum_{j=2}^{\ell}{\alpha_{j,\ell}^w} m_{j-1}} \prod_{i=1}^\ell (b_i)_{m_i+\sum_{j=2}^{i}{\alpha_{j,i}^w} m_{j-1}}}{\prod_{i: w_i=0} (c_i)_{m_i-m_{i-1}-\sum_{j=2}^{i-1}{\alpha_{j,i-1}^w} m_{j-1}} \prod_{i:w_i=1} (d_i)_{m_i+\sum_{j=2}^{i}{\alpha_{j,i}^w} m_{j-1}} (e)_{m_1}} \prod_{i=1}^{\ell}\frac{x_i^{m_i}}{m_i!} \,,
\end{multline}
where $w=w_2\dots w_\ell$ is a word of length $\ell-1$ with $n$ 0s and $\ell-1-n$ 1s, $\vec{b}$ is of length $\ell$, $\vec{c}$ is of length $n$, and $\vec{d}$ is of length $\ell-1-n$. For convenience, we label the components of $\vec{c}$ (respectively $\vec{d}$) with the positions of the 0s (respectively 1s) in the word $w$, namely $\vec{c} = (c_{i_1},...,c_{i_n})$ with $2\leqslant i_1<\dots<i_n\leqslant \ell$ such that $w_{i_1} = \dots = w_{i_n} = 0$, and similarly for~$\vec{d}$. The notation $\alpha_{j,i}^w$ is defined in \eqref{alphajkw}.

\paragraph{Conformal Double-Box Integral.}
\begin{align}
	&\CC{1}{a_1,a_2;b}{c_1,c_2}{x_1,x_2,x_3}  \\
	&\quad = \frac{A_0(\sfrac{c_1}{2})A_0(\sfrac{c_2}{2})}{A_0(\sfrac{a_1}{2})A_0(\sfrac{a_2}{2})A_0(\sfrac{b}{2})}
	\sum_{m_1,m_2,m_3=0}^{\infty}  \frac{(a_1)_{m_1} (a_2)_{m_3}(b)_{m_1+m_3-m_2}}{(c_1)_{m_1-m_2} (c_2)_{m_3-m_2}} \prod_{i=1}^3 \frac{x_i^{m_i}}{m_i!}\, , \nonumber\\
	&\CC{2}{a_1,a_2;b}{c_1,c_2}{x_1,x_2,x_3} \\
	&\quad = \frac{A_0(\sfrac{c_1}{2})A_0(\sfrac{c_2}{2})}{A_0(\sfrac{a_1}{2})A_0(\sfrac{a_2}{2})A_0(\sfrac{b}{2})}
	\sum_{m_1,m_2,m_3=0}^{\infty}  \frac{(a_1)_{m_1+m_2} (a_2)_{m_2+m_3}(b)_{m_1+m_2+m_3}}{(c_1)_{m_1+m_2} (c_2)_{m_2+m_3}}  \prod_{i=1}^3 \frac{x_i^{m_i}}{m_i!}\, , \nonumber\\
	&\CC{3}{a_1,a_2;b_1,b_2}{c}{x_1,x_2,x_3}  \\ 
	&\quad = \frac{A_0(\sfrac{c}{2})}{A_0(\sfrac{a_1}{2})A_0(\sfrac{a_2}{2})A_0(\sfrac{b_1}{2})A_0(\sfrac{b_2}{2})} 	\sum_{m_1,m_2,m_3=0}^{\infty}  \frac{ (a_1)_{m_1} (a_2)_{m_3} (b_1)_{m_1+m_2} (b_2)_{m_2+m_3}}{(c)_{m_1+m_2+m_3}} \prod_{i=1}^3 \frac{x_i^{m_i}}{m_i!}\, , \nonumber\\
	&\CC{4}{a;b_1,b_2}{c;d}{x_1,x_2,x_3} \\
	&\quad = \frac{A_0(\sfrac{c}{2})A_0(\sfrac{d}{2})}{A_0(\sfrac{a}{2})A_0(\sfrac{b_1}{2})A_0(\sfrac{b_2}{2})}
	\sum_{m_1,m_2,m_3=0}^{\infty}   \frac{(a)_{m_1}(b_1)_{m_1+m_3} (b_2)_{m_2+m_3}}{(c)_{m_1-m_2} (d)_{m_2+m_3}} \prod_{i=1}^3 \frac{x_i^{m_i}}{m_i!}\, . \nonumber
\end{align}

\section{Spectral Transform Computations}
\label{app:spectranscomps}
In this appendix we provide several details with regard to the spectral transform method introduced in \Secref{sec:spectransf}.

\subsection{Triangle Integral}
\label{app:triangle}

The three-point integral \eqref{star 3} makes sense when $\Re(a_i)<1/2$ (to avoid divergences when $x_0\to x_i$) and $\Re(a_1+a_2+a_3)>1/2$ (to avoid divergences when $|x_0|\to +\infty$). Other values of the parameters are reached by analytic continutation. When the integral is conformal, i.e.\ for $a_1 + a_2 + a_3 = 1$, the result is trivially obtained via the star-triangle identity \eqref{star-triangle}. For generic values of the propagator powers, we can replace one of the propagators, say $|x_{30}|^{-2a_3}$, with its spectral representation \eqref{SoV prop inf}. This gives
\begin{equation}
	I_3 =  \sum_{\ep = 0}^1\!\int\! \frac{1}{|x_{20}|^{2a_2}} \int_{\mathbb{R} + \ii \eta} \! \frac{\sgn^\ep(x_{31} x_{01})}{|x_{13}|^{-2\ii u} |x_{01}|^{2(a_1+a_3+\ii u)}} \frac{A_0(a_3)}{A_{\ep}(-\ii u) A_{\ep}(a_3+\ii u)} \frac{\dd u}{2\sqrt{\pi}} \frac{\dd x_0}{\sqrt{\pi}}\, ,
\end{equation}
which makes sense when $\Re(a_3) >0$ and $\eta\in ]0;\Re(a_3)[$, in addition to the previous constraints. We now want to apply Fubini's theorem to change the order of the integrals. However, the integrand is integrable only if $\Re(a_3)<0$ so we first write
\begin{multline}\label{3-pt intermediate}
	I_3 = \sum_{\ep = 0}^1\!\int\! \frac{1}{|x_{20}|^{2a_2}} \int_{\mathbb{R} + \ii \eta} \! \frac{\sgn^\ep(x_{31} x_{01})}{|x_{13}|^{-2\ii u} |x_{01}|^{2(a_1+a_3+\ii u)}} \frac{A_0(a_3)}{A_{\ep}(-\ii u) A_{\ep}(a_3+\ii u)} \frac{\dd u}{2\sqrt{\pi}} \frac{\dd x_0}{\sqrt{\pi}}\, ,\\
	+ \int\! \frac{\pi^{-1/2} \dd x_0}{|x_{20}|^{2a_2} |x_{13}|^{2a_3} |x_{01}|^{2a_1}}
\end{multline}
which holds when $-1/2<\Re(a_3)<0$ and $\eta\in ]0;\Re(a_3)+1/2[$, in addition to the constraints coming from \eqref{star 3}. The second line corresponds to the residue of the pole in $\ii a_3$ that is now below the integration contour, whereas it was above it in the previous expression. The integrand of the first line is now integrable:
\begin{equation}
	\int_{\mathbb{R}^2} \frac{|A_{\ep}(1/2-\eta+\ii u) A_{\ep}(1/2-a_3+\eta-\ii u)|}{|x_{20}|^{2\Re(a_2)} |x_{01}|^{2(\Re(a_1+a_3)-\eta)}} \dd u \dd x_0 <\infty
\end{equation}
provided $\Re(a_3)<0$ (for convergence of the integral over $u$), $\Re(a_2)<1/2$, $\Re(a_1+a_3)-\eta<1/2$, and $\Re(a_1+a_2+a_3) - \eta >1/2$ (for the integral over $x_0$).\footnote{This is clearly compatible with the constraints coming from \eqref{3-pt intermediate}.} Hence, we may change the order of the integrals in \eqref{3-pt intermediate}. We then perform the integral over $x_0$ using the chain relation \eqref{chain}, thus obtaining
\begin{multline}
	I_3 = \frac{A_0(a_2) A_0(a_3)}{|x_{12}|^{2(a_1+a_2+a_3)-1}} \sum_{\ep = 0}^1 \sgn^\ep(\chi) \int_{\mathbb{R} + \ii \eta}\!\!\!\! \frac{A_{\ep}(a_1+a_3+\ii u) A_{\ep}(1-a_1-a_2-a_3-\ii u)}{A_{\ep}(-\ii u) A_{\ep}(a_3+\ii u)} \frac{|\chi|^{2\ii u} \dd u}{2\sqrt{\pi}}\\
	+ \frac{A_0(a_1) A_0(a_2) A_0(1-a_1-a_2)}{|x_{13}|^{2a_3} |x_{12}|^{2(a_1+a_2)-1}}\, ,
\end{multline}
where we introduced the ratio $\chi = x_{13}/x_{12}$. Finally, we deform the contour to cross the pole in $\ii a_3$. This cancels the term in the second line. We may then take $\Re(a_3)>0$ and return to a horizontal contour so that
\begin{equation}
	I_3 = \frac{A_0(a_2) A_0(a_3)}{|x_{12}|^{2(a_1+a_2+a_3)-1}} \sum_{\ep = 0}^1 \sgn^\ep(\chi) \int_{\mathbb{R} + \ii \eta}\!\!\!\! \frac{A_{\ep}(a_1+a_3+\ii u) A_{\ep}(1-a_1-a_2-a_3-\ii u)}{A_{\ep}(-\ii u) A_{\ep}(a_3+\ii u)} \frac{|\chi|^{2\ii u} \dd u}{2\sqrt{\pi}}\, ,
\end{equation}
for $\Re(a_3)>0$, $\Re(a_1+a_3)-\eta<1/2$, $\Re(a_1+a_2+a_3) - \eta >1/2$, and $\eta>0$, which is simply stating that the poles of $A_\ep(...\pm\ii u)$ are all below/above the integration contour. Assuming that $|\chi|<1$, we may close the contour in the lower half-plane. There are two series of simple poles: one in $-\ii(\ep/2+\mathbb{N})$ and one in $-\ii((1+\ep)/2-a_1-a_3+\mathbb{N})$. Summing over their residues yields the final result \eqref{3-pt result}.

\subsection{Box Integral}

Replacing the two propagators connected to $x_3$ and $x_4$ by their spectral representations \eqref{SoV prop inf} with auxiliary point $x_1$ leads to
\begin{multline}
	I_4 = \frac{A_0(a_2) A_0(a_3) A_0(a_4)}{|x_{12}|^{2a_1+2a_2-1}|x_{13}|^{2a_3}|x_{14}|^{2a_4}} \sum_{\ep_1,\ep_2} \sgn^{\ep_1}(\chi_1) \sgn^{\ep_2}(\chi'_2) \! \int_{\mathbb{R} + \ii \eta_1} \!\! |\chi_1|^{2\ii u_1} \!\int_{\mathbb{R} + \ii \eta_2} \!\! |\chi'_2|^{2\ii u_2} \\
	\times \frac{A_{[\ep_1+\ep_2]}(a_1-\ii(u_1+u_2)) A_{[\ep_1+\ep_2]}(1-a_1-a_2 + \ii(u_1+u_2))}{A_{\ep_1}(-\ii u_1) A_{\ep_1}(a_3+\ii u_1) A_{\ep_2}(-\ii u_2) A_{\ep_2}(a_4+\ii u_2)} \frac{\dd u_1 \dd u_2}{(2\sqrt{\pi})^2}\, .
\end{multline}
where the ratios are
\begin{equation}
	\chi_1 = \frac{x_{12}}{x_{13}} \quad \text{and}\quad \chi'_2 = \chi_1\chi_2= \frac{x_{12}}{x_{14}}\, .
\end{equation}
Let us assume that $|\chi_1|<1$ and $|\chi_2|<1$. We then compute the integrals as sums over residues and obtain the result \eqref{star4Result}.

\subsection{H-Integral  (Two-Loop Triangle Track)}

We have
\begin{multline}
	I_{2,2} = \frac{A_0(b) A_0(a_3) A_0(a_4)}{|x_{12}|^{2(b+\sum_i a_i-1)}} \sum_{\ep_1,\ep_2} (-1)^{\ep_1\ep_2} \sgn^{\ep_1}(\chi_1) \sgn^{\ep_2}(\chi_2) \! \int_{\mathbb{R} + \ii \eta_1} \!\! |\chi_1|^{2\ii u_1} \!\int_{\mathbb{R} + \ii \eta_2} \!\! |\chi_2|^{2\ii u_2} \\
	\times \frac{A_{\ep_1}(a_2+a_3+\ii u_1) A_{[\ep_1+\ep_2]}(3/2-b-\sum_i a_i-\ii u_1 - \ii u_2) A_{\ep_2}(a_1+a_4+\ii u_2)}{A_{\ep_1}(-\ii u_1) A_{\ep_1}(a_3+\ii u_1) A_{\ep_2}(-\ii u_2) A_{\ep_2}(a_4+\ii u_2)} \frac{\dd u_1 \dd u_2}{(2\sqrt{\pi})^2}\, ,
\end{multline}
where the ratios are  $\chi_1 = x_{23}/x_{21}$, $\chi_2 = x_{14}/x_{12}$. We can now compute the integrals by deforming the contours depending on the values of the ratios. For instance, if $|\chi_1|+|\chi_2|<1$, we arrive at the result \eqref{H explicit}.

\subsection{Triangle-Box Integral}

One of the simplest SoV representations we could derive reads
\begin{multline}
	I_{2,3} = \frac{A_0(b) \prod_{i=3}^5 A_0(a_i)}{|x_{12}|^{2(\sum_{i=1}^5 a_i + b-1)}} \prod_{j=1}^3 \sum_{\ep_j} \int_{\mathbb{R}+\ii \eta_j} \frac{\dd u_j}{2\sqrt{\pi}} \frac{\sgn^{\ep_j}(\chi'_j)  |\chi'_j|^{2\ii u_j}}{A_{\ep_j}(-\ii u_j) A_{\ep_j}(a_{j+2}+\ii u_j)} (-1)^{(\ep_1+\ep_2)\ep_3} \\
	\times A_{[\ep_1+\ep_2]}(a_2 +a_3 + a_4 + \ii u_1+\ii u_2) A_{\ep_3}(a_1 + a_5 + \ii u_3) A_{[\sum_j\ep_j]}(1+\tilde{b} - {\textstyle\sum_{i}} a_i - \ii {\textstyle\sum_{j}} u_j)\, ,
\end{multline}
in terms of the following ratios:
\begin{equation}
	\chi'_1 = \chi_1\chi_2 = \frac{x_{23}}{x_{21}}\, ,\quad \chi'_2 = \chi_2 = \frac{x_{24}}{x_{21}}\, ,\quad \chi'_3 = \chi_3 = \frac{x_{15}}{x_{12}}\, .
\end{equation}
Let us assume that
\begin{equation}
	|\chi_j|<1\,, \quad \text{i.e.\ that} \quad |\chi'_1|<|\chi'_2|<1\quad \text{and}\quad |\chi'_3|<1\, .
\end{equation}
Then, we can perform the integrals to find the result \eqref{eq:resultTriangleBox}.

\subsection{Double-Box Integral (Two-Loop Train Track)}

One of the simplest SoV representations we could derive reads
\begin{multline}
	I_{3,3} = \frac{A_0(b) \prod_{i=3}^6 A_0(a_i)}{|x_{12}|^{2(\sum_{i=1}^6 a_i + b-1)}} \prod_{j=1}^4 \sum_{\ep_j} \int_{\mathbb{R}+\ii \eta_j} \frac{\dd u_j}{2\sqrt{\pi}} \frac{\sgn^{\ep_j}(\chi'_j)  |\chi'_j|^{2\ii u_j}}{A_{\ep_j}(-\ii u_j) A_{\ep_j}(a_{j+2}+\ii u_j)} (-1)^{(\ep_1+\ep_2)(\ep_3+\ep_4)} \\
	\times A_{[\ep_1+\ep_2]}(a_2 +a_3 + a_4 + \ii(u_1+u_2)) A_{[\ep_3+\ep_4]}(a_1 + a_5 + a_6 + \ii(u_3+u_4))\\
	\times A_{[\sum_j\ep_j]}(1+\tilde{b} - \textstyle{\sum_i} a_i - \ii \textstyle{\sum_j} u_j)\, ,
\end{multline}
in terms of the following ratios:
\begin{equation}
	\chi'_1 = \chi_1\chi_2 = \frac{x_{23}}{x_{21}}\, ,\quad \chi'_2 = \frac{x_{24}}{x_{21}}\, ,\quad \chi'_3 = \chi_3\chi_4 = \frac{x_{15}}{x_{12}}\, ,\quad \chi'_4 = \frac{x_{16}}{x_{12}}\, .
\end{equation}
Let us assume that
\begin{equation}
	|\chi_j|<1\,, \quad \text{i.e.\ that} \quad |\chi'_1|<|\chi'_2|<1\quad \text{and}\quad |\chi'_3|<|\chi'_4|<1\, .
\end{equation}
Then, we can compute the integrals to find the result \eqref{eq:resultDoubleBox}.

\subsection{Generic Polygon Integrals}
\label{app:spectralpolygon}

In order to compute $I_{n}$, we begin by replacing $n-2$ propagators with their spectral representation \eqref{SoV prop inf}. More precisely, we apply \eqref{SoV prop inf} with $(x_1,x_2,x_3) \to (x_3,x_0,x_1)$, $(x_4,x_0,x_1)$, $\dots,(x_n,x_0,x_1)$. After this step, the integral over the $x_0$ can be computed using the chain relation \eqref{chain}. We arrive at
\begin{multline}
	I_{n} = \frac{\prod_{j=2}^n A_0(a_j)}{|x_{12}|^{2(a_1+a_2) - 1} \prod_{j=3}^n |x_{1j}|^{2a_j}} \prod_{j=1}^{n-2} \sum_{\ep_j = 0}^1 \int_{\mathbb{R}+\ii\eta} \frac{\dd u_j}{2\sqrt{\pi}} \frac{s^{\ep_j}(\chi'_j)  |\chi'_j|^{2\ii u_j}}{A_{\ep_j}(-\ii u_j) A_{\ep_j}(a_{j+2}+\ii u_j)}\\
	\times A_{[\sum_j \ep_j]}(a_1 - \ii{\textstyle \sum}_j u_j) A_{[\sum_j \ep_j]}(1-a_1-a_2+ \ii{\textstyle \sum}_j u_j) \, ,
\end{multline}
where the ratios are	$\chi'_j = x_{12}/x_{1,j+2}$. Performing the change of spectral variables $(\ep_j,u_j) \to ([\ep_j+\ep_{j+1}], u_j - u_{j+1})$ for $j\leqslant n-3$, we arrive at
\begin{multline}
	I_{n} = \frac{\prod_{j=2}^n A_0(a_j)}{|x_{12}|^{2(a_1+a_2) - 1} \prod_{j=3}^n |x_{1j}|^{2a_j}} \prod_{j=1}^{n-2} \sum_{\ep_j = 0}^1 \int_{\mathbb{R}+\ii\eta_j} \frac{(2\sqrt{\pi})^{-1}\dd u_j\, s^{\ep_j}(\chi_j)  |\chi_j|^{2\ii u_j}}{A_{[\ep_j+\ep_{j+1}]}(\ii u_{j+1,j}) A_{[\ep_j+\ep_{j+1}]}(a_{j+2}+\ii u_{j,j+1})}\\
	\times A_{\ep_1}(a_1 - \ii u_1) A_{\ep_1}(1-a_1-a_2+ \ii u_1) \, ,
\end{multline}
where the ratios are	 $\chi_j = x_{1,j+1}/x_{1,j+2}$, and we introduced $\kappa_{n-1} = u_{n-1} = 0$ for convenience. Note that the small parameters $\eta_j$ must be ordered according to $\eta_1>\dots>\eta_{n-2}$ for the series of poles to lie completely on one side of the integration contours.

We assume that all ratios are smaller than $1$ in absolute value and compute the integrals in the order $u_1,\dots,u_{n-2}$. There will only be $n-1$ terms in the final result, corresponding to the indicials \eqref{indicials polygon}, as we now explain. The main point to notice is the following: if for the integral over $u_j$ we pick the residues coming from $A^{-1}_{[\ep_j+\ep_{j+1}]}(\ii u_{j+1,j})$, then only the residues coming from $A^{-1}_{[\ep_{j+1}+\ep_{j+2}]}(\ii u_{j+2,j+1})$ will survive in the integral over $u_{j+1}$. Since we evaluated at $u_j = u_{j+1} -\ii(\dots)$, it would seem that some residues from $A^{-1}_{[\ep_{j-1}+\ep_{j}]}(a_{j+1}+\ii u_{j-1,j}) = A^{-1}_{[\ep_{j-1}+\ep_{j}]}(a_{j+1}+\ii u_{j-1,j+1}-\dots)$ could also contribute to the integral over $u_{j+1}$. However, they are exactly cancelled by the contributions coming from picking the poles of $A^{-1}_{[\ep_{j-1}+\ep_{j}]}(a_{j+1}+\ii u_{j-1,j})$ for the integral over $u_j$ and then those from $A^{-1}_{[\ep_j+\ep_{j+1}]}(\ii u_{j+1,j})$ for the integral over $u_{j+1}$. Hence, there are only $n-1$ ways to pick the residues that survive, they correspond to
\begin{equation}
	A^{-1}_{[\ep_1+\ep_2]}(\ii u_{21}),A^{-1}_{[\ep_2+\ep_3]}(\ii u_{32}),\dots,A^{-1}_{[\ep_{n-2}+\ep_{n-1}]}(\ii u_{n-1,n-2})
\end{equation}
or to
\begin{equation}
	A_{\ep_1}(1-a_1-a_2+ \ii u_1),A^{-1}_{[\ep_2+\ep_3]}(\ii u_{32}),\dots,A^{-1}_{[\ep_{n-2}+\ep_{n-1}]}(\ii u_{n-1,n-2})
\end{equation}
or to
\begin{multline}
	A_{\ep_1}(1-a_1-a_2+ \ii u_1),A^{-1}_{[\ep_1+\ep_2]}(a_3+\ii u_{12}),\dots,A^{-1}_{[\ep_{j-2}+\ep_{j-1}]}(a_j+\ii u_{j-2,j-1}),\\
	A^{-1}_{[\ep_j+\ep_{j+1}]}(\ii u_{j+1,j}),A^{-1}_{[\ep_{j+1}+\ep_{j+2}]}(\ii u_{j+2,j+1}),\dots,A^{-1}_{[\ep_{n-2}+\ep_{n-1}]}(\ii u_{n-1,n-2})
\end{multline}
for any $j\in \{3,n-2\}$, or to
\begin{equation}
	A_{\ep_1}(1-a_1-a_2+ \ii u_1),A^{-1}_{[\ep_1+\ep_2]}(a_3+\ii u_{12}),\dots, A^{-1}_{[\ep_{n-2}+\ep_{n-1}]}(a_{n-1}+\ii u_{n-3,n-2}) \, .
\end{equation}

\subsection{Generic Triangle-Track Integrals}
\label{app:spectraltriangle}

In order to compute $I_{2,1^{\ell-2},2}$, we begin by replacing $\ell$ propagators with their spectral representation \eqref{SoV prop inf}. More precisely, we apply \eqref{SoV prop inf} with $(x_1,x_2,x_3) \to (x_1,y_1,x_{\ell+2})$, $(y_1,y_2,x_{\ell+1})$, $\dots,(y_{\ell-1},y_\ell,x_3)$. After this step, the integrals over the $y_j$'s can all be computed using the chain relation \eqref{chain}. We arrive at
\begin{multline}
	I_{2,1^{\ell-2},2} = V_n(x_1,\dots,x_{\ell+2}) A_0(a_1) A_0(a_2) \prod_{j=1}^{\ell-1}A_0(b_j) \prod_{j=1}^{\ell} \sum_{\ep_j = 0}^1 s^{\ep_j}(\chi_j) \int_{\mathbb{R}+\ii\eta} \frac{\dd u_j}{2\sqrt{\pi}} |\chi_j|^{2\ii u_j}\\
	\times \frac{A_{\ep_1}(\tilde{a}_1-\ii u_\ell) A_{\ep_{\ell}}(\tilde{a}_2+\tilde{a}_3+\ii u_1)}{\prod_{j=1}^{\ell} A_{\ep_j}(- \ii u_j) A_{\ep_j}(\tilde{a}_{j+2} + \ii u_j)}\prod_{j=2}^{\ell} (-1)^{\ep_j\ep_{j-1}} A_{[\ep_j + \ep_{j-1}]}\left(\tilde{a}_{j+2}+\tilde{b}_{\ell+1-j} + \ii u_{j,j-1}\right)  \, ,
\end{multline}
where the prefactor is
\begin{equation}
	V_n(x_1,\dots,x_{\ell+2}) =  |x_{23}|^{1-2a_2-2a_3} \prod_{j=4}^{\ell+2} |x_{j-1,j}|^{1-2a_j-2b_{\ell+3-j}} |x_{1,\ell+2}|^{-2a_1}\, ,
\end{equation}
and the ratios are
\begin{equation}
	\chi_{j} = \frac{x_{j+2,j+1}}{x_{j+2,j+3}}\quad \text{for}\quad 1\leqslant j\leqslant \ell\, ,
\end{equation}
where we identify $x_{\ell+3} = x_1$. Let us assume that all the ratios are small enough in absolute value. We can thus close all the contours in the lower half-plane, and given the form of the integrand, it seems reasonable to compute the integrals in the order $u_1,\dots,u_\ell$. Given the location of the poles, we immediately find the indicials \eqref{indicials triangle-track}. The full result contains $2^{\ell}$ terms. For instance, the term that arises if we only pick the residues at $u_j = -\ii(\ep_j/2+\mathbb{N})$ is straightforward to read. It is simply
\begin{equation}
	\sum_{p_1,\dots,p_{\ell} = 0}^{+\infty} \prod_{j=1}^{\ell} \frac{(2\tilde{a}_{\ell+3-j})_{p_j}}{p_j!} \prod_{j=1}^{\ell-1} \frac{(-\chi_{j+1})^{p_{j+1}}}{(2(\tilde{a}_{\ell+3-j}+\tilde{b}_j))_{p_j-p_{j+1}}} \frac{\chi^{p_1}_1 (2a_1)_{p_1}}{(2(\tilde{a}_2+\tilde{a}_3))_{p_{\ell}}}
\end{equation}
up to an overall prefactor. This is exactly the term $w=0\dots0$ of the sum in \eqref{eq:triangleTrackResult}. More generally, one can check that taking into account all the residues, we arrive exactly at \eqref{eq:triangleTrackResult}.


\section{Proof of Triangle-Track Differential Equations}
\label{sec:ProofTriangleTrackEq}

In this appendix, we prove \eqref{eq:TriangleTrackDiffEq} by applying the differential operator in \eqref{eq:exampleop} to the ansatz 
\eqref{eq:DecompositionTriangleTracks}, i.e.\ we apply the operator ($\partial_i=\partial_{x_i}$)
\begin{equation}
	\frac{2}{i} \levo{P}_{\textrm{bridge}} := \sum_{m=k+3}^{\ell+3}\sum_{r=2}^{k+1}\levo{P}_{mr} +\left(\! N_2-2\! \sum_{i=4}^{k}b_{\ell +3-i}\right)\!\sum_{m=k+3}^{\ell +3}\!\partial_m-\left(\! N_1-2\! \sum_{i=k+1}^{\ell +2}b_{\ell +3-i}\right)\sum_{r=2}^{k+1}\partial_r
\end{equation}
to $V(x_1,\dots,x_{\ell+2})\phi(\chi_1,\dots,\chi_{\ell})$, where
\begin{equation}
	\chi_{j} = \frac{x_{j+1,j+2}}{x_{j+3,j+2}}\quad \text{for}\quad 1\leqslant j\leqslant \ell\, ,
\end{equation}
and
\begin{align}
	V(x_1,\dots,x_{\ell+2}) = \prod_{j=3}^{\ell+3}|x_{j-1,j}|^{1-2a_j-2b_{\ell +3-j}}\equiv \prod_{j=3}^{\ell+3}|x_{j-1,j}|^{\rho_j},
\end{align}
with
\begin{align}
	b_\ell = a_2,\quad b_0 = \frac{1}{2},\quad a_{\ell +3} = a_1\,,\quad \rho_j=1-2a_j-2b_{\ell +3-j}\,.
\end{align}
Note that the summation indices $m$ and $r$ now run over legs to the left ($m$) and to the right ($r$) of the bridge vertex $y_{\ell +1-k}$, which is connected to the external point $x_{k+2}$. All indices are understood modulo $\ell +2 = n$ and we employ the following labeling conventions:
\begin{equation*}
	\includegraphicsbox{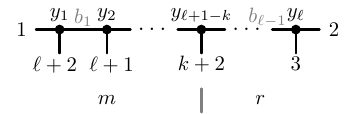}\label{fig:TriangleTrackProofconvention}\quad.
\end{equation*}

Let us rewrite the above bridge operator as
\begin{align}
	\frac{2}{i}\levo{P}_{\textrm{bridge}} = \left(\widetilde{A} - \sum_{r=2}^{k+1}x_r\partial_r\right)\sum_{m=k+3}^{\ell +3}\partial_m-\left(\widetilde{B}-\sum_{m=k+3}^{\ell +3}x_m\partial_m\right)\sum_{r=2}^{k+1}\partial_r,
\end{align}
with
\begin{equation}
	\widetilde{A} = k-1-2\sum_{i=4}^{k+2}b_{\ell +3-i}-2\sum_{r=2}^{k+1}a_r\,,\qquad\widetilde{B} =\ell -k-2\sum_{i=k+3}^{\ell +2}b_{\ell +3-i}-2\sum_{m=k+3}^{\ell +3}a_m,
\end{equation}
where we plugged in $N_1 = \ell -k$ and $N_2 = k-1$. We now apply the operator to the product $V(x_1,\dots,x_{\ell+2})\phi(\chi_{1},\dots,\chi_\ell)$ and then sort by order of derivatives in $\phi$, dropping kinematical dependencies: 
\begin{equation}\label{eq:levopbridgeProductRule}
	\frac{2}{i}\levo{P}_{\textrm{bridge}}V\phi = \frac{2}{i} \left[\levo{P}_{\textrm{bridge}}V\right] \phi + V \sum_{m=k+3}^{\ell +3} \sum_{r=2}^{k+1} (x_m\!-x_r)\partial_m\partial_r\phi + \widetilde{C}\, .
\end{equation}
Here 
\begin{align}\label{eq:TildeCdef}
	\widetilde{C} = &\quad+\left(\widetilde{A}V-\sum_{r=2}^{k+1}x_m\partial_rV\right)\sum_{m=k+3}^{\ell +3}\partial_m\phi -\sum_{m=k+3}^{\ell +3}\partial_mV\sum_{r=2}^{k+1}x_r\partial_r\phi \nonumber\\
	&\quad- \left(\widetilde{B}V-\sum_{m=k+3}^{\ell +3}x_m\partial_mV\right)\sum_{r=2}^{k+1}\partial_r\phi+\sum_{r=2}^{k+1}\partial_rV\sum_{m=k+3}^{\ell +3}x_m\partial_m\phi\,.
\end{align}
We can now systematically simplify expressions that are telescopic sums:
\begin{align}\label{eq:simplificationqV}
	\frac{1}{V}\sum_{r=2}^{k+1}\partial_rV &= -\frac{\rho_{k+2}}{x_{k+2,k+1}},\qquad \frac{1}{V}\sum_{r=2}^{k+1}x_r\partial_rV  = \sum_{r=3}^{k+1}\rho_{r}-\frac{\rho_{k+2}x_{k+1}}{x_{k+2,k+1}}\,,\nonumber\\
	\frac{1}{V}\sum_{m=k+3}^{\ell +3}\partial_m V &= \frac{\rho_{k+3}}{x_{k+3,k+2}}\,,\qquad
	\frac{1}{V}\sum_{m=k+3}^{\ell +3}x_m\partial_mV = \frac{\rho_{k+3}x_{k+3}}{x_{k+3,k+2}}+\sum_{k+2}^{\ell +3}\rho_{m}\,.
\end{align}
This implies
\begin{align}\label{eq:simplificationsTilde}
	\widetilde{A}V-\!\sum_{r=2}^{k+1}x_r\partial_rV = V\!\left(\frac{\rho_{k+2}x_{k+1}}{x_{k+2,k+1}} - 2b_{\ell +1-k}\right)\,,\ 	\widetilde{B}V-\!\!\sum_{m=k+3}^{\ell +3}x_m\partial_mV =	 -V\frac{\rho_{k+3}x_{k+2}}{x_{k+3,k+2}}\,.
\end{align}
Notice that none of the $|x_{j-1,j}|^{\rho_j}$ factors in $V(x_1,\dots,x_{\ell+2})$ can be affected at the same time by a derivative from the $m\geqslant k+3$ sum and and a derivative from the $r\leqslant k+1$ sum. Therefore, we can use the expressions \eqref{eq:simplificationqV} and \eqref{eq:simplificationsTilde} and plug them straightforwardly into \eqref{eq:levopbridgeProductRule} and \eqref{eq:TildeCdef}. The first term on the right-hand side of \eqref{eq:levopbridgeProductRule} thus becomes
\begin{align}\label{eq:resultnoderiv}
	\frac{2}{i}\levo{P}_{\textrm{bridge}}V &= \left[\left(\frac{\rho_{k+2}x_{k+1}}{x_{k+2,k+1}} -2b_{\ell +1-k} \right)\frac{\rho_{k+3}}{x_{k+3,k+2}} - \frac{\rho_{k+3}x_{k+2}}{x_{k+3,k+2}}\frac{\rho_{k+2}}{x_{k+2,k+1}}\right]V \nonumber\\
	&= \frac{V}{x_{k+3,k+2}}\left(1-2a_{k +2}\right)\left(2a_{k+3}+2b_{\ell -k}-1\right)\, ,
\end{align}
where we used the explicit form of the exponents $\rho_j$.

Summations over derivatives of $\phi$ simplify as well. The chain rule gives 
\begin{align}
	\sum_{r=2}^{k+1}\partial_{r}\phi = \sum_{r=2}^{k+1}\sum_{i=1}^{\ell}(\partial_{x_r}\chi_i)\partial_{\chi_i}\phi,
\end{align}
where for a given $r$ the $i$-summation contains at most three terms due to the definition of the ratios, see \eqref{eq:varsTriangleTrack}. Performing first the sum over $r$, we realise that only the terms for $i\in\{k-1,k\}$ remain. Indeed, if all three non-zero derivatives of a ratio are summed up, the sum vanishes: since the $\chi_i$ are translation invariant, the generator of translations ($\partial_{i+1}+\partial_{i+2}+\partial_{i+3}$) annihilates $\chi_i$.  
A similar cancellation happens for terms of the form $\sum_{r=2}^{k+1}x_r\partial_r\phi\,,$ where we argue with dilatation invariance of $\chi$. The results are simply
\begin{align}\label{eq:simplificationsPhi}
	\sum_{r=2}^{k+1}\partial_r\phi &= \frac{1}{x_{k+2,k+1}} \left[\chi_{k-1}\partial_{\chi_{k-1}}\phi - \chi_k\partial_{\chi_k}\phi\right]\,,\nonumber\\
	\sum_{m=k+3}^{\ell +3}\partial_m\phi &= \frac{1}{x_{k+3,k+2}} \left[\chi_{k+1}\partial_{\chi_{k+1}}\phi - \chi_k\partial_{\chi_k}\phi\right]\,,\nonumber\\
	\sum_{r=2}^{k+1}x_r\partial_r\phi &= \frac{1}{x_{k+2,k+1}} \left[x_{k+2}\chi_{k-1}\partial_{\chi_{k-1}}\phi - x_{k+1} \chi_k\partial_{\chi_k}\phi\right]\,,\nonumber\\
	\sum_{m=k+3}^{\ell +3}x_m\partial_m\phi &= \frac{1}{x_{k+3,k+2}} \left[x_{k+2}\chi_{k+1}\partial_{\chi_{k+1}}\phi - x_{k+3} \chi_k\partial_{\chi_k}\phi\right]\,.
\end{align}
Using these identities sequentially, we can write
\begin{align}\label{eq:resultsecondorder}
	\sum_{m=k+3}^{\ell +3}\! \sum_{r=2}^{k+1} (x_m\!\!-x_r)\partial_m\partial_r\phi &= \frac{1}{x_{k+3,k+2}\, x_{k+2,k+1}} \Big[(x_{k+2}\chi_{k+1}\partial_{\chi_{k+1}}\!\!\! - x_{k+3} \chi_k\partial_{\chi_k}) (\chi_{k-1}\partial_{\chi_{k-1}}\!\!\! - \chi_k\partial_{\chi_k}) \nonumber\\
	&\qquad\qquad\qquad\  - (x_{k+2}\chi_{k-1}\partial_{\chi_{k-1}}\!\!\! - x_{k+1} \chi_k\partial_{\chi_k}) (\chi_{k+1}\partial_{\chi_{k+1}}\!\!\! - \chi_k\partial_{\chi_k})\Big]\phi \nonumber\\
	&= \frac{1}{x_{k+3,k+2}} \Big[\chi_{k-1}\partial_{\chi_{k-1}}\!\!-\chi_k\chi_{k+1}\partial_{\chi_{k+1}}\!\!+\chi_k(\chi_k-1)\partial_{\chi_k}\Big]\partial_{\chi_k}\phi\nonumber\\
	&\qquad\qquad\qquad\qquad\qquad\qquad\quad\  +\frac{x_{k+3,k+1}}{x_{k+3,k+2}\, x_{k+2,k+1}} \chi_k\partial_{\chi_k}\phi\, .
\end{align}
The last term of this equation, which contains a single derivative of $\phi$, combines with $\widetilde{C}$ defined in \eqref{eq:TildeCdef}.
Plugging \eqref{eq:simplificationqV}, \eqref{eq:simplificationsTilde}, and \eqref{eq:simplificationsPhi}, we get
\begin{align}\label{eq:TildeC}
	\widetilde{C} + \frac{V x_{k+3,k+1} \chi_k\partial_{\chi_k}\phi}{x_{k+3,k+2}\, x_{k+2,k+1}} &= \frac{V}{x_{k+3,k+2}\, x_{k+2,k+1}}\Big[(\rho_{k+2}x_{k+1}-2b_{\ell +1-k}x_{k+2,k+1})(\chi_{k+1}\partial_{\chi_{k+1}}\!\! - \chi_k\partial_{\chi_k})\nonumber\\
	& - \rho_{k+3} (x_{k+2}\chi_{k-1}\partial_{\chi_{k-1}}\!\! - x_{k+1} \chi_k\partial_{\chi_k}) + \rho_{k+3}x_{k+2} (\chi_{k-1}\partial_{\chi_{k-1}}\!\! - \chi_k\partial_{\chi_k})\nonumber\\
	&-\rho_{k+2} (x_{k+2}\chi_{k+1}\partial_{\chi_{k+1}}\!\! - x_{k+3} \chi_k\partial_{\chi_k}) + x_{k+3,k+1}\chi_k\partial_{\chi_k} \Big]\phi\nonumber\\
	&= \frac{V}{x_{k+3,k+2}}\Big[(2(a_{k+2}\!+b_{\ell+1-k}\!-1)+(1-2a_{k+2}\!+2a_{k+3}\!+2b_{\ell-k})\chi_k)\partial_{\chi_k}\nonumber\\
	&\qquad\qquad\qquad\qquad\qquad\qquad\qquad\  + (2a_{k+2}\!-1)\chi_{k+1}\partial_{\chi_{k+1}}\Big]\phi\,.
\end{align}
Combining \eqref{eq:resultnoderiv}, \eqref{eq:resultsecondorder}, and \eqref{eq:TildeC}, we arrive at \eqref{eq:TriangleTrackDiffEq}.

\section{\texorpdfstring{$\levo{P}$}{Phat}-Symmetries From Aomoto--Gelfand Differential \\Equations}
\label{sec:PHatFromAGDEs}

In this appendix we will show how the various non-local symmetries proven in \cite{Loebbert:2024qbw} follow from the Aomoto--Gelfand (AG) hypergeometric equations for one-dimensional Feynman integrals. We have already shown this in some examples in the main text. Hence we are only left with showing the (generalized) end-vertex symmetry as well as the generalized version of the bridge-vertex symmetry.

\paragraph{Generalized End-Vertex Symmetry.}
Consider some Feynman graph with a subtree $A$, i.e., a collection of vertices $a_1,\dots a_N$ connected amongst each other in a tree like fashion and to external points. We denote the external propagators connected to vertex $a_i$ and the corresponding columns in the associated $Z$ matrix by $X_{a_i}$. Now let this tree be connected to a vertex $b$ through a propagator $\alpha_1$ from $a_1$ to $b$ with no further connections between $A$ and the rest of the graph.

Let us record the equations \eqref{eq:AGDE1} for the pairs of rows $(a_k,a_k)$ for $k=1$ and $k>1$
\begin{align}
	&\left[ \sum_{i\in X_{a_1}}\partial_{z_{a_1i}}-\partial_{z_{a_1\alpha_1}} +\sum_{i=2}^N\sigma_{\alpha_i}(a_1)\partial_{z_{a_1\alpha_i}}+1  
	\right]F=0 \,, \label{eq:gevSymmABEq1}\\
	&\left[ \sum_{i\in X_{a_k}}\partial_{z_{a_ki}}+\sum_{i=2}^N\sigma_{\alpha_i}(a_k)\partial_{z_{a_k\alpha_i}}+1  
	\right]F=0 \,,  \label{eq:gevSymmABEq2}
\end{align}
as well as the pairs of rows $(a_k,b)$ for $k=1$ and $k>1$
\begin{align}
	&\left[ \sum_{i\in X_{a_1}}\partial_{z_{bi}}-\partial_{z_{b\alpha_1}}+\sum_{i=2}^N\sigma_{\alpha_i}(a_1)\partial_{z_{b\alpha_i}} \right]F=0 \,, \\
	&\left[ \sum_{i\in X_{a_k}}\partial_{z_{bi}}+\sum_{i=2}^N\sigma_{\alpha_i}(a_k)\partial_{z_{b\alpha_i}} \right]F=0 \,.
\end{align}
Here $\sigma_{\alpha}(a)$ is $0$ if $a$ is not an end-point of the propagator $\alpha$ and otherwise yields $\pm 1$ depending on the orientation of the propagator. Furthermore we need the equations \eqref{eq:AGDE2} for the columns $\alpha_k$, again for $k=1$ and $k>1$
\begin{align}
	&\left[\partial_{z_{b\alpha_1}}-\partial_{z_{a_1\alpha_1}}+2b_1\right]F =0 \,, \\
	&\left[\sum_{a\in A} \sigma_{\alpha_k}(a)\partial_{z_{a\alpha_k}}+2b_k\right]F=0 \,.
\end{align}
Here $b_i$ refers to the propagator power of $\alpha_i$.

Note that the sum of equation \eqref{eq:gevSymmABEq1} and equations \eqref{eq:gevSymmABEq2} for all $k=2,\dots ,N$ yields
\begin{equation}
	\sum_{a\in A}\sum_{i\in X_a}\partial_{z_{bi}}F=\partial_{z_{b\alpha_1}}F-\sum_{k=2}^N \partial_{z_{b\alpha_k}}F\sum_{a\in A}\sigma_{\alpha_k}(a)=\partial_{z_{b\alpha_1}}F \,,
\end{equation}
since all propagators $\alpha_k$ with $k>1$ start and end in the subtree $A$. Using this we can now compute similarly to before
\begin{align}
	&\sum_{a\in A} \sum_{i\in X_a}\sum_{j\in X_b}\left( \partial_{z_{0j}}\partial_{ai}-\partial_{z_{0i}}\partial_{z_{bj}} \right)F \nonumber \\
	&=\sum_{j\in X_b}\left(\partial_{z_{b\alpha_1}}-N+2\sum_{j=1}^N b_i\right)\partial_{z_{0j}}F- \sum_{a\in A}\sum_{i\in X_a}\sum_{j\in X_b}\partial_{z_{0i}}\partial_{z_{bj}}F \\
	&=\sum_{a\in A} \sum_{i\in X_a}\sum_{j\in X_b}\left( \partial_{z_{0j}}\partial_{z_{bi}}-\partial_{z_{0i}}\partial_{z_{bj}} \right)F+\left(2\sum_{k=1}^N b_k-N \right)\sum_{j\in X_b}\partial_{z_{0j}} F \nonumber \,.
\end{align}
As before, the first sum vanishes due the to equations \eqref{eq:AGDE3}. We can identify the left hand side with the sum of the $\levo{P}$-operators $\levo{P}_{ab}$ acting on the vertices $a\in A$ and $b$, while the right hand side yields the momentum operator $P_b$ acting on vertex $b$. We have hence indeed recovered the generalized end-vertex symmetry of \cite{Loebbert:2024qbw}
\begin{equation}
	\left[ \sum_{a\in A}\levo{P}_{ab}+\frac{1}{2}\left(N-2\sum_{i=1}^N b_i \right)P_b \right]F=0 \,.
\end{equation}

\paragraph{Generalized Bridge-Vertex Symmetry.}
Finally let us consider the generalized bridge-vertex symmetry. To this end consider some (bridge) vertex $b$ connected to two subtrees $A,C$. These are made up of internal vertices $a_1,\dots ,a_{N_1}$, $c_1,\dots c_{N_2}$ and internal propagators between these denoted by $\alpha_2,\dots \alpha_{N_1}$, $\gamma_2,\dots \gamma_{N_2}$, respectively. These two subtrees are connected to $b$ via propagators $\alpha_1$ from $b$ to $a_1$ and $\gamma_1$ from $b$ to $c_1$. 

As before we need the differential equations  \eqref{eq:AGDE1} for the pairs of rows $(a_1,a_1)$, $(a_k,a_k)$, $(a_1,b)$, $(a_k,b)$, for $k>1$, which take precisely the same form as given above with $N$ replaced by $N_1$. We furthermore need the same set of equations for $a_i$ replaced by $c_i$ (and $N_1$ replaced by $N_2$). Finally we need the equations  \eqref{eq:AGDE2} for columns $\alpha_1,\alpha_k$ which again takes exactly the same form as above, as well as the corresponding equations for subtree $C$. Here we denote the powers of the propagators $\alpha_i$ by $b_i$ and those of the $\gamma_i$ by $b_i'$. As above it then follows that
\begin{equation}
	\sum_{a\in A}\sum_{i\in X_a}\partial_{z_{bi}}F=\partial_{z_{b\alpha_1}}F, \qquad \sum_{c\in C}\sum_{i\in X_c}\partial_{z_{ci}}F=\partial_{c\gamma_1}F \,.
\end{equation}
Using this and proceeding as before, we can find
\begin{align}
	&\sum_{a\in A}\sum_{i\in X_a}\partial_{z_{ai}}F=\left[ \sum_{a\in A}\sum_{i\in X_a}\partial_{z_{bi}} +2\sum_{i=1}^{N_1} b_i-N_1 \right]F \,, \\
	&\sum_{c\in C}\sum_{j\in X_c}\partial_{z_{cj}}F=\left[ \sum_{c\in C}\sum_{j\in X_c}\partial_{z_{bj}} +2\sum_{i=1}^{N_2} b_i'-N_2 \right]F \,.
\end{align}
These equations allow us to compute
\begin{align}
	&\sum_{a\in A}\sum_{c\in C}\sum_{i\in X_a}\sum_{j\in X_c}\left( \partial_{z_{0j}}\partial_{z_{ai}}-\partial_{z_{0i}}\partial_{z_{cj}} \right) F \nonumber\\
	&= \sum_{a\in A}\sum_{c\in C}\sum_{i\in X_a}\sum_{j\in X_c}\left( \partial_{z_{0j}}\partial_{z_{bi}}-\partial_{z_{0i}}\partial_{z_{bj}} \right) F \\
	&\qquad +\left( 2\sum_{i=1}^{N_1}b_i-N_1 \right)\sum_{c\in C}\sum_{j\in X_c}\partial_{z_{0j}}F 
	-\left( 2\sum_{i=1}^{N_2}b_i'-N_2 \right)\sum_{a\in A}\sum_{j\in X_a}\partial_{z_{0j}}F \nonumber \,.
\end{align}
We can identify the left hand side as the level-one momentum generator $\levo{P}_{AC}$ acting on the subtrees $A,C$. The first term on the right hand side vanishes by \eqref{eq:AGDE3}, while the other two sums yield the momentum generators $\gen{P}_C, \gen{P}_A$ acting on the subtrees $C$ and $A$, respectively. 

\end{appendix}


\bibliographystyle{nb}
\bibliography{1DTreeIntegrals}

\end{document}